\documentclass[showpacs,amssymb,notitlepage,pre,8pt,nofootinbib,eqsecnum,letter]{revtex4-1}

\usepackage{bm}

\usepackage[english]{babel}
\usepackage{amsmath,amsfonts,amssymb,amsthm,latexsym}

\usepackage{verbatim}
\usepackage[utf8]{inputenc}
\usepackage{pst-all}
\usepackage{pstricks}
\usepackage[dvips]{graphicx}
\usepackage{psfrag}
\usepackage{subfig}
\usepackage{enumitem}
\usepackage{cleveref}
\usepackage{color}

\newcommand{\parde}[3]{\ensuremath{\frac{\partial^{#1}{#2}}{\partial{#3}^{#1}}}}

\newcommand{\suchthat}{\mathrel{\ooalign{$\ni$\cr\kern-1pt$-$\kern-6.5pt$-$}}}

\newpsobject{grilla}{psgrid}{subgriddiv=1,griddots=10,gridlabels=6pt}

\def\({\left(}
\def\){\right)}
\def\[{\left[}
\def\]{\right]}
\def\|l{\left|}
\def\|{\right|}
\def\<{\left <}
\def\>{\right>}
\newcommand{\Rpar}[1]{\left(#1\right)}

\newcommand{\Bpar}[1]{\left[#1\right]}
\newcommand{\Vpar}[1]{\left\vert #1\right\vert}
\newcommand{\VVpar}[1]{\left\Vert #1\right\Vert}

\newcommand{\Kpar}[1]{\left\{#1\right\}}
\newcommand{\Whole}[1]{\left\lfloor #1\right\rfloor}
\newcommand{\Floor}[1]{\Whole{#1}}
\newcommand{\Ceil}[1]{\left\lceil #1\right\rceil}
\renewcommand{\r}{\mathbf{r}}

\newcommand{\ceil}[1]{\left\lceil #1 \right\rceil}

\def\emp{\begin{equation}}
\def\emps{\begin{equation*}}
\def\fin{\end{equation}} 
\def\fins{\end{equation*}}
\def\eemp{\begin{equation}\begin{aligned}} 
\def\eemps{\begin{equation*}\begin{aligned}}
\def\ffin{\end{aligned}\end{equation}} 
\def\ffins{\end{aligned}\end{equation*}}
\newcommand{\ii}{\hat{\imath}}
\newcommand{\jj}{\hat{\jmath}}

\newcommand{\VEC}[1]{\mathbf{#1}}
\newcommand{\Ttilde}[1]{\widetilde{\mathbf{#1}}}

\newcommand{\defeq}{\mathrel{\mathop:}=}

\newcommand{\dif}{\text{d}}

\newcommand{\Mean}[1]{\left\langle#1\right\rangle}
\newcommand{\Mat}[2]{\mathsf{#1}_{#2}}

\newcommand{\eff}{\text{eff}}

\newcommand{\gammafun}[1]{\mathbf{\Gamma}\Rpar{#1}}

\newcommand{\superscript}[1]{\ensuremath{^{\textrm{#1}}}}

\renewcommand{\TH}[0]{\superscript{th}}
\newcommand{\ST}[0]{\superscript{st}}

\newcommand{\ImTc}[5]{\begin{figure}[htp!]
\centering
\includegraphics[width=#1\textwidth]{#2.eps}\\
\caption[#5]{\small #4}
\label{fig:#3}
\end{figure}}

\newcommand{\ImT}[4]{\begin{figure}[htp!]
\begin{minipage}[c]{0.80\linewidth}
\centering
\includegraphics[width=#1\textwidth]{#2}\\
\caption{\small #4}
\label{fig:#3}
\end{minipage}
\end{figure}}

\newcommand{\ImTR}[5]{\begin{figure}[h!]
\begin{minipage}[c]{0.80\linewidth}
\centering
\includegraphics[width=#1\textwidth,angle=#5]{#2}\\
\caption{\small #4}
\label{fig:#3}
\end{minipage}
\end{figure}}

\newcommand{\IImT}[9]{%
    \def\tempa{#1}%
    \def\tempb{#2}%
    \def\tempc{#3}%
    \def\tempd{#4}%
    \def\tempe{#5}%
    \def\tempf{#6}%
    \def\tempg{#7}%
    \def\temph{#8}%
    \def\tempi{#9}%
    \IImTcontinued
}
\newcommand{\IImTcontinued}[1]{\begin{figure}
  \centering
  \subfloat[][\tempd]{%
  \includegraphics[width=\tempa\linewidth]{\tempb.eps}
  \label{fig:\tempc}}%
  \quad%
  \subfloat[][\temph]{%
  \includegraphics[width=\tempe\linewidth]{\tempf.eps}
  \label{fig:\tempg}}%
  \caption{\small #1}
  \label{fig:\tempi}
\end{figure}}

\definecolor{orange}{rgb}{1,0.5,0}
\definecolor{OliveGreen}{rgb}{0,0.6,0}
\definecolor{lightblue}{rgb}{.90,.95,1}
\definecolor{darkblue}{rgb}{0,0,0.5}

\definecolor{airforceblue}{rgb}{0.36, 0.54, 0.66}
\definecolor{aliceblue}{rgb}{0.94, 0.97, 1.0}
\definecolor{alizarin}{rgb}{0.82, 0.1, 0.26}
\definecolor{almond}{rgb}{0.94, 0.87, 0.8}
\definecolor{amaranth}{rgb}{0.9, 0.17, 0.31}
\definecolor{amber}{rgb}{1.0, 0.75, 0.0}
\definecolor{amber(sae/ece)}{rgb}{1.0, 0.49, 0.0}
\definecolor{americanrose}{rgb}{1.0, 0.01, 0.24}
\definecolor{amethyst}{rgb}{0.6, 0.4, 0.8}
\definecolor{anti-flashwhite}{rgb}{0.95, 0.95, 0.96}
\definecolor{antiquebrass}{rgb}{0.8, 0.58, 0.46}
\definecolor{antiquefuchsia}{rgb}{0.57, 0.36, 0.51}
\definecolor{antiquewhite}{rgb}{0.98, 0.92, 0.84}
\definecolor{ao}{rgb}{0.0, 0.0, 1.0}
\definecolor{ao(english)}{rgb}{0.0, 0.5, 0.0}
\definecolor{applegreen}{rgb}{0.55, 0.71, 0.0}
\definecolor{apricot}{rgb}{0.98, 0.81, 0.69}
\definecolor{aqua}{rgb}{0.0, 1.0, 1.0}
\definecolor{aquamarine}{rgb}{0.5, 1.0, 0.83}
\definecolor{armygreen}{rgb}{0.29, 0.33, 0.13}
\definecolor{arsenic}{rgb}{0.23, 0.27, 0.29}
\definecolor{arylideyellow}{rgb}{0.91, 0.84, 0.42}
\definecolor{ashgrey}{rgb}{0.7, 0.75, 0.71}
\definecolor{asparagus}{rgb}{0.53, 0.66, 0.42}
\definecolor{atomictangerine}{rgb}{1.0, 0.6, 0.4}
\definecolor{auburn}{rgb}{0.43, 0.21, 0.1}
\definecolor{aureolin}{rgb}{0.99, 0.93, 0.0}
\definecolor{aurometalsaurus}{rgb}{0.43, 0.5, 0.5}
\definecolor{awesome}{rgb}{1.0, 0.13, 0.32}
\definecolor{azure(colorwheel)}{rgb}{0.0, 0.5, 1.0}
\definecolor{azure(web)(azuremist)}{rgb}{0.94, 1.0, 1.0}
\definecolor{babyblue}{rgb}{0.54, 0.81, 0.94}
\definecolor{babyblueeyes}{rgb}{0.63, 0.79, 0.95}
\definecolor{babypink}{rgb}{0.96, 0.76, 0.76}
\definecolor{ballblue}{rgb}{0.13, 0.67, 0.8}
\definecolor{bananamania}{rgb}{0.98, 0.91, 0.71}
\definecolor{bananayellow}{rgb}{1.0, 0.88, 0.21}
\definecolor{battleshipgrey}{rgb}{0.52, 0.52, 0.51}
\definecolor{bazaar}{rgb}{0.6, 0.47, 0.48}
\definecolor{beaublue}{rgb}{0.74, 0.83, 0.9}
\definecolor{beaver}{rgb}{0.62, 0.51, 0.44}
\definecolor{beige}{rgb}{0.96, 0.96, 0.86}
\definecolor{bisque}{rgb}{1.0, 0.89, 0.77}
\definecolor{bistre}{rgb}{0.24, 0.17, 0.12}
\definecolor{bittersweet}{rgb}{1.0, 0.44, 0.37}
\definecolor{black}{rgb}{0.0, 0.0, 0.0}
\definecolor{blanchedalmond}{rgb}{1.0, 0.92, 0.8}
\definecolor{bleudefrance}{rgb}{0.19, 0.55, 0.91}
\definecolor{blizzardblue}{rgb}{0.67, 0.9, 0.93}
\definecolor{blond}{rgb}{0.98, 0.94, 0.75}
\definecolor{blue}{rgb}{0.0, 0.0, 1.0}
\definecolor{blue(munsell)}{rgb}{0.0, 0.5, 0.69}
\definecolor{blue(ncs)}{rgb}{0.0, 0.53, 0.74}
\definecolor{blue(pigment)}{rgb}{0.2, 0.2, 0.6}
\definecolor{blue(ryb)}{rgb}{0.01, 0.28, 1.0}
\definecolor{bluebell}{rgb}{0.64, 0.64, 0.82}
\definecolor{bluegray}{rgb}{0.4, 0.6, 0.8}
\definecolor{blue-green}{rgb}{0.0, 0.87, 0.87}
\definecolor{blue-violet}{rgb}{0.54, 0.17, 0.89}
\definecolor{blush}{rgb}{0.87, 0.36, 0.51}
\definecolor{bole}{rgb}{0.47, 0.27, 0.23}
\definecolor{bondiblue}{rgb}{0.0, 0.58, 0.71}
\definecolor{bostonuniversityred}{rgb}{0.8, 0.0, 0.0}
\definecolor{brandeisblue}{rgb}{0.0, 0.44, 1.0}
\definecolor{brass}{rgb}{0.71, 0.65, 0.26}
\definecolor{brickred}{rgb}{0.8, 0.25, 0.33}
\definecolor{brightcerulean}{rgb}{0.11, 0.67, 0.84}
\definecolor{brightgreen}{rgb}{0.4, 1.0, 0.0}
\definecolor{brightlavender}{rgb}{0.75, 0.58, 0.89}
\definecolor{brightmaroon}{rgb}{0.76, 0.13, 0.28}
\definecolor{brightpink}{rgb}{1.0, 0.0, 0.5}
\definecolor{brightturquoise}{rgb}{0.03, 0.91, 0.87}
\definecolor{brightube}{rgb}{0.82, 0.62, 0.91}
\definecolor{brilliantlavender}{rgb}{0.96, 0.73, 1.0}
\definecolor{brilliantrose}{rgb}{1.0, 0.33, 0.64}
\definecolor{brinkpink}{rgb}{0.98, 0.38, 0.5}
\definecolor{britishracinggreen}{rgb}{0.0, 0.26, 0.15}
\definecolor{bronze}{rgb}{0.8, 0.5, 0.2}
\definecolor{brown(traditional)}{rgb}{0.59, 0.29, 0.0}
\definecolor{brown(web)}{rgb}{0.65, 0.16, 0.16}
\definecolor{bubblegum}{rgb}{0.99, 0.76, 0.8}
\definecolor{bubbles}{rgb}{0.91, 1.0, 1.0}
\definecolor{buff}{rgb}{0.94, 0.86, 0.51}
\definecolor{bulgarianrose}{rgb}{0.28, 0.02, 0.03}
\definecolor{burgundy}{rgb}{0.5, 0.0, 0.13}
\definecolor{burlywood}{rgb}{0.87, 0.72, 0.53}
\definecolor{burntorange}{rgb}{0.8, 0.33, 0.0}
\definecolor{burntsienna}{rgb}{0.91, 0.45, 0.32}
\definecolor{burntumber}{rgb}{0.54, 0.2, 0.14}
\definecolor{byzantine}{rgb}{0.74, 0.2, 0.64}
\definecolor{byzantium}{rgb}{0.44, 0.16, 0.39}
\definecolor{cadet}{rgb}{0.33, 0.41, 0.47}
\definecolor{cadetblue}{rgb}{0.37, 0.62, 0.63}
\definecolor{cadetgrey}{rgb}{0.57, 0.64, 0.69}
\definecolor{cadmiumgreen}{rgb}{0.0, 0.42, 0.24}
\definecolor{cadmiumorange}{rgb}{0.93, 0.53, 0.18}
\definecolor{cadmiumred}{rgb}{0.89, 0.0, 0.13}
\definecolor{cadmiumyellow}{rgb}{1.0, 0.96, 0.0}
\definecolor{calpolypomonagreen}{rgb}{0.12, 0.3, 0.17}
\definecolor{cambridgeblue}{rgb}{0.64, 0.76, 0.68}
\definecolor{camel}{rgb}{0.76, 0.6, 0.42}
\definecolor{camouflagegreen}{rgb}{0.47, 0.53, 0.42}
\definecolor{canaryyellow}{rgb}{1.0, 0.94, 0.0}
\definecolor{candyapplered}{rgb}{1.0, 0.03, 0.0}
\definecolor{candypink}{rgb}{0.89, 0.44, 0.48}
\definecolor{capri}{rgb}{0.0, 0.75, 1.0}
\definecolor{caputmortuum}{rgb}{0.35, 0.15, 0.13}
\definecolor{cardinal}{rgb}{0.77, 0.12, 0.23}
\definecolor{caribbeangreen}{rgb}{0.0, 0.8, 0.6}
\definecolor{carmine}{rgb}{0.59, 0.0, 0.09}
\definecolor{carminepink}{rgb}{0.92, 0.3, 0.26}
\definecolor{carminered}{rgb}{1.0, 0.0, 0.22}
\definecolor{carnationpink}{rgb}{1.0, 0.65, 0.79}
\definecolor{carnelian}{rgb}{0.7, 0.11, 0.11}
\definecolor{carolinablue}{rgb}{0.6, 0.73, 0.89}
\definecolor{carrotorange}{rgb}{0.93, 0.57, 0.13}
\definecolor{ceil}{rgb}{0.57, 0.63, 0.81}
\definecolor{celadon}{rgb}{0.67, 0.88, 0.69}
\definecolor{celestialblue}{rgb}{0.29, 0.59, 0.82}
\definecolor{cerise}{rgb}{0.87, 0.19, 0.39}
\definecolor{cerisepink}{rgb}{0.93, 0.23, 0.51}
\definecolor{cerulean}{rgb}{0.0, 0.48, 0.65}
\definecolor{ceruleanblue}{rgb}{0.16, 0.32, 0.75}
\definecolor{chamoisee}{rgb}{0.63, 0.47, 0.35}
\definecolor{champagne}{rgb}{0.97, 0.91, 0.81}
\definecolor{charcoal}{rgb}{0.21, 0.27, 0.31}
\definecolor{chartreuse(traditional)}{rgb}{0.87, 1.0, 0.0}
\definecolor{chartreuse(web)}{rgb}{0.5, 1.0, 0.0}
\definecolor{cherryblossompink}{rgb}{1.0, 0.72, 0.77}
\definecolor{chestnut}{rgb}{0.8, 0.36, 0.36}
\definecolor{chocolate(traditional)}{rgb}{0.48, 0.25, 0.0}
\definecolor{chocolate(web)}{rgb}{0.82, 0.41, 0.12}
\definecolor{chromeyellow}{rgb}{1.0, 0.65, 0.0}
\definecolor{cinereous}{rgb}{0.6, 0.51, 0.48}
\definecolor{cinnabar}{rgb}{0.89, 0.26, 0.2}
\definecolor{cinnamon}{rgb}{0.82, 0.41, 0.12}
\definecolor{citrine}{rgb}{0.89, 0.82, 0.04}
\definecolor{classicrose}{rgb}{0.98, 0.8, 0.91}
\definecolor{cobalt}{rgb}{0.0, 0.28, 0.67}
\definecolor{cocoabrown}{rgb}{0.82, 0.41, 0.12}
\definecolor{columbiablue}{rgb}{0.61, 0.87, 1.0}
\definecolor{coolblack}{rgb}{0.0, 0.18, 0.39}
\definecolor{coolgrey}{rgb}{0.55, 0.57, 0.67}
\definecolor{copper}{rgb}{0.72, 0.45, 0.2}
\definecolor{copperrose}{rgb}{0.6, 0.4, 0.4}
\definecolor{coquelicot}{rgb}{1.0, 0.22, 0.0}
\definecolor{coral}{rgb}{1.0, 0.5, 0.31}
\definecolor{coralpink}{rgb}{0.97, 0.51, 0.47}
\definecolor{coralred}{rgb}{1.0, 0.25, 0.25}
\definecolor{cordovan}{rgb}{0.54, 0.25, 0.27}
\definecolor{corn}{rgb}{0.98, 0.93, 0.36}
\definecolor{cornellred}{rgb}{0.7, 0.11, 0.11}
\definecolor{cornflowerblue}{rgb}{0.39, 0.58, 0.93}
\definecolor{cornsilk}{rgb}{1.0, 0.97, 0.86}
\definecolor{cosmiclatte}{rgb}{1.0, 0.97, 0.91}
\definecolor{cottoncandy}{rgb}{1.0, 0.74, 0.85}
\definecolor{cream}{rgb}{1.0, 0.99, 0.82}
\definecolor{crimson}{rgb}{0.86, 0.08, 0.24}
\definecolor{crimsonglory}{rgb}{0.75, 0.0, 0.2}
\definecolor{cyan}{rgb}{0.0, 1.0, 1.0}
\definecolor{cyan(process)}{rgb}{0.0, 0.72, 0.92}
\definecolor{daffodil}{rgb}{1.0, 1.0, 0.19}
\definecolor{dandelion}{rgb}{0.94, 0.88, 0.19}
\definecolor{darkblue}{rgb}{0.0, 0.0, 0.55}
\definecolor{darkbrown}{rgb}{0.4, 0.26, 0.13}
\definecolor{darkbyzantium}{rgb}{0.36, 0.22, 0.33}
\definecolor{darkcandyapplered}{rgb}{0.64, 0.0, 0.0}
\definecolor{darkcerulean}{rgb}{0.03, 0.27, 0.49}
\definecolor{darkchampagne}{rgb}{0.76, 0.7, 0.5}
\definecolor{darkchestnut}{rgb}{0.6, 0.41, 0.38}
\definecolor{darkcoral}{rgb}{0.8, 0.36, 0.27}
\definecolor{darkcyan}{rgb}{0.0, 0.55, 0.55}
\definecolor{darkelectricblue}{rgb}{0.33, 0.41, 0.47}
\definecolor{darkgoldenrod}{rgb}{0.72, 0.53, 0.04}
\definecolor{darkgray}{rgb}{0.66, 0.66, 0.66}
\definecolor{darkgreen}{rgb}{0.0, 0.2, 0.13}
\definecolor{darkjunglegreen}{rgb}{0.1, 0.14, 0.13}
\definecolor{darkkhaki}{rgb}{0.74, 0.72, 0.42}
\definecolor{darklava}{rgb}{0.28, 0.24, 0.2}
\definecolor{darklavender}{rgb}{0.45, 0.31, 0.59}
\definecolor{darkmagenta}{rgb}{0.55, 0.0, 0.55}
\definecolor{darkmidnightblue}{rgb}{0.0, 0.2, 0.4}
\definecolor{darkolivegreen}{rgb}{0.33, 0.42, 0.18}
\definecolor{darkorange}{rgb}{1.0, 0.55, 0.0}
\definecolor{darkorchid}{rgb}{0.6, 0.2, 0.8}
\definecolor{darkpastelblue}{rgb}{0.47, 0.62, 0.8}
\definecolor{darkpastelgreen}{rgb}{0.01, 0.75, 0.24}
\definecolor{darkpastelpurple}{rgb}{0.59, 0.44, 0.84}
\definecolor{darkpastelred}{rgb}{0.76, 0.23, 0.13}
\definecolor{darkpink}{rgb}{0.91, 0.33, 0.5}
\definecolor{darkpowderblue}{rgb}{0.0, 0.2, 0.6}
\definecolor{darkraspberry}{rgb}{0.53, 0.15, 0.34}
\definecolor{darkred}{rgb}{0.55, 0.0, 0.0}
\definecolor{darksalmon}{rgb}{0.91, 0.59, 0.48}
\definecolor{darkscarlet}{rgb}{0.34, 0.01, 0.1}
\definecolor{darkseagreen}{rgb}{0.56, 0.74, 0.56}
\definecolor{darksienna}{rgb}{0.24, 0.08, 0.08}
\definecolor{darkslateblue}{rgb}{0.28, 0.24, 0.55}
\definecolor{darkslategray}{rgb}{0.18, 0.31, 0.31}
\definecolor{darkspringgreen}{rgb}{0.09, 0.45, 0.27}
\definecolor{darktan}{rgb}{0.57, 0.51, 0.32}
\definecolor{darktangerine}{rgb}{1.0, 0.66, 0.07}
\definecolor{darktaupe}{rgb}{0.28, 0.24, 0.2}
\definecolor{darkterracotta}{rgb}{0.8, 0.31, 0.36}
\definecolor{darkturquoise}{rgb}{0.0, 0.81, 0.82}
\definecolor{darkviolet}{rgb}{0.58, 0.0, 0.83}
\definecolor{dartmouthgreen}{rgb}{0.05, 0.5, 0.06}
\definecolor{davysgrey}{rgb}{0.33, 0.33, 0.33}
\definecolor{debianred}{rgb}{0.84, 0.04, 0.33}
\definecolor{deepcarmine}{rgb}{0.66, 0.13, 0.24}
\definecolor{deepcarminepink}{rgb}{0.94, 0.19, 0.22}
\definecolor{deepcarrotorange}{rgb}{0.91, 0.41, 0.17}
\definecolor{deepcerise}{rgb}{0.85, 0.2, 0.53}
\definecolor{deepchampagne}{rgb}{0.98, 0.84, 0.65}
\definecolor{deepchestnut}{rgb}{0.73, 0.31, 0.28}
\definecolor{deepfuchsia}{rgb}{0.76, 0.33, 0.76}
\definecolor{deepjunglegreen}{rgb}{0.0, 0.29, 0.29}
\definecolor{deeplilac}{rgb}{0.6, 0.33, 0.73}
\definecolor{deepmagenta}{rgb}{0.8, 0.0, 0.8}
\definecolor{deeppeach}{rgb}{1.0, 0.8, 0.64}
\definecolor{deeppink}{rgb}{1.0, 0.08, 0.58}
\definecolor{deepsaffron}{rgb}{1.0, 0.6, 0.2}
\definecolor{deepskyblue}{rgb}{0.0, 0.75, 1.0}
\definecolor{denim}{rgb}{0.08, 0.38, 0.74}
\definecolor{desert}{rgb}{0.76, 0.6, 0.42}
\definecolor{desertsand}{rgb}{0.93, 0.79, 0.69}
\definecolor{dimgray}{rgb}{0.41, 0.41, 0.41}
\definecolor{dodgerblue}{rgb}{0.12, 0.56, 1.0}
\definecolor{dogwoodrose}{rgb}{0.84, 0.09, 0.41}
\definecolor{dollarbill}{rgb}{0.52, 0.73, 0.4}
\definecolor{drab}{rgb}{0.59, 0.44, 0.09}
\definecolor{dukeblue}{rgb}{0.0, 0.0, 0.61}
\definecolor{earthyellow}{rgb}{0.88, 0.66, 0.37}
\definecolor{ecru}{rgb}{0.76, 0.7, 0.5}
\definecolor{eggplant}{rgb}{0.38, 0.25, 0.32}
\definecolor{eggshell}{rgb}{0.94, 0.92, 0.84}
\definecolor{egyptianblue}{rgb}{0.06, 0.2, 0.65}
\definecolor{electricblue}{rgb}{0.49, 0.98, 1.0}
\definecolor{electriccrimson}{rgb}{1.0, 0.0, 0.25}
\definecolor{electriccyan}{rgb}{0.0, 1.0, 1.0}
\definecolor{electricgreen}{rgb}{0.0, 1.0, 0.0}
\definecolor{electricindigo}{rgb}{0.44, 0.0, 1.0}
\definecolor{electriclavender}{rgb}{0.96, 0.73, 1.0}
\definecolor{electriclime}{rgb}{0.8, 1.0, 0.0}
\definecolor{electricpurple}{rgb}{0.75, 0.0, 1.0}
\definecolor{electricultramarine}{rgb}{0.25, 0.0, 1.0}
\definecolor{electricviolet}{rgb}{0.56, 0.0, 1.0}
\definecolor{electricyellow}{rgb}{1.0, 1.0, 0.0}
\definecolor{emerald}{rgb}{0.31, 0.78, 0.47}
\definecolor{etonblue}{rgb}{0.59, 0.78, 0.64}
\definecolor{fallow}{rgb}{0.76, 0.6, 0.42}
\definecolor{falured}{rgb}{0.5, 0.09, 0.09}
\definecolor{fandango}{rgb}{0.71, 0.2, 0.54}
\definecolor{fashionfuchsia}{rgb}{0.96, 0.0, 0.63}
\definecolor{fawn}{rgb}{0.9, 0.67, 0.44}
\definecolor{feldgrau}{rgb}{0.3, 0.36, 0.33}
\definecolor{ferngreen}{rgb}{0.31, 0.47, 0.26}
\definecolor{ferrarired}{rgb}{1.0, 0.11, 0.0}
\definecolor{fielddrab}{rgb}{0.42, 0.33, 0.12}
\definecolor{firebrick}{rgb}{0.7, 0.13, 0.13}
\definecolor{fireenginered}{rgb}{0.81, 0.09, 0.13}
\definecolor{flame}{rgb}{0.89, 0.35, 0.13}
\definecolor{flamingopink}{rgb}{0.99, 0.56, 0.67}
\definecolor{flavescent}{rgb}{0.97, 0.91, 0.56}
\definecolor{flax}{rgb}{0.93, 0.86, 0.51}
\definecolor{floralwhite}{rgb}{1.0, 0.98, 0.94}
\definecolor{fluorescentorange}{rgb}{1.0, 0.75, 0.0}
\definecolor{fluorescentpink}{rgb}{1.0, 0.08, 0.58}
\definecolor{fluorescentyellow}{rgb}{0.8, 1.0, 0.0}
\definecolor{folly}{rgb}{1.0, 0.0, 0.31}
\definecolor{forestgreen(traditional)}{rgb}{0.0, 0.27, 0.13}
\definecolor{forestgreen(web)}{rgb}{0.13, 0.55, 0.13}
\definecolor{frenchbeige}{rgb}{0.65, 0.48, 0.36}
\definecolor{frenchblue}{rgb}{0.0, 0.45, 0.73}
\definecolor{frenchlilac}{rgb}{0.53, 0.38, 0.56}
\definecolor{frenchrose}{rgb}{0.96, 0.29, 0.54}
\definecolor{fuchsia}{rgb}{1.0, 0.0, 1.0}
\definecolor{fuchsiapink}{rgb}{1.0, 0.47, 1.0}
\definecolor{fulvous}{rgb}{0.86, 0.52, 0.0}
\definecolor{fuzzywuzzy}{rgb}{0.8, 0.4, 0.4}
\definecolor{gainsboro}{rgb}{0.86, 0.86, 0.86}
\definecolor{gamboge}{rgb}{0.89, 0.61, 0.06}
\definecolor{ghostwhite}{rgb}{0.97, 0.97, 1.0}
\definecolor{ginger}{rgb}{0.69, 0.4, 0.0}
\definecolor{glaucous}{rgb}{0.38, 0.51, 0.71}
\definecolor{gold(metallic)}{rgb}{0.83, 0.69, 0.22}
\definecolor{gold(web)(golden)}{rgb}{1.0, 0.84, 0.0}
\definecolor{goldenbrown}{rgb}{0.6, 0.4, 0.08}
\definecolor{goldenpoppy}{rgb}{0.99, 0.76, 0.0}
\definecolor{goldenyellow}{rgb}{1.0, 0.87, 0.0}
\definecolor{goldenrod}{rgb}{0.85, 0.65, 0.13}
\definecolor{grannysmithapple}{rgb}{0.66, 0.89, 0.63}
\definecolor{gray}{rgb}{0.5, 0.5, 0.5}
\definecolor{gray(html/cssgray)}{rgb}{0.5, 0.5, 0.5}
\definecolor{gray(x11gray)}{rgb}{0.75, 0.75, 0.75}
\definecolor{gray-asparagus}{rgb}{0.27, 0.35, 0.27}
\definecolor{green(colorwheel)(x11green)}{rgb}{0.0, 1.0, 0.0}
\definecolor{green(html/cssgreen)}{rgb}{0.0, 0.5, 0.0}
\definecolor{green(munsell)}{rgb}{0.0, 0.66, 0.47}
\definecolor{green(ncs)}{rgb}{0.0, 0.62, 0.42}
\definecolor{green(pigment)}{rgb}{0.0, 0.65, 0.31}
\definecolor{green(ryb)}{rgb}{0.4, 0.69, 0.2}
\definecolor{green-yellow}{rgb}{0.68, 1.0, 0.18}
\definecolor{grullo}{rgb}{0.66, 0.6, 0.53}
\definecolor{guppiegreen}{rgb}{0.0, 1.0, 0.5}
\definecolor{halayaube}{rgb}{0.4, 0.22, 0.33}
\definecolor{hanblue}{rgb}{0.27, 0.42, 0.81}
\definecolor{hanpurple}{rgb}{0.32, 0.09, 0.98}
\definecolor{hansayellow}{rgb}{0.91, 0.84, 0.42}
\definecolor{harlequin}{rgb}{0.25, 1.0, 0.0}
\definecolor{harvardcrimson}{rgb}{0.79, 0.0, 0.09}
\definecolor{harvestgold}{rgb}{0.85, 0.57, 0.0}
\definecolor{heartgold}{rgb}{0.5, 0.5, 0.0}
\definecolor{heliotrope}{rgb}{0.87, 0.45, 1.0}
\definecolor{hollywoodcerise}{rgb}{0.96, 0.0, 0.63}
\definecolor{honeydew}{rgb}{0.94, 1.0, 0.94}
\definecolor{hookersgreen}{rgb}{0.0, 0.44, 0.0}
\definecolor{hotmagenta}{rgb}{1.0, 0.11, 0.81}
\definecolor{hotpink}{rgb}{1.0, 0.41, 0.71}
\definecolor{huntergreen}{rgb}{0.21, 0.37, 0.23}
\definecolor{iceberg}{rgb}{0.44, 0.65, 0.82}
\definecolor{icterine}{rgb}{0.99, 0.97, 0.37}
\definecolor{inchworm}{rgb}{0.7, 0.93, 0.36}
\definecolor{indiagreen}{rgb}{0.07, 0.53, 0.03}
\definecolor{indianred}{rgb}{0.8, 0.36, 0.36}
\definecolor{indianyellow}{rgb}{0.89, 0.66, 0.34}
\definecolor{indigo(dye)}{rgb}{0.0, 0.25, 0.42}
\definecolor{indigo(web)}{rgb}{0.29, 0.0, 0.51}
\definecolor{internationalkleinblue}{rgb}{0.0, 0.18, 0.65}
\definecolor{internationalorange}{rgb}{1.0, 0.31, 0.0}
\definecolor{iris}{rgb}{0.35, 0.31, 0.81}
\definecolor{isabelline}{rgb}{0.96, 0.94, 0.93}
\definecolor{islamicgreen}{rgb}{0.0, 0.56, 0.0}
\definecolor{ivory}{rgb}{1.0, 1.0, 0.94}
\definecolor{jade}{rgb}{0.0, 0.66, 0.42}
\definecolor{jasper}{rgb}{0.84, 0.23, 0.24}
\definecolor{jazzberryjam}{rgb}{0.65, 0.04, 0.37}
\definecolor{jonquil}{rgb}{0.98, 0.85, 0.37}
\definecolor{junebud}{rgb}{0.74, 0.85, 0.34}
\definecolor{junglegreen}{rgb}{0.16, 0.67, 0.53}
\definecolor{kellygreen}{rgb}{0.3, 0.73, 0.09}
\definecolor{khaki(html/css)(khaki)}{rgb}{0.76, 0.69, 0.57}
\definecolor{khaki(x11)(lightkhaki)}{rgb}{0.94, 0.9, 0.55}
\definecolor{lasallegreen}{rgb}{0.03, 0.47, 0.19}
\definecolor{languidlavender}{rgb}{0.84, 0.79, 0.87}
\definecolor{lapislazuli}{rgb}{0.15, 0.38, 0.61}
\definecolor{laserlemon}{rgb}{1.0, 1.0, 0.13}
\definecolor{lava}{rgb}{0.81, 0.06, 0.13}
\definecolor{lavender(floral)}{rgb}{0.71, 0.49, 0.86}
\definecolor{lavender(web)}{rgb}{0.9, 0.9, 0.98}
\definecolor{lavenderblue}{rgb}{0.8, 0.8, 1.0}
\definecolor{lavenderblush}{rgb}{1.0, 0.94, 0.96}
\definecolor{lavendergray}{rgb}{0.77, 0.76, 0.82}
\definecolor{lavenderindigo}{rgb}{0.58, 0.34, 0.92}
\definecolor{lavendermagenta}{rgb}{0.93, 0.51, 0.93}
\definecolor{lavendermist}{rgb}{0.9, 0.9, 0.98}
\definecolor{lavenderpink}{rgb}{0.98, 0.68, 0.82}
\definecolor{lavenderpurple}{rgb}{0.59, 0.48, 0.71}
\definecolor{lavenderrose}{rgb}{0.98, 0.63, 0.89}
\definecolor{lawngreen}{rgb}{0.49, 0.99, 0.0}
\definecolor{lemon}{rgb}{1.0, 0.97, 0.0}
\definecolor{lemonchiffon}{rgb}{1.0, 0.98, 0.8}
\definecolor{lightapricot}{rgb}{0.99, 0.84, 0.69}
\definecolor{lightblue}{rgb}{0.68, 0.85, 0.9}
\definecolor{lightbrown}{rgb}{0.71, 0.4, 0.11}
\definecolor{lightcarminepink}{rgb}{0.9, 0.4, 0.38}
\definecolor{lightcoral}{rgb}{0.94, 0.5, 0.5}
\definecolor{lightcornflowerblue}{rgb}{0.6, 0.81, 0.93}
\definecolor{lightcyan}{rgb}{0.88, 1.0, 1.0}
\definecolor{lightfuchsiapink}{rgb}{0.98, 0.52, 0.9}
\definecolor{lightgoldenrodyellow}{rgb}{0.98, 0.98, 0.82}
\definecolor{lightgray}{rgb}{0.83, 0.83, 0.83}
\definecolor{lightgreen}{rgb}{0.56, 0.93, 0.56}
\definecolor{lightkhaki}{rgb}{0.94, 0.9, 0.55}
\definecolor{lightmauve}{rgb}{0.86, 0.82, 1.0}
\definecolor{lightpastelpurple}{rgb}{0.69, 0.61, 0.85}
\definecolor{lightpink}{rgb}{1.0, 0.71, 0.76}
\definecolor{lightsalmon}{rgb}{1.0, 0.63, 0.48}
\definecolor{lightsalmonpink}{rgb}{1.0, 0.6, 0.6}
\definecolor{lightseagreen}{rgb}{0.13, 0.7, 0.67}
\definecolor{lightskyblue}{rgb}{0.53, 0.81, 0.98}
\definecolor{lightslategray}{rgb}{0.47, 0.53, 0.6}
\definecolor{lighttaupe}{rgb}{0.7, 0.55, 0.43}
\definecolor{lightthulianpink}{rgb}{0.9, 0.56, 0.67}
\definecolor{lightyellow}{rgb}{1.0, 1.0, 0.88}
\definecolor{lilac}{rgb}{0.78, 0.64, 0.78}
\definecolor{lime(colorwheel)}{rgb}{0.75, 1.0, 0.0}
\definecolor{lime(web)(x11green)}{rgb}{0.0, 1.0, 0.0}
\definecolor{limegreen}{rgb}{0.2, 0.8, 0.2}
\definecolor{lincolngreen}{rgb}{0.11, 0.35, 0.02}
\definecolor{linen}{rgb}{0.98, 0.94, 0.9}
\definecolor{liver}{rgb}{0.33, 0.29, 0.31}
\definecolor{lust}{rgb}{0.9, 0.13, 0.13}
\definecolor{macaroniandcheese}{rgb}{1.0, 0.74, 0.53}
\definecolor{magenta}{rgb}{1.0, 0.0, 1.0}
\definecolor{magenta(dye)}{rgb}{0.79, 0.08, 0.48}
\definecolor{magenta(process)}{rgb}{1.0, 0.0, 0.56}
\definecolor{magicmint}{rgb}{0.67, 0.94, 0.82}
\definecolor{magnolia}{rgb}{0.97, 0.96, 1.0}
\definecolor{mahogany}{rgb}{0.75, 0.25, 0.0}
\definecolor{maize}{rgb}{0.98, 0.93, 0.37}
\definecolor{majorelleblue}{rgb}{0.38, 0.31, 0.86}
\definecolor{malachite}{rgb}{0.04, 0.85, 0.32}
\definecolor{manatee}{rgb}{0.59, 0.6, 0.67}
\definecolor{mangotango}{rgb}{1.0, 0.51, 0.26}
\definecolor{maroon(html/css)}{rgb}{0.5, 0.0, 0.0}
\definecolor{maroon(x11)}{rgb}{0.69, 0.19, 0.38}
\definecolor{mauve}{rgb}{0.88, 0.69, 1.0}
\definecolor{mauvetaupe}{rgb}{0.57, 0.37, 0.43}
\definecolor{mauvelous}{rgb}{0.94, 0.6, 0.67}
\definecolor{mayablue}{rgb}{0.45, 0.76, 0.98}
\definecolor{meatbrown}{rgb}{0.9, 0.72, 0.23}
\definecolor{mediumaquamarine}{rgb}{0.4, 0.8, 0.67}
\definecolor{mediumblue}{rgb}{0.0, 0.0, 0.8}
\definecolor{mediumcandyapplered}{rgb}{0.89, 0.02, 0.17}
\definecolor{mediumcarmine}{rgb}{0.69, 0.25, 0.21}
\definecolor{mediumchampagne}{rgb}{0.95, 0.9, 0.67}
\definecolor{mediumelectricblue}{rgb}{0.01, 0.31, 0.59}
\definecolor{mediumjunglegreen}{rgb}{0.11, 0.21, 0.18}
\definecolor{mediumlavendermagenta}{rgb}{0.8, 0.6, 0.8}
\definecolor{mediumorchid}{rgb}{0.73, 0.33, 0.83}
\definecolor{mediumpersianblue}{rgb}{0.0, 0.4, 0.65}
\definecolor{mediumpurple}{rgb}{0.58, 0.44, 0.86}
\definecolor{mediumred-violet}{rgb}{0.73, 0.2, 0.52}
\definecolor{mediumseagreen}{rgb}{0.24, 0.7, 0.44}
\definecolor{mediumslateblue}{rgb}{0.48, 0.41, 0.93}
\definecolor{mediumspringbud}{rgb}{0.79, 0.86, 0.54}
\definecolor{mediumspringgreen}{rgb}{0.0, 0.98, 0.6}
\definecolor{mediumtaupe}{rgb}{0.4, 0.3, 0.28}
\definecolor{mediumtealblue}{rgb}{0.0, 0.33, 0.71}
\definecolor{mediumturquoise}{rgb}{0.28, 0.82, 0.8}
\definecolor{mediumviolet-red}{rgb}{0.78, 0.08, 0.52}
\definecolor{melon}{rgb}{0.99, 0.74, 0.71}
\definecolor{midnightblue}{rgb}{0.1, 0.1, 0.44}
\definecolor{midnightgreen(eaglegreen)}{rgb}{0.0, 0.29, 0.33}
\definecolor{mikadoyellow}{rgb}{1.0, 0.77, 0.05}
\definecolor{mint}{rgb}{0.24, 0.71, 0.54}
\definecolor{mintcream}{rgb}{0.96, 1.0, 0.98}
\definecolor{mintgreen}{rgb}{0.6, 1.0, 0.6}
\definecolor{mistyrose}{rgb}{1.0, 0.89, 0.88}
\definecolor{moccasin}{rgb}{0.98, 0.92, 0.84}
\definecolor{modebeige}{rgb}{0.59, 0.44, 0.09}
\definecolor{moonstoneblue}{rgb}{0.45, 0.66, 0.76}
\definecolor{mordantred19}{rgb}{0.68, 0.05, 0.0}
\definecolor{mossgreen}{rgb}{0.68, 0.87, 0.68}
\definecolor{mountainmeadow}{rgb}{0.19, 0.73, 0.56}
\definecolor{mountbattenpink}{rgb}{0.6, 0.48, 0.55}
\definecolor{mulberry}{rgb}{0.77, 0.29, 0.55}
\definecolor{mustard}{rgb}{1.0, 0.86, 0.35}
\definecolor{myrtle}{rgb}{0.13, 0.26, 0.12}
\definecolor{msugreen}{rgb}{0.09, 0.27, 0.23}
\definecolor{nadeshikopink}{rgb}{0.96, 0.68, 0.78}
\definecolor{napiergreen}{rgb}{0.16, 0.5, 0.0}
\definecolor{naplesyellow}{rgb}{0.98, 0.85, 0.37}
\definecolor{navajowhite}{rgb}{1.0, 0.87, 0.68}
\definecolor{navyblue}{rgb}{0.0, 0.0, 0.5}
\definecolor{neoncarrot}{rgb}{1.0, 0.64, 0.26}
\definecolor{neonfuchsia}{rgb}{1.0, 0.25, 0.39}
\definecolor{neongreen}{rgb}{0.22, 0.88, 0.08}
\definecolor{non-photoblue}{rgb}{0.64, 0.87, 0.93}
\definecolor{oceanboatblue}{rgb}{0.0, 0.47, 0.75}
\definecolor{ochre}{rgb}{0.8, 0.47, 0.13}
\definecolor{officegreen}{rgb}{0.0, 0.5, 0.0}
\definecolor{oldgold}{rgb}{0.81, 0.71, 0.23}
\definecolor{oldlace}{rgb}{0.99, 0.96, 0.9}
\definecolor{oldlavender}{rgb}{0.47, 0.41, 0.47}
\definecolor{oldmauve}{rgb}{0.4, 0.19, 0.28}
\definecolor{oldrose}{rgb}{0.75, 0.5, 0.51}
\definecolor{olive}{rgb}{0.5, 0.5, 0.0}
\definecolor{olivedrabN3}{rgb}{0.42, 0.56, 0.14}
\definecolor{olivedrabN7}{rgb}{0.24, 0.2, 0.12}
\definecolor{olivine}{rgb}{0.6, 0.73, 0.45}
\definecolor{onyx}{rgb}{0.06, 0.06, 0.06}
\definecolor{operamauve}{rgb}{0.72, 0.52, 0.65}
\definecolor{orange(colorwheel)}{rgb}{1.0, 0.5, 0.0}
\definecolor{orange(ryb)}{rgb}{0.98, 0.6, 0.01}
\definecolor{orange(webcolor)}{rgb}{1.0, 0.65, 0.0}
\definecolor{orangepeel}{rgb}{1.0, 0.62, 0.0}
\definecolor{orange-red}{rgb}{1.0, 0.27, 0.0}
\definecolor{orchid}{rgb}{0.85, 0.44, 0.84}
\definecolor{otterbrown}{rgb}{0.4, 0.26, 0.13}
\definecolor{outerspace}{rgb}{0.25, 0.29, 0.3}
\definecolor{outrageousorange}{rgb}{1.0, 0.43, 0.29}
\definecolor{oxfordblue}{rgb}{0.0, 0.13, 0.28}
\definecolor{oucrimsonred}{rgb}{0.6, 0.0, 0.0}
\definecolor{pakistangreen}{rgb}{0.0, 0.4, 0.0}
\definecolor{palatinateblue}{rgb}{0.15, 0.23, 0.89}
\definecolor{palatinatepurple}{rgb}{0.41, 0.16, 0.38}
\definecolor{paleaqua}{rgb}{0.74, 0.83, 0.9}
\definecolor{paleblue}{rgb}{0.69, 0.93, 0.93}
\definecolor{palebrown}{rgb}{0.6, 0.46, 0.33}
\definecolor{palecarmine}{rgb}{0.69, 0.25, 0.21}
\definecolor{palecerulean}{rgb}{0.61, 0.77, 0.89}
\definecolor{palechestnut}{rgb}{0.87, 0.68, 0.69}
\definecolor{palecopper}{rgb}{0.85, 0.54, 0.4}
\definecolor{palecornflowerblue}{rgb}{0.67, 0.8, 0.94}
\definecolor{palegold}{rgb}{0.9, 0.75, 0.54}
\definecolor{palegoldenrod}{rgb}{0.93, 0.91, 0.67}
\definecolor{palegreen}{rgb}{0.6, 0.98, 0.6}
\definecolor{palemagenta}{rgb}{0.98, 0.52, 0.9}
\definecolor{palepink}{rgb}{0.98, 0.85, 0.87}
\definecolor{paleplum}{rgb}{0.8, 0.6, 0.8}
\definecolor{palered-violet}{rgb}{0.86, 0.44, 0.58}
\definecolor{palerobineggblue}{rgb}{0.59, 0.87, 0.82}
\definecolor{palesilver}{rgb}{0.79, 0.75, 0.73}
\definecolor{palespringbud}{rgb}{0.93, 0.92, 0.74}
\definecolor{paletaupe}{rgb}{0.74, 0.6, 0.49}
\definecolor{paleviolet-red}{rgb}{0.86, 0.44, 0.58}
\definecolor{pansypurple}{rgb}{0.47, 0.09, 0.29}
\definecolor{papayawhip}{rgb}{1.0, 0.94, 0.84}
\definecolor{parisgreen}{rgb}{0.31, 0.78, 0.47}
\definecolor{pastelblue}{rgb}{0.68, 0.78, 0.81}
\definecolor{pastelbrown}{rgb}{0.51, 0.41, 0.33}
\definecolor{pastelgray}{rgb}{0.81, 0.81, 0.77}
\definecolor{pastelgreen}{rgb}{0.47, 0.87, 0.47}
\definecolor{pastelmagenta}{rgb}{0.96, 0.6, 0.76}
\definecolor{pastelorange}{rgb}{1.0, 0.7, 0.28}
\definecolor{pastelpink}{rgb}{1.0, 0.82, 0.86}
\definecolor{pastelpurple}{rgb}{0.7, 0.62, 0.71}
\definecolor{pastelred}{rgb}{1.0, 0.41, 0.38}
\definecolor{pastelviolet}{rgb}{0.8, 0.6, 0.79}
\definecolor{pastelyellow}{rgb}{0.99, 0.99, 0.59}
\definecolor{patriarch}{rgb}{0.5, 0.0, 0.5}
\definecolor{paynesgrey}{rgb}{0.25, 0.25, 0.28}
\definecolor{peach}{rgb}{1.0, 0.9, 0.71}
\definecolor{peach-orange}{rgb}{1.0, 0.8, 0.6}
\definecolor{peachpuff}{rgb}{1.0, 0.85, 0.73}
\definecolor{peach-yellow}{rgb}{0.98, 0.87, 0.68}
\definecolor{pear}{rgb}{0.82, 0.89, 0.19}
\definecolor{pearl}{rgb}{0.94, 0.92, 0.84}
\definecolor{peridot}{rgb}{0.9, 0.89, 0.0}
\definecolor{periwinkle}{rgb}{0.8, 0.8, 1.0}
\definecolor{persianblue}{rgb}{0.11, 0.22, 0.73}
\definecolor{persiangreen}{rgb}{0.0, 0.65, 0.58}
\definecolor{persianindigo}{rgb}{0.2, 0.07, 0.48}
\definecolor{persianorange}{rgb}{0.85, 0.56, 0.35}
\definecolor{peru}{rgb}{0.8, 0.52, 0.25}
\definecolor{persianpink}{rgb}{0.97, 0.5, 0.75}
\definecolor{persianplum}{rgb}{0.44, 0.11, 0.11}
\definecolor{persianred}{rgb}{0.8, 0.2, 0.2}
\definecolor{persianrose}{rgb}{1.0, 0.16, 0.64}
\definecolor{persimmon}{rgb}{0.93, 0.35, 0.0}
\definecolor{phlox}{rgb}{0.87, 0.0, 1.0}
\definecolor{phthaloblue}{rgb}{0.0, 0.06, 0.54}
\definecolor{phthalogreen}{rgb}{0.07, 0.21, 0.14}
\definecolor{piggypink}{rgb}{0.99, 0.87, 0.9}
\definecolor{pinegreen}{rgb}{0.0, 0.47, 0.44}
\definecolor{pink}{rgb}{1.0, 0.75, 0.8}
\definecolor{pink-orange}{rgb}{1.0, 0.6, 0.4}
\definecolor{pinkpearl}{rgb}{0.91, 0.67, 0.81}
\definecolor{pinksherbet}{rgb}{0.97, 0.56, 0.65}
\definecolor{pistachio}{rgb}{0.58, 0.77, 0.45}
\definecolor{platinum}{rgb}{0.9, 0.89, 0.89}
\definecolor{plum(traditional)}{rgb}{0.56, 0.27, 0.52}
\definecolor{plum(web)}{rgb}{0.8, 0.6, 0.8}
\definecolor{portlandorange}{rgb}{1.0, 0.35, 0.21}
\definecolor{powderblue(web)}{rgb}{0.69, 0.88, 0.9}
\definecolor{princetonorange}{rgb}{1.0, 0.56, 0.0}
\definecolor{prune}{rgb}{0.44, 0.11, 0.11}
\definecolor{prussianblue}{rgb}{0.0, 0.19, 0.33}
\definecolor{psychedelicpurple}{rgb}{0.87, 0.0, 1.0}
\definecolor{puce}{rgb}{0.8, 0.53, 0.6}
\definecolor{pumpkin}{rgb}{1.0, 0.46, 0.09}
\definecolor{purple(html/css)}{rgb}{0.5, 0.0, 0.5}
\definecolor{purple(munsell)}{rgb}{0.62, 0.0, 0.77}
\definecolor{purple(x11)}{rgb}{0.63, 0.36, 0.94}
\definecolor{purpleheart}{rgb}{0.41, 0.21, 0.61}
\definecolor{purplemountainmajesty}{rgb}{0.59, 0.47, 0.71}
\definecolor{purplepizzazz}{rgb}{1.0, 0.31, 0.85}
\definecolor{purpletaupe}{rgb}{0.31, 0.25, 0.3}
\definecolor{radicalred}{rgb}{1.0, 0.21, 0.37}
\definecolor{raspberry}{rgb}{0.89, 0.04, 0.36}
\definecolor{raspberryglace}{rgb}{0.57, 0.37, 0.43}
\definecolor{raspberrypink}{rgb}{0.89, 0.31, 0.61}
\definecolor{raspberryrose}{rgb}{0.7, 0.27, 0.42}
\definecolor{rawumber}{rgb}{0.51, 0.4, 0.27}
\definecolor{razzledazzlerose}{rgb}{1.0, 0.2, 0.8}
\definecolor{razzmatazz}{rgb}{0.89, 0.15, 0.42}
\definecolor{red}{rgb}{1.0, 0.0, 0.0}
\definecolor{red(munsell)}{rgb}{0.95, 0.0, 0.24}
\definecolor{red(ncs)}{rgb}{0.77, 0.01, 0.2}
\definecolor{red(pigment)}{rgb}{0.93, 0.11, 0.14}
\definecolor{red(ryb)}{rgb}{1.0, 0.15, 0.07}
\definecolor{red-brown}{rgb}{0.65, 0.16, 0.16}
\definecolor{red-violet}{rgb}{0.78, 0.08, 0.52}
\definecolor{redwood}{rgb}{0.67, 0.31, 0.32}
\definecolor{regalia}{rgb}{0.32, 0.18, 0.5}
\definecolor{richblack}{rgb}{0.0, 0.25, 0.25}
\definecolor{richbrilliantlavender}{rgb}{0.95, 0.65, 1.0}
\definecolor{richcarmine}{rgb}{0.84, 0.0, 0.25}
\definecolor{richelectricblue}{rgb}{0.03, 0.57, 0.82}
\definecolor{richlavender}{rgb}{0.67, 0.38, 0.8}
\definecolor{richlilac}{rgb}{0.71, 0.4, 0.82}
\definecolor{richmaroon}{rgb}{0.69, 0.19, 0.38}
\definecolor{riflegreen}{rgb}{0.25, 0.28, 0.2}
\definecolor{robineggblue}{rgb}{0.0, 0.8, 0.8}
\definecolor{rose}{rgb}{1.0, 0.0, 0.5}
\definecolor{rosebonbon}{rgb}{0.98, 0.26, 0.62}
\definecolor{roseebony}{rgb}{0.4, 0.3, 0.28}
\definecolor{rosegold}{rgb}{0.72, 0.43, 0.47}
\definecolor{rosemadder}{rgb}{0.89, 0.15, 0.21}
\definecolor{rosepink}{rgb}{1.0, 0.4, 0.8}
\definecolor{rosequartz}{rgb}{0.67, 0.6, 0.66}
\definecolor{rosetaupe}{rgb}{0.56, 0.36, 0.36}
\definecolor{rosevale}{rgb}{0.67, 0.31, 0.32}
\definecolor{rosewood}{rgb}{0.4, 0.0, 0.04}
\definecolor{rossocorsa}{rgb}{0.83, 0.0, 0.0}
\definecolor{rosybrown}{rgb}{0.74, 0.56, 0.56}
\definecolor{royalazure}{rgb}{0.0, 0.22, 0.66}
\definecolor{royalblue(traditional)}{rgb}{0.0, 0.14, 0.4}
\definecolor{royalblue(web)}{rgb}{0.25, 0.41, 0.88}
\definecolor{royalfuchsia}{rgb}{0.79, 0.17, 0.57}
\definecolor{royalpurple}{rgb}{0.47, 0.32, 0.66}
\definecolor{ruby}{rgb}{0.88, 0.07, 0.37}
\definecolor{ruddy}{rgb}{1.0, 0.0, 0.16}
\definecolor{ruddybrown}{rgb}{0.73, 0.4, 0.16}
\definecolor{ruddypink}{rgb}{0.88, 0.56, 0.59}
\definecolor{rufous}{rgb}{0.66, 0.11, 0.03}
\definecolor{russet}{rgb}{0.5, 0.27, 0.11}
\definecolor{rust}{rgb}{0.72, 0.25, 0.05}
\definecolor{sacramentostategreen}{rgb}{0.0, 0.34, 0.25}
\definecolor{saddlebrown}{rgb}{0.55, 0.27, 0.07}
\definecolor{safetyorange(blazeorange)}{rgb}{1.0, 0.4, 0.0}
\definecolor{saffron}{rgb}{0.96, 0.77, 0.19}
\definecolor{st.patricksblue}{rgb}{0.14, 0.16, 0.48}
\definecolor{salmon}{rgb}{1.0, 0.55, 0.41}
\definecolor{salmonpink}{rgb}{1.0, 0.57, 0.64}
\definecolor{sand}{rgb}{0.76, 0.7, 0.5}
\definecolor{sanddune}{rgb}{0.59, 0.44, 0.09}
\definecolor{sandstorm}{rgb}{0.93, 0.84, 0.25}
\definecolor{sandybrown}{rgb}{0.96, 0.64, 0.38}
\definecolor{sandytaupe}{rgb}{0.59, 0.44, 0.09}
\definecolor{sangria}{rgb}{0.57, 0.0, 0.04}
\definecolor{sapgreen}{rgb}{0.31, 0.49, 0.16}
\definecolor{sapphire}{rgb}{0.03, 0.15, 0.4}
\definecolor{satinsheengold}{rgb}{0.8, 0.63, 0.21}
\definecolor{scarlet}{rgb}{1.0, 0.13, 0.0}
\definecolor{schoolbusyellow}{rgb}{1.0, 0.85, 0.0}
\definecolor{screamingreen}{rgb}{0.46, 1.0, 0.44}
\definecolor{seagreen}{rgb}{0.18, 0.55, 0.34}
\definecolor{sealbrown}{rgb}{0.2, 0.08, 0.08}
\definecolor{seashell}{rgb}{1.0, 0.96, 0.93}
\definecolor{selectiveyellow}{rgb}{1.0, 0.73, 0.0}
\definecolor{sepia}{rgb}{0.44, 0.26, 0.08}
\definecolor{shadow}{rgb}{0.54, 0.47, 0.36}
\definecolor{shamrockgreen}{rgb}{0.0, 0.62, 0.38}
\definecolor{shockingpink}{rgb}{0.99, 0.06, 0.75}
\definecolor{sienna}{rgb}{0.53, 0.18, 0.09}
\definecolor{silver}{rgb}{0.75, 0.75, 0.75}
\definecolor{sinopia}{rgb}{0.8, 0.25, 0.04}
\definecolor{skobeloff}{rgb}{0.0, 0.48, 0.45}
\definecolor{skyblue}{rgb}{0.53, 0.81, 0.92}
\definecolor{skymagenta}{rgb}{0.81, 0.44, 0.69}
\definecolor{slateblue}{rgb}{0.42, 0.35, 0.8}
\definecolor{slategray}{rgb}{0.44, 0.5, 0.56}
\definecolor{smalt(darkpowderblue)}{rgb}{0.0, 0.2, 0.6}
\definecolor{smokeytopaz}{rgb}{0.58, 0.25, 0.03}
\definecolor{smokyblack}{rgb}{0.06, 0.05, 0.03}
\definecolor{snow}{rgb}{1.0, 0.98, 0.98}
\definecolor{spirodiscoball}{rgb}{0.06, 0.75, 0.99}
\definecolor{splashedwhite}{rgb}{1.0, 0.99, 1.0}
\definecolor{springbud}{rgb}{0.65, 0.99, 0.0}
\definecolor{springgreen}{rgb}{0.0, 1.0, 0.5}
\definecolor{steelblue}{rgb}{0.27, 0.51, 0.71}
\definecolor{stildegrainyellow}{rgb}{0.98, 0.85, 0.37}
\definecolor{straw}{rgb}{0.89, 0.85, 0.44}
\definecolor{sunglow}{rgb}{1.0, 0.8, 0.2}
\definecolor{sunset}{rgb}{0.98, 0.84, 0.65}
\definecolor{tan}{rgb}{0.82, 0.71, 0.55}
\definecolor{tangelo}{rgb}{0.98, 0.3, 0.0}
\definecolor{tangerine}{rgb}{0.95, 0.52, 0.0}
\definecolor{tangerineyellow}{rgb}{1.0, 0.8, 0.0}
\definecolor{taupe}{rgb}{0.28, 0.24, 0.2}
\definecolor{taupegray}{rgb}{0.55, 0.52, 0.54}
\definecolor{teagreen}{rgb}{0.82, 0.94, 0.75}
\definecolor{tearose(orange)}{rgb}{0.97, 0.51, 0.47}
\definecolor{tearose(rose)}{rgb}{0.96, 0.76, 0.76}
\definecolor{teal}{rgb}{0.0, 0.5, 0.5}
\definecolor{tealblue}{rgb}{0.21, 0.46, 0.53}
\definecolor{tealgreen}{rgb}{0.0, 0.51, 0.5}
\definecolor{tawny}{rgb}{0.8, 0.34, 0.0}
\definecolor{terracotta}{rgb}{0.89, 0.45, 0.36}
\definecolor{thistle}{rgb}{0.85, 0.75, 0.85}
\definecolor{thulianpink}{rgb}{0.87, 0.44, 0.63}
\definecolor{ticklemepink}{rgb}{0.99, 0.54, 0.67}
\definecolor{tiffanyblue}{rgb}{0.04, 0.73, 0.71}
\definecolor{tigerseye}{rgb}{0.88, 0.55, 0.24}
\definecolor{timberwolf}{rgb}{0.86, 0.84, 0.82}
\definecolor{titaniumyellow}{rgb}{0.93, 0.9, 0.0}
\definecolor{tomato}{rgb}{1.0, 0.39, 0.28}
\definecolor{toolbox}{rgb}{0.45, 0.42, 0.75}
\definecolor{tractorred}{rgb}{0.99, 0.05, 0.21}
\definecolor{trolleygrey}{rgb}{0.5, 0.5, 0.5}
\definecolor{tropicalrainforest}{rgb}{0.0, 0.46, 0.37}
\definecolor{trueblue}{rgb}{0.0, 0.45, 0.81}
\definecolor{tuftsblue}{rgb}{0.28, 0.57, 0.81}
\definecolor{tumbleweed}{rgb}{0.87, 0.67, 0.53}
\definecolor{turkishrose}{rgb}{0.71, 0.45, 0.51}
\definecolor{turquoise}{rgb}{0.19, 0.84, 0.78}
\definecolor{turquoiseblue}{rgb}{0.0, 1.0, 0.94}
\definecolor{turquoisegreen}{rgb}{0.63, 0.84, 0.71}
\definecolor{tuscanred}{rgb}{0.51, 0.21, 0.21}
\definecolor{twilightlavender}{rgb}{0.54, 0.29, 0.42}
\definecolor{tyrianpurple}{rgb}{0.4, 0.01, 0.24}
\definecolor{uablue}{rgb}{0.0, 0.2, 0.67}
\definecolor{uared}{rgb}{0.85, 0.0, 0.3}
\definecolor{ube}{rgb}{0.53, 0.47, 0.76}
\definecolor{uclablue}{rgb}{0.33, 0.41, 0.58}
\definecolor{uclagold}{rgb}{1.0, 0.7, 0.0}
\definecolor{ufogreen}{rgb}{0.24, 0.82, 0.44}
\definecolor{ultramarine}{rgb}{0.07, 0.04, 0.56}
\definecolor{ultramarineblue}{rgb}{0.25, 0.4, 0.96}
\definecolor{ultrapink}{rgb}{1.0, 0.44, 1.0}
\definecolor{umber}{rgb}{0.39, 0.32, 0.28}
\definecolor{unitednationsblue}{rgb}{0.36, 0.57, 0.9}
\definecolor{unmellowyellow}{rgb}{1.0, 1.0, 0.4}
\definecolor{upforestgreen}{rgb}{0.0, 0.27, 0.13}
\definecolor{upmaroon}{rgb}{0.48, 0.07, 0.07}
\definecolor{upsdellred}{rgb}{0.68, 0.09, 0.13}
\definecolor{urobilin}{rgb}{0.88, 0.68, 0.13}
\definecolor{usccardinal}{rgb}{0.6, 0.0, 0.0}
\definecolor{uscgold}{rgb}{1.0, 0.8, 0.0}
\definecolor{utahcrimson}{rgb}{0.83, 0.0, 0.25}
\definecolor{vanilla}{rgb}{0.95, 0.9, 0.67}
\definecolor{vegasgold}{rgb}{0.77, 0.7, 0.35}
\definecolor{venetianred}{rgb}{0.78, 0.03, 0.08}
\definecolor{verdigris}{rgb}{0.26, 0.7, 0.68}
\definecolor{vermilion}{rgb}{0.89, 0.26, 0.2}
\definecolor{veronica}{rgb}{0.63, 0.36, 0.94}
\definecolor{violet}{rgb}{0.56, 0.0, 1.0}
\definecolor{violet(colorwheel)}{rgb}{0.5, 0.0, 1.0}
\definecolor{violet(ryb)}{rgb}{0.53, 0.0, 0.69}
\definecolor{violet(web)}{rgb}{0.93, 0.51, 0.93}
\definecolor{viridian}{rgb}{0.25, 0.51, 0.43}
\definecolor{vividauburn}{rgb}{0.58, 0.15, 0.14}
\definecolor{vividburgundy}{rgb}{0.62, 0.11, 0.21}
\definecolor{vividcerise}{rgb}{0.85, 0.11, 0.51}
\definecolor{vividtangerine}{rgb}{1.0, 0.63, 0.54}
\definecolor{vividviolet}{rgb}{0.62, 0.0, 1.0}
\definecolor{warmblack}{rgb}{0.0, 0.26, 0.26}
\definecolor{wenge}{rgb}{0.39, 0.33, 0.32}
\definecolor{wheat}{rgb}{0.96, 0.87, 0.7}
\definecolor{white}{rgb}{1.0, 1.0, 1.0}
\definecolor{whitesmoke}{rgb}{0.96, 0.96, 0.96}
\definecolor{wildblueyonder}{rgb}{0.64, 0.68, 0.82}
\definecolor{wildstrawberry}{rgb}{1.0, 0.26, 0.64}
\definecolor{wildwatermelon}{rgb}{0.99, 0.42, 0.52}
\definecolor{wisteria}{rgb}{0.79, 0.63, 0.86}
\definecolor{xanadu}{rgb}{0.45, 0.53, 0.47}
\definecolor{yaleblue}{rgb}{0.06, 0.3, 0.57}
\definecolor{yellow}{rgb}{1.0, 1.0, 0.0}
\definecolor{yellow(munsell)}{rgb}{0.94, 0.8, 0.0}
\definecolor{yellow(ncs)}{rgb}{1.0, 0.83, 0.0}
\definecolor{yellow(process)}{rgb}{1.0, 0.94, 0.0}
\definecolor{yellow(ryb)}{rgb}{1.0, 1.0, 0.2}
\definecolor{yellow-green}{rgb}{0.6, 0.8, 0.2}
\definecolor{zaffre}{rgb}{0.0, 0.08, 0.66}
\definecolor{zinnwalditebrown}{rgb}{0.17, 0.09, 0.03}

\newcommand{\juan}[1]{{\color{black} #1}}

\begin{document}

\title{Counter-ion density profile around a charged disk: from the weak to the strong association regime}

\author{Juan Pablo Mallarino}
\email{jp.mallarino50@uniandes.edu.co}
\affiliation{Facultad de Ciencias -- Laboratorio Computacional HPC, Universidad de los Andes - Bogot\'a, Colombia}
\author{Gabriel T\'ellez}
\email{gtellez@uniandes.edu.co}
\affiliation{Departamento de F\'{\i}sica, Universidad de los Andes - Bogot\'a, Colombia}

\date{\today}

\begin{abstract}
We present a comprehensive study of the two dimensional one component plasma in the cell model with charged boundaries. Starting from weak couplings through a convenient approximation of the interacting potential we were able to obtain an analytic formulation to the problem deriving the partition function,  density profile,  contact densities and integrated profiles that compared well with the numerical data from Monte-Carlo simulations. Additionally,  we derived the exact solution for the special cases of $\Xi = 1, 2, 3, \dots$ finding a correspondence between those from weak couplings and the latter. Furthermore,  we investigated the strong coupling regime taking into consideration the Wigner formulation. Elaborating on this,  we obtained the profile to leading order,  computed the contact density values as compared to those derived in an earlier work on the contact theorem. We formulated adequately the strong coupling regime for this system that differed from previous formulations. Ultimately,  we calculated the first order corrections and compared them against numerical results from our simulations with very good agreement; these results compared equally well in the planar limit,  whose results are well known.
\end{abstract}

\maketitle

\section{Introduction}
\label{sec:intro}

In this work we present a thorough analysis of the condensation
phenomenon of counter-ions around a charged disk.  The model under
consideration is a two-dimensional (2D) system formed by an impenetrable
disk of charge $Q_1$ surrounded by ions dispersed freely in a larger
disk with external charged boundary $Q_2$ and no dielectric discontinuities
between the regions delimited by the geometry. The problem resembles an
annulus with particles moving freely between the inner and outer radii
as in \cref{fig:2D_system}. The $N$ ions have, respectively, a charge
$-q$ in such a way that neutrality yields, 
\emp Q_1+Q_2=N\, q.
\label{eq:neutrality}
\fin
We will assume a point-like geometry for the free charges, which is not a problem due to electrostatic repulsion alone. This model is seemingly the one component plasma (2D-OCP) with a small variation. First the neutralizing charge is not distributed homogeneously in the background and second the inner core is impenetrable.

\ImT{0.43}{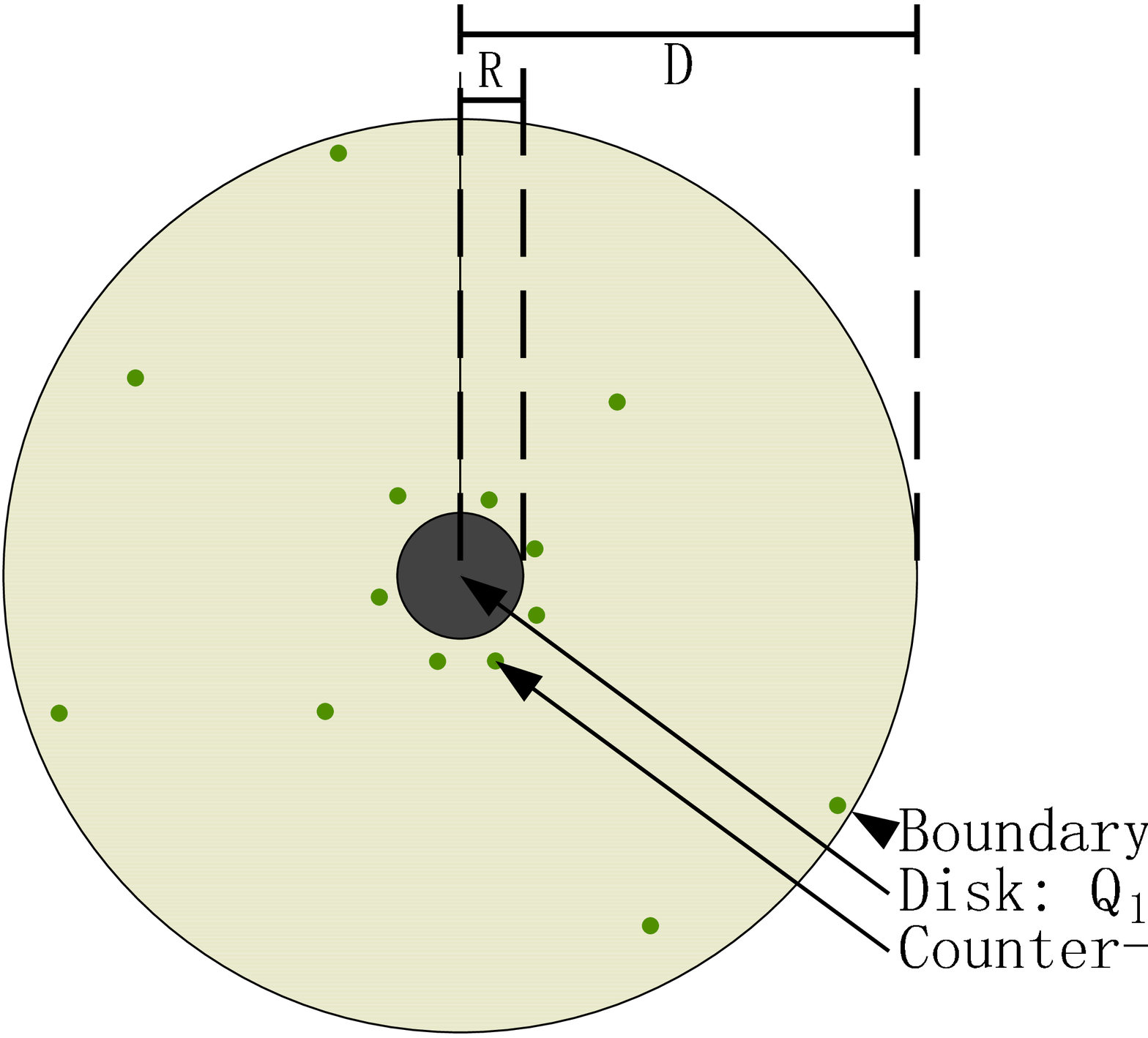}{2D_system}{The 2D cylindrical cell model. The disk with charge $Q_1$ and radius $R$ is surrounded by counter-ions with charge $-q$ within $R$ and the external boundary at $D$ with charge $Q_2$. The interior of the disk and the exterior of the cell have the same dielectric material.}

This is the two-dimensional analog of the Manning counter-ion
condensation phenomenon around charged cylinders~\cite{manning:924,
  manning:934, manning:3249}.  Unlike the three-dimensional (3D) situation
where the Coulomb potential shapes as $1/\Vpar{\VEC{r}}$, the
partition function for two-dimensional Coulomb systems is written as a
product of contributions which, in some cases, may be computed
exactly. That and the logarithmic nature of the potential motivated
the theoretical computation of the abundant static and dynamic
properties of electrolytes for two-dimensional systems.

The interaction between two unit charges separated by a distance $r$
is given by the two-dimensional Coulomb potential $-\log (r/L)$, where
$L$ is an irrelevant arbitrary length scale. We are interested in the
equilibrium thermal properties of the system at a temperature $T$. As
usual, we define $\beta=1/(k_B T)$ where $k_B$ is the Boltzmann
constant. There are three important dimensionless parameters which
caracterize the system. The 2D equivalent of the Manning parameter is
$\xi=\beta q Q_1/2$, which caracterizes the strengh of the Coulomb
coupling between the inner disk and the counter-ions. The Coulomb
coupling between counter-ions is $\Xi=\beta q^2/2$. Here the
familiarized reader may have noticed that for the two dimensional
systems it is accustomed to use $\Gamma$ for the notation of the
coupling constant~\cite{PhysRevE.49.5623, tellez1999exact,
  forrester1983two, 0305-4470-18-9-023, cornu:2444}; the previous defined coupling equates the
standardized one as $\Xi=\Gamma/2$, but we have kept $\Xi$
motivated by a discussion with the three dimensional case.  
Finally the third parameter, defined as $\Delta=\log (D/R)$, which
measures in log-scale the size of the system.

To our problem, we acknowledge the published works from
\citet{PhysRevLett.95.185703,PhysRevE.73.056105} and
\citet{PhysRevE.73.010501}, which have covered thoroughly the basics
of the two dimensional construction of the cell model. They observed
that condensation obeys the well-known threshold at $\xi=1$ where
below this value there is none. \juan{In mean field, it is known that a disk or
cylinder with a dimensionless charge inferior to unity is unable to bind
counter-ions. When the charge is above unity, it attracts an ion cloud 
in such a way that it neutralizes partially the disk/cylinder so that 
the effective dimensionless charge of the disk/cylinder and the cloud 
is unity. Hence, the fraction of condensed ions in the cloud corresponds 
to the excess above unity by the total, also known as the Manning fraction
$f_M=1-1/\xi$.} They also showed that the system with
two dimensions differs from the three-dimensional one in several
ways. First of all, when the outer boundary is uncharged $Q_2=0$, the
relationship between the coupling parameter and the Manning parameter
is dictated by neutrality, which, unlike in 3D where they both enjoyed
independence\footnote{
	The dimensionless parameters in 3D are as follows:
	\emps
	\xi = l_b\lambda q,
	\fins
	\emps
	\Xi = \frac{\xi^2}{\frac{\lambda}{q}R},
	\fins
	with $l_b$ the Bjerrum length, $\lambda$ the cylinder's line charge, $R$ its radii and $q$ the counter-ion charge.
}, it fixes one or the other in such a way that, \emp
\Xi\, =\, \xi/N,
\label{eq:intro_xi_Xi_N}
\fin
as will be discussed in \cref{sec:basics}. 

Additionally, if we were to look at a system with a fixed number of particles going from a temperature above the critical temperature for condensation, i.e. $\xi<1$, and allow it undergo a cooling process, the energy, heat capacity and other quantities will present successive transitions due to iterative localization phenomena occurring when a counter-ion is condensed~\cite{PhysRevLett.95.185703,PhysRevE.73.056105,PhysRevE.73.010501}.

The interesting regime for the two dimensional construction is that
which corresponds to small number of counter-ions. Due to
\cref{eq:intro_xi_Xi_N}, a large number $N$ of ions is equivalent to
$\Xi\to0$ which is unquestionably the mean field regime. Therefore, as
mentioned before by \citet{PhysRevLett.95.185703}, our interest stands
in the range for weak and strong couplings \juan{controlled by the ratio
between $\xi$ and the number of counter-ions.}

The outline for the following work begins with the presentation of the
two dimensional model (Sec.~\ref{sec:basics}), followed in
Sec.~\ref{sec:method1} by an analysis of the regime when $\Xi<1$ (or
$\Gamma<2$), where we extend previous works by
\citet{PhysRevE.73.010501,PhysRevLett.95.185703,PhysRevE.73.056105,PhysRevE.85.011119}. Then
follows, in Secs.~\ref{sec:method2} and \ref{sec:interm}, the study of
the special cases $\Xi=1,2,3,\dots$ (or $\Gamma=2,4,6,\dots$), which
are exactly solvable using expansions of powers of Vandermonde
determinants in basis of symmetric or antisymmetric monomials
\citep{PhysRevE.49.5623,tellez1999exact,vsamaj2004two,tellez2012expanded}. Through
these analytic models we will be able to obtain the profiles and the
energy along with all the quantities associated with
them. Furthermore, we will present a model for condensation
in each case.

Finally, we study, in Sec.~\ref{sec:SC}, the strong coupling
situation when $\Xi\gg1$ (or $\Gamma\gg2$) with special attention to
condensation and evaluate the profile and other statistical
properties. Ultimately, we will retake the contact theorem to obtain
the value of the densities at contact comparing to the values derived
by the models and numerical Monte Carlo simulations.

\section{The model}
\label{sec:basics}

The Hamiltonian for the system considering the logarithmic Coulomb potential interaction between charges as the solution to the Poisson equation $\nabla_{\VEC{r}}^{2}v(\VEC{r})=2\pi\delta(\VEC{r})$ reads as,
\eemp
\mathcal{H}\defeq\beta H=&\beta Q_1q\sum_{j=1}^{N}\log\frac{\vert\VEC{r}_j\vert}{L}+N\beta Q_2q\log\frac{D}{L}-\beta q^2\sum_{1\leq j<k\leq N}\log\left\vert\frac{\VEC{r}_j-\VEC{r}_k}{L}\right\vert\\
&-\beta Q_1Q_2\log\frac{D}{L}-\frac{\beta Q_{1}^{2}}{2}\log\frac{R}{L}-\frac{\beta Q_{2}^{2}}{2}\log\frac{D}{L},
\label{eq:H_orig}
\ffin where $L$ is an arbitrary reference length. The position vector,
with respect to the center of the disk, of the particle number $j$ is
denoted by $\VEC{r}_j$. It will prove convenient to use polar
coordinates $r_j=|\VEC{r}_j|$ and $\theta_j$. \juan{Notice that due to Gauss'
theorem the contribution to the Hamiltonian of the exterior charge $Q_2$
interacting with $Q_1$ and counter-ions is independent of the positions,
yielding a constant. For that matter, this system's behavior depends
on $Q_1-Nq$, which can take arbitrary values.}

\juan{We may introduce the standardized dimensionless notation
  suggested by \citet{PhysRevE.73.056105,PhysRevLett.95.185703} by
  rescaling all distances with the Gouy-Chapmann length ($\mu=R/\xi$),
  i.e. $\widetilde{r}=r/\mu$, but this will only be important
  extending the analysis to the planar limit at which the Gouy length
  is significant; to this problem, all radial distances will be
  divided by either $R$ or $D$. Substituting the Manning parameter as
  $\xi=(\beta\, Q_1\, q/2)$ and the coupling $\Xi=\beta\, q^2/2$ (or
  $\Gamma=\beta q^2$) we obtain, using \cref{eq:neutrality},
\emp
\beta Q_{2}^2 = 2\frac{\Bpar{N\Xi-\xi}^2}{\Xi}
\label{eq:Q2_reduced_units}
\fin
}
\eemp
\mathcal{H}=&2\xi\sum_{j=1}^{N}\log\left\vert\frac{\VEC{r}_j}{R}\right\vert-2\Xi\sum_{1\leq j<k\leq N}\log\left\vert\frac{\VEC{r}_j-\VEC{r}_k}{R}\right\vert+E_0,
\label{eq:H_reduced_units}
\ffin
where,
\emp
E_0=\frac{\Bpar{N\Xi-\xi}^2}{\Xi}\Delta+N\Xi\log\frac{R}{L}\, =\, \frac{\xi_{B}^{2}}{\Xi}\Delta+N\Xi\log\frac{R}{L},
\fin
with $\xi_B$ a parameter equivalent to $\xi$ that speaks of the dimensionless charge accumulated in the exterior boundary. Neutrality reads then in terms of these set of parameters as,
\emp
\xi+\xi_B = N\Xi\quad \text{and}\quad \xi_B=\frac{Q_2}{Q_1}\xi.
\label{eq:neutrality_params}
\fin

A remarkable feature is the relationship, through neutrality with neutral
exterior boundary, between the coupling and the number of particles in
the absence of external charge ($Q_2=0$). Therefore, as in \citet{PhysRevLett.95.185703}, we recover the mean field regime as $\Xi\to0$ (or $\Gamma\to0$) which amounts to say that $N\to\infty$ at constant $\xi$, as was shown by \citet{PhysRevE.73.010501}.

However, the situation for finite $N$ is quite different. \citet{PhysRevLett.95.185703} showed how the energy, heat capacity and the order parameter presented a series of transitions that were absent in the three dimensional case. The underpinnings of this process are at the condensation of ions when reducing the temperature.

In order to see this, we will investigate the problem in three
phases. The first, we will focus on $\Xi<1$ (or $\Gamma<2$) through a
method proposed by \citet{PhysRevE.73.010501,PhysRevE.85.011119}. Then
we will look at the integer $\Xi$ (or even $\Gamma$) cases which admit
a analytic solutions as a prelude to the $\Xi>1$ case, or the strong
coupling regime with some interesting features due to curvature.

In all the different coupling regimes considered, we will compare our
analytical predictions to Monte Carlo simulation data, obtained with a
code developped by one of the authors (JPM). The code uses the
so-called centrifugal sampling~\cite{PhysRevE.73.056105}, that uses
$y=\log(r/R)$ as variable, necessary to sample large box sizes ($D\gg
R$). Also, besides the usual Monte Carlo moves, the codes implements
moves that exchanges particles between the condensed and un-condensed
populations ($y \longmapsto \Delta - y$), necessary to properly sample the
configuration space~\cite{doi:10.1021/jp311873a}. \juan{The ions are point-like
particles of a single type in order to avoid the introduction of hard-core
interactions. Considering ions with size is a perspective
for a future work due to the non-trivial behavior} emerging from steric effects.
Regarding the sampling steps, data was collected after proper thermalization
for as long as $10^8$ steps.

\section{The weakly coupled case or $\Xi<1$}
\label{sec:method1}

Our analysis begins in the $\Xi<1$ case which is better recalled as the weakly coupled case. We indicated that $N\to\infty$ recovers mean field which is not particularly interesting. However, when the number of counter-ions is small and the coupling too, some interesting phenomena occur. This has been described before as a transition due to the condensation of an ion, which is not smooth as in the three dimensional case. In order to see this, let us evaluate the partition function $\mathcal{Z}(\Xi,\xi,N,\Delta)$ assuming that \juan{the box's radius is much larger than that of the disk (large $\Delta$)}. Here, the logarithmic interaction term can be written conveniently \citep{PhysRevE.73.010501,PhysRevE.85.011119},
\eemp
2\log\vert\VEC{r}_1-\VEC{r}_2\vert=&\log\VVpar{\VEC{r}_1}+\log\VVpar{\VEC{r}_2}\\
&+\log\Bpar{2\cosh\Rpar{\log\VVpar{\VEC{r}_1}-\log\VVpar{\VEC{r}_2}}-2\cos\theta_{12}},
\label{eq:log_potential}
\ffin
where $\theta_{12}=\theta_1-\theta_2$. \juan{In the weakly coupled limit 
angular correlations are neglectible thus we can construct an effective
interacting potential averaging over the angle in such a way that,
\emp
\Kpar{2\log\vert\VEC{r}_1-\VEC{r}_2\vert}_{\eff}\defeq\frac{1}{2\pi}\int_{0}^{2\pi}2\log\vert\VEC{r}_1-\VEC{r}_2\vert \dif\theta = 2\log\Vpar{\VEC{r}_>},
\fin
where $\Vpar{\VEC{r}_>}\equiv \text{max}(\Vpar{\VEC{r}_1},\Vpar{\VEC{r}_2})$; consequently,}
\emp
2\log\vert\VEC{r}_1-\VEC{r}_2\vert\simeq2\log\Vpar{\VEC{r}_>}.
\label{eq:log_approx}
\fin
Keep in mind that this approximation is not valid for high couplings
or \emph{small} box sizes where angular correlations are not
negligible (\emph{i.e.} $D\to R$ would represent another problem, a ring of counter-ions in two dimensions). Under these assumptions, the Hamiltonian (\ref{eq:H_reduced_units}) reads,
\emp
\mathcal{H}\approx\frac{2\xi}{\Xi}\sum_{j=1}^{N}y_j-2\sum_{1\leq j<k\leq N}y_{>}^{(j,k)}+E_0,
\fin
with $y_j=\Xi\log(r_j/R)$ and $y_{>}^{(j,k)}$ the greatest of $y_j$
and $y_k$. The partition function is then conveniently written in
terms of the $\{y_k\}$ variables as,\footnote{With the partition
  function defined as
	\emps
	\mathcal{Z}=\frac{1}{N!}\int\dif^{2N}r\, e^{-\mathcal{H}\Rpar{\VEC{r}_1,\VEC{r}_2,\dots,\VEC{r}_N}}.
	\fins
	Notice that the excess free energy is then
	\emps
	\widetilde{F}_{exc}=-\log\mathcal{Z}+\log\Rpar{V^N}
        \,.
	\fins
}
\emp
\mathcal{Z}=\frac{1}{N!}\Rpar{\frac{R^{2}}{\Xi}}^N\int\dif^{N}y
\,\dif^{N}\theta\, e^{-\mathcal{H}+\frac{2}{\Xi}\sum_{j=1}^{N}y_j}.
\label{eq:Z_weak_couplings}
\fin
Via the transformation of coordinates we redefine the energy $\mathcal{H}^\prime\defeq\mathcal{H}-\frac{2}{\Xi}\sum_jy_j\, -E_0$ containing all the important behavior of the system. There is an analytic route to evaluate the partition function as commanded by \citet{PhysRevE.73.010501}. The following procedure follows closely that which was done by the abovementioned authors as part of the presentation of the problem in this context. In order to integrate the Boltzmann factor over the coordinate's phase space we separate the intervals from least to greatest. We know that each partitioning of the phase space that is an ordered arrangement of the $\{y_k\}$'s corresponds to a simple permutation of the \emph{base order} (denoted by {\bf [BO]}): $y_1<y_2<\dots<y_N$. Then, \cref{eq:Z_weak_couplings} yields,
\emp
\mathcal{Z}=e^{-E_0}\Rpar{\frac{2\pi\, R^{2}}{\Xi}}^N\int_{0}^{\Xi\Delta}\dif y_N\int_{0}^{y_N}\dif y_{N-1}\dots\int_{0}^{y_2}\dif y_1\, e^{-\mathcal{H}^\prime}.
\label{eq:Z_weak_couplings_final}
\fin

Burak {\it et. al.} realized that the Boltzmann factor could be written conveniently as a product of functions and the form of the integral involved is a successive convolution of functions. Using the Laplace transformation we can evaluate the partition function analytically. Arranging the set of $\{y_j\}$'s by the {\bf [BO]} $\mathcal{H}^\prime$ reads in simplified form,
\eemp
\mathcal{H}^\prime=\sum_{j=0}^{N}a_j(y_{j+1}-y_j)-a_N\Xi\Delta=\sum_{j=1}^{N}(a_{j-1}-a_j)y_{j},\\
\label{eq:Hprime_weak_couplings_simplified}
\ffin
where $y_0=0$ and $y_{N+1}=\Xi\Delta$, and the positive constants $\{a_j\}$ are conveniently defined as,
\eemp
a_j=\Bpar{j-\Rpar{\frac{\xi-1}{\Xi}+\frac{1}{2}}}^2\\
a_{j-1}-a_j=2\frac{\xi-1}{\Xi}-2(j-1)\\
\label{eq:a_rec_rel}
\ffin

The choice for the set of $\{a_j\}$ is not unique since we have a
total of $N+1$ variables and $N$ conditions; then an arbitrary choice
for any term will define the rest. Anticipating the following steps,
choosing each of the terms positive will be convenient in the calculation of
the Laplace transformation of the partition function. Note that $a_j$
has a minimum sitting on 
\begin{equation}
j_{\text{min}}=\sqrt{a_0}   =
\frac{\xi-1}{\Xi}+\frac{1}{2}
\,,
\end{equation}
thus telling that the smallest term of the set is that which $j$ is the integer closest to $j_{\text{min}}$.

Through this construction we are now able to evaluate the partition function writing the integral as a convolution of functions $\{f_j\}$. From \cref{eq:Z_weak_couplings_final},
\eemp
\mathcal{Z}=&\Rpar{\frac{2\pi\, R^{2}}{\Xi}}^N\, e^{a_N\Xi\Delta-E_0}\\
&\times\int_{0}^{\Xi\Delta}\dif y_N\int_{0}^{y_N}\dif y_{N-1}\dots\int_{0}^{y2}\dif y_1\, \prod_{j=0}^{N}f_j\Rpar{y_{j+1}-y_j}\\
=&\Rpar{\frac{2\pi\, R^{2}}{\Xi}}^N\, e^{a_N\Xi\Delta-E_0}\Bpar{f_N\otimes f_{N-1}\otimes\cdots\otimes f_1\otimes f_0}(\Xi\Delta),
\label{eq:finZ2}
\ffin
\juan{with $f_j(x)=e^{-a_jx}$, whereas the Laplace transform of the convolution part is then,}
\eemp
\mathcal{T}_{\Kpar{f_N\otimes\cdots\otimes f_0}}^{\Bpar{\Xi\Delta}}(s)=&\int_{0}^{\infty}\Bpar{f_N\otimes f_{N-1}\otimes\cdots\otimes f_1\otimes f_0}(\Xi\Delta)\, e^{-s(\Xi\Delta)}\dif(\Xi\Delta)\\
=&\prod_{j=0}^{N}\frac{1}{s+a_j}\\
\label{eq:laplace_Z_simple}
\ffin

The milestone of this analysis is the approximation of the interacting
potential term, valid for a large box size
($\Delta\gg 1$). A small box size surfaces other effects which cannot
be neglected. The route of the simplification comes from transforming
the two-dimensional gas into a one-dimensional mean interacting strip
of ions \citep{PhysRevE.73.010501}. 

As we proceed to invert the Laplace transform to obtain the partition
function, factors of the form $e^{-a_j\Xi\Delta}$ will emerge in the
function where the dominating contributions will come from the
smallest of the $\{a_j\}$'s; this fact enforces choosing a positive
value for $a_j$. Hence, the prevailing term is the smallest of the set,
indicated by $j^\star$, and closest to $\sqrt{a_0}$, or,\footnote{The
  $\Whole{\, }$ and $\Ceil{\, }$ notation is reserved for the floor
  and ceiling functions.}  
\eemp
j_{inf}^{\star}=\Whole{\frac{f_MN}{1+\frac{\xi_B}{\xi}}+\frac{1}{2}}&\leq
j^\star\leq\Ceil{\frac{f_MN}{1+\frac{\xi_B}{\xi}}+\frac{1}{2}}
=j_{sup}^{\star}
.  
\ffin
However, choosing the lower or upper bounds depends if \eemps
a_{j_{inf}^{\star}}\leq a_{j_{sup}^{\star}} \, \, \Rightarrow\, \,
j^\star=\Whole{\frac{f_MN}{1+\frac{\xi_B}{\xi}}+\frac{1}{2}}\\ a_{j_{inf}^{\star}}>a_{j_{sup}^{\star}}
\, \, \Rightarrow\, \,
j^\star=\Ceil{\frac{f_MN}{1+\frac{\xi_B}{\xi}}+\frac{1}{2}}.\\ \ffins

Since $j_{sup}^{\star}=j_{inf}^{\star}+1$ we can rewrite the previous
condition using \cref{eq:a_rec_rel} in such a way that the sign of
$\Rpar{(\xi-1)/\Xi-j_{inf}^{\star}}$ will determine the minimal
parameter; a representation of this situation is given in
\cref{fig:j_star_step}. We can summarize, \eemp
j^{\star}=\Whole{\frac{f_MN}{1+\frac{\xi_B}{\xi}}}+1.
\label{eq:jstar}
\ffin

\ImT{0.3}{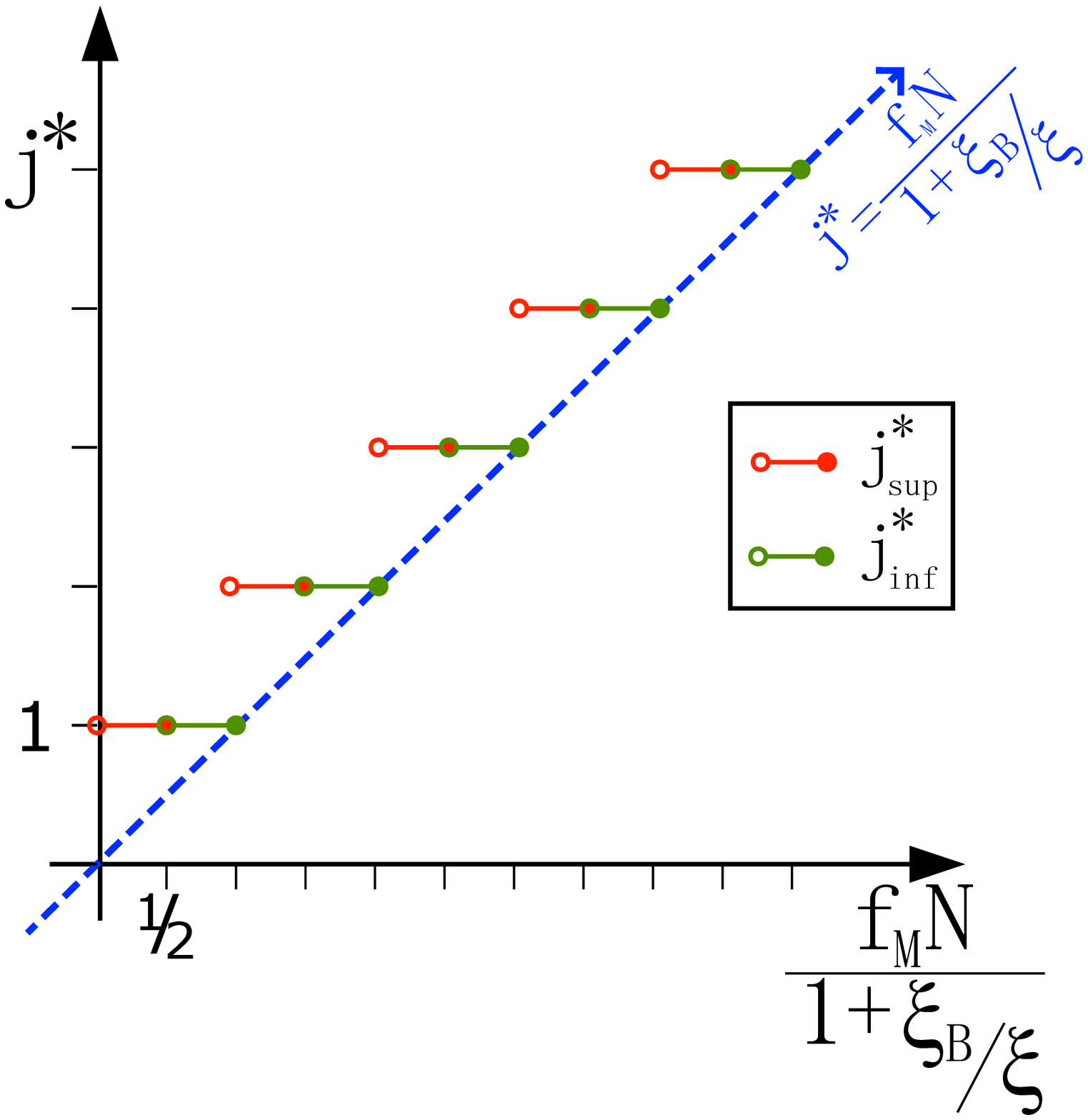}{j_star_step}{The value for $j^\star$ as a function of ${f_MN}/(1+\xi_B/\xi)$. Notice the solution is the composition of the red and the green parts which tells us that $j^\star=\Whole{{f_MN}/(1+\xi_B/\xi)}+1$ and signals the discontinuities precisely at the points where ${f_MN}/(1+\xi_B/\xi)$ is a whole number.}

The mathematical relationship for $j^\star$ hides even so a more important fact about this particular system in relation to condensation. Inspecting the aforementioned equation, $N/(1+{\xi_B}/{\xi})$ corresponds to the {\it number} of counter-ions that neutralize the center disk charge (namely $N_n$); or, by using \cref{eq:neutrality_params}, $N_n = \xi/\Xi = N - \xi_B/\Xi$ (i.e. $N_n = N - 2\xi_B/\Gamma$). Furthermore, we know from previous works in mean field \citep{PhysRevE.73.010501,PhysRevLett.95.185703,PhysRevE.73.056105,doi:10.1021/jp311873a} that the ratio of condensed ions is the well-known Manning relationship $f_M$. In other words, from \cref{eq:jstar},
\eemp
\frac{j^\star}{N_n}\overset{N\to\infty}{=}f_M,
\label{eq:Manning2D-jstar}
\ffin
thus recovering the celebrated Manning condensation fraction in the thermodynamic limit. As a result, $j^\star$ will be intrinsically related to the number of condensed ions; this will be further clarified in the following sections.

\subsection{The partition function}
\label{subsec:partition_Z}

In order to invert the transform for the partition function
$\mathcal{Z}$ we need to revise three possible cases according to the
set of $\Kpar{a_j}$. These cases depend on the values of $\Xi$ and
$\xi$. In fact, given that the $a_j=(j-c)^2$, with
$c=\frac{\xi-1}{\Xi}+\frac{1}{2}$, the set will be degenerate if given
two values $k$ and $j$, $a_j = a_k$. That is the case of either $j =
k$ or that $2c\in\mathbb{Z}$. In other words, $c$ could be either a
semi or a whole number. In terms of our variables $\xi$, $\xi_B$,
$\Xi$ and $N$ it implies that \eemp \frac{2}{\Xi}\Rpar{1+\xi_B}& \in
\mathbb{N}\\ \text{or}\\
\frac{2}{\Xi} + 2(N-N_n)& \in \mathbb{N}\\
\label{eq:degeneracy_condition}
\ffin
There are three cases that correspond to different kinds of solutions as follows:
\begin{enumerate}[leftmargin=2.5cm,rightmargin=2.5cm,label=(\Roman*)]
\item The non-degenerate case
\label{subsec:nondegeneracy}

\ImTR{0.27}{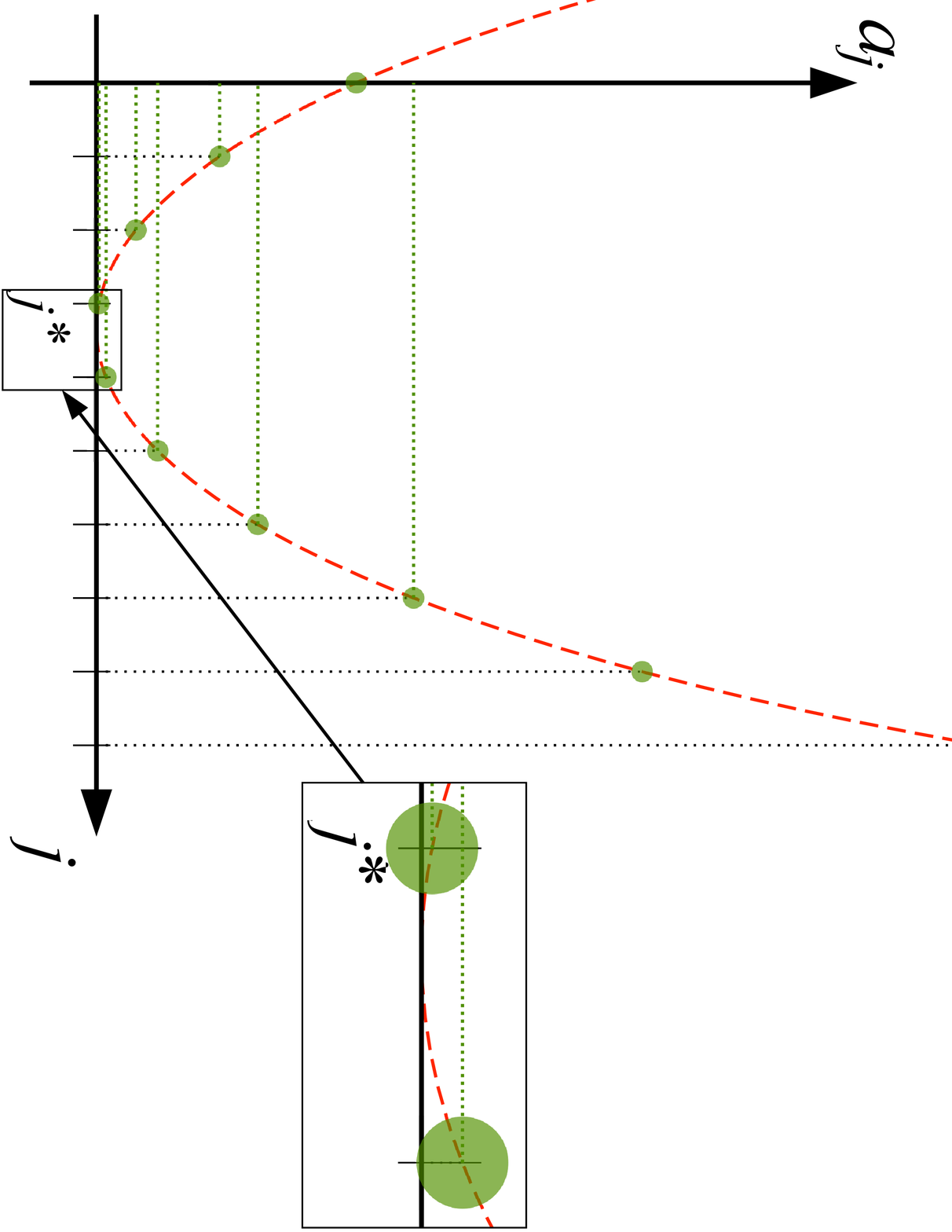}{non-deg-case}{Representation of $a_j$ as a function of $j$ with $j^\star$ the location of the minimum of the $\Kpar{a_j}$ for the non-degenerate case}{90}

For this case, the values of $a_j$ are non-degenerate as shown in \cref{fig:non-deg-case}. Given that we expect a simple separation of the product terms from \cref{eq:laplace_Z_simple} into sums as,
\eemp
\mathcal{T}_{\Kpar{f_N\otimes\cdots\otimes f_0}}^{\Bpar{\Xi\Delta}}(s)=&\prod_{k=0}^{N}\frac{1}{s+a_k}=\sum_{k=0}^{N}C_k\frac{1}{s+a_k},\\
\label{eq:laplace_Z_non-deg}
\ffin
with coefficients $C_k=\prod_{l=0,l\neq k}^{N}\frac{1}{a_l-a_k}$ simplified to,\footnote{A key remark on the notation used to avoid confusion between the coupling constant $\Gamma$ and the Gamma function $\gammafun{x}$}
\emp
C_k=\frac{2(-1)^{k}}{k!\Rpar{N-k}!}\frac{\Rpar{k-\Bpar{\frac{\xi-1}{\Xi}+\frac{1}{2}}}\, \gammafun{k-2\Bpar{\frac{\xi-1}{\Xi}+\frac{1}{2}}}}{\gammafun{N+1+k-2\Bpar{\frac{\xi-1}{\Xi}+\frac{1}{2}}}}
\label{eq:coef_Cj}
\fin
which inverted gives for the partition function,
\eemp
\mathcal{Z}=&\Rpar{\frac{2\pi\, R^{2}}{\Xi}}^N\, e^{a_N\Xi\Delta-E_0}\sum_{k=0}^{N}C_k\, e^{-a_k\Xi\Delta}.\\
\label{eq:exact_Z_non-deg}
\ffin
If the size of the box is large such that $\Xi\Delta\to\infty$, then the partition function scales as,
\eemp
\mathcal{Z}\, \underset{\Xi\Delta\to\infty}{\sim}\, C_{j^\star}e^{-a_{\j^\star}\Xi\Delta}.\\
\label{eq:approx_Z_non-deg}
\ffin
\item The degenerate case: $2(1+\xi_B)/\Xi$ even
\label{subsec:degeneracy_case1}

\ImTR{0.27}{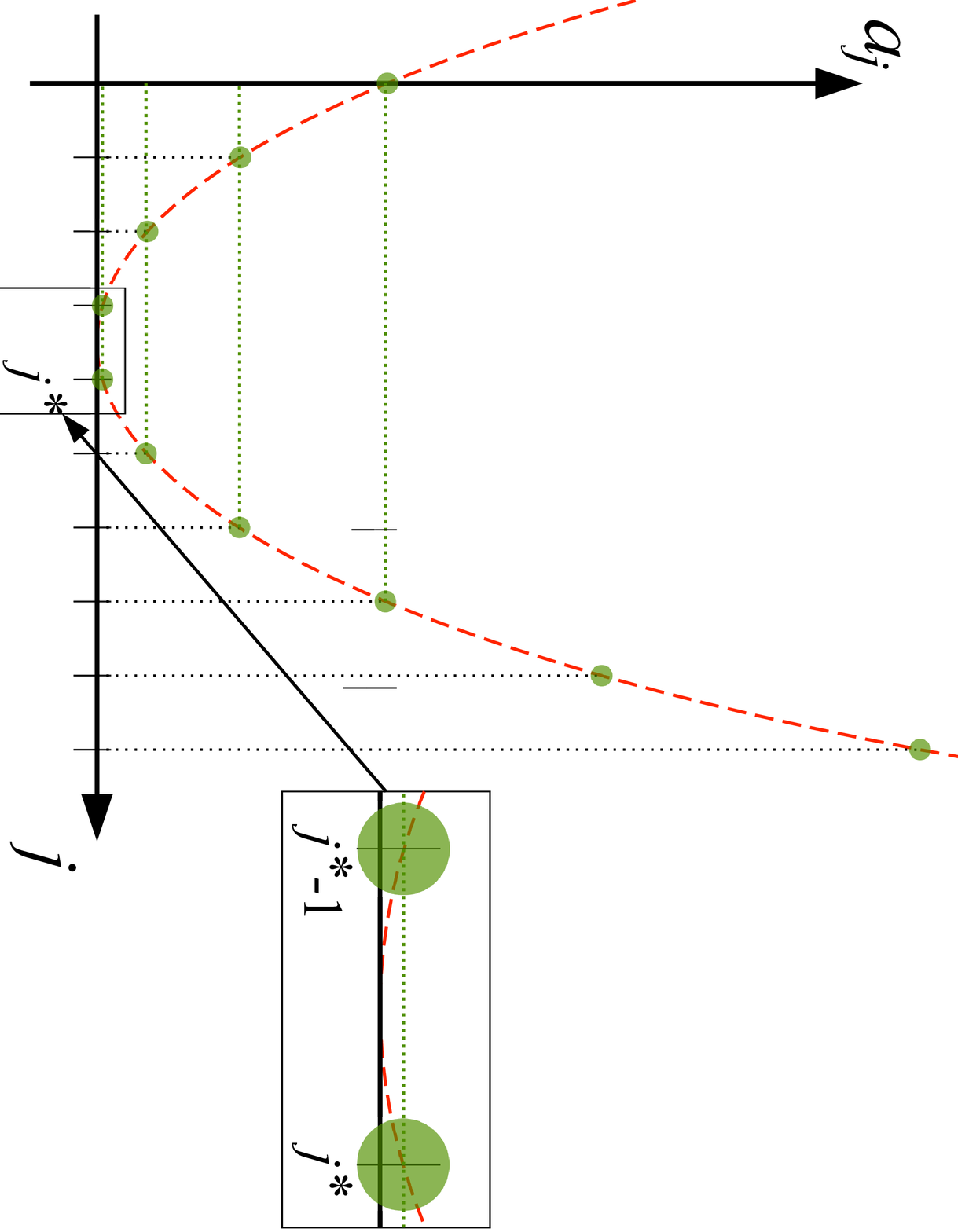}{even-deg-case}{Representation of $a_j$ as a function of $j$ with $j^\star$ the location of the minimum of the $\Kpar{a_j}$ for the degenerate case with $2(1+\xi_B)/\Xi$ an even number.}{90}

The even degeneracy case is represented graphically in \cref{fig:even-deg-case} where the value of $j^{\star}$ is degenerate and the number of degenerate $a_j$'s equates the minimum between $(j^\star-1)$ and $(N-j^\star+1)$. If $\{j^{\dag}\}$ denotes the set of degenerate values, then the partition function reads from \cref{eq:laplace_Z_simple},\footnote{The $\star$ symbol represents a product or sum where only one of the two degenerate indeces ($k$ or $k^\dag$ such that $a_k=a_{k^\dag}$) is taken into account.}
\eemp
\mathcal{T}_{\Kpar{f_N\otimes\cdots\otimes f_0}}^{\Bpar{\Xi\Delta}}(s)=&\Bpar{\prod_{k\notin\{j^\dag\}}\frac{1}{s+a_k}}\Bpar{{\prod_{k\in\{j^\dag\}}}^\star\frac{1}{\Rpar{s+a_k}^2}}\\
=&\sum_{k\notin\{j^\dag\}}\Kpar{\prod_{l=0,l\neq k}^{N}\frac{1}{a_l-a_k}}\frac{1}{s+a_k}\\
&+{\sum_{k\in\{j^\dag\}}}^\star\Kpar{\prod_{l=0,l\neq k,k^\dag}^{N}\frac{1}{a_l-a_k}}\Bpar{\frac{1}{\Rpar{s+a_k}^2}-\Rpar{\sum_{l=0,l\neq k,k^\dag}^{N}\frac{1}{a_l-a_k}}\frac{1}{s+a_k}}\\
=&\sum_{k\notin\{j^\dag\}}C_k\, \frac{1}{s+a_k}+{\sum_{k\in\{j^\dag\}}}^\star C_{k,k^\dag}\, \Bpar{\frac{1}{\Rpar{s+a_k}^2}-S_{k,k^\dag}\frac{1}{s+a_k}},\\
\label{eq:laplace_Z_deg_case1}
\ffin
with $C_{k,k^\dag}=\prod_{l=0,l\neq k,k^\dag}^{N}\frac{1}{a_l-a_k}$ and $S_{k,k^\dag}=\sum_{l=0,l\neq k,k^\dag}^{N}\frac{1}{a_l-a_k}$ coefficients simplified to,
\emp
C_{k,k^\dag}=-\frac{(-1)^{k+k^\dag}\Rpar{k-k^\dag}^2}{k!(k^\dag)!(N-k)!(N-k^\dag)!},
\label{eq:coef_Cjjdag}
\fin
and\footnote{We have used the $\Phi_0\Rpar{\, }$ notation for the Digamma function.}
\eemp
S_{k,k^\dag}=\frac{1}{k-k^\dag}&\left\{\frac{2}{k-k^\dag}+\Phi_0\Rpar{N-k+1}-\Phi_0\Rpar{N-k^\dag+1}+\Phi_0\Rpar{k^\dag+1}-\Phi_0\Rpar{k+1}\right\}.
\label{eq:coef_Sjjdag}
\ffin

Ultimately, the partition function,
\eemp
\mathcal{Z}=&\Rpar{\frac{2\pi\, R^{2}}{\Xi}}^N\, e^{a_N\Xi\Delta-E_0}\\
&\times\Rpar{\sum_{k\notin\{j^\dag\}}C_k\, e^{-a_k\Xi\Delta}+{\sum_{k\in\{j^\dag\}}}^\star C_{k,k^\dag}\, \Bpar{\Xi\Delta-S_{k,k^\dag}}\, e^{-a_k\Xi\Delta}}.\\
\label{eq:exact_Z_even-deg}
\ffin

\item The degenerate case: $2(1+\xi_B)/\Xi$ odd
\label{subsec:degeneracy_case2}

\ImT{0.27}{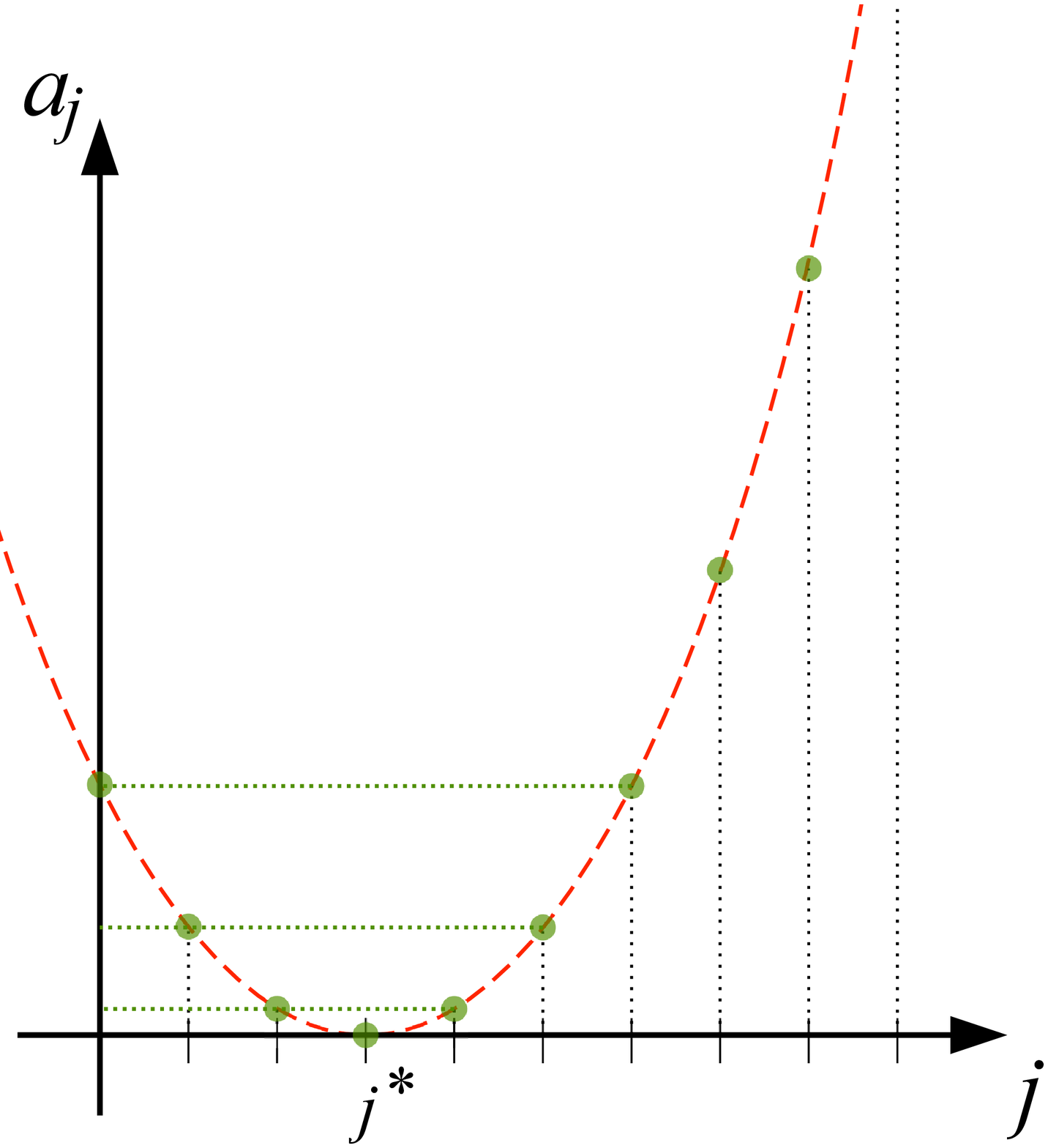}{odd-deg-case}{Representation of $a_j$ as a function of $j$ with $j^\star$ the location of the minimum of the $\Kpar{a_j}$ for the degenerate case with $2(1+\xi_B)/\Xi$ an odd number}

Different from the previous case, $j^\star$ is not degenerate as shown in \cref{fig:odd-deg-case} and so we would expect that the partition function gives,

\eemp
\mathcal{Z}=&\Rpar{\frac{2\pi\, R^{2}}{\Xi}}^N\, e^{a_N\Xi\Delta-E_0}\\
&\times\Rpar{D_{j^\star}+\sum_{k\notin\{j^\dag\}\wedge j^\star}C_k\, e^{-a_k\Xi\Delta}+{\sum_{k\in\{j^\dag\}}}^\star C_{k,k^\dag}\, \Bpar{\Xi\Delta-S_{k,k^\dag}}\, e^{-a_k\Xi\Delta}},\\
\label{eq:exact_Z_odd-deg}
\ffin
with
\emp
D_{k}=\Rpar{\frac{1}{k!(N-k)!}}^2.
\fin
\end{enumerate}

The partition function is given by three different expressions,
(\ref{eq:exact_Z_non-deg}), (\ref{eq:exact_Z_even-deg}) and
(\ref{eq:exact_Z_odd-deg}) depending if $4(1+\xi_B)/\Xi$ is a whole
number or not. Since the free energy has to be continuous, the
partition function requires the same property. \juan{Despite that fact,}
it is interesting to verify that we can recover (\ref{eq:exact_Z_even-deg})
and (\ref{eq:exact_Z_odd-deg}) from an appropriate limit
of~(\ref{eq:exact_Z_non-deg}). This is done in \cref{sec:PF_deg}.
The rationale proceeding to this argument is to use the non-degenerate
partition function from \cref{eq:exact_Z_non-deg} for all future
calculations.

\subsection{Density profile}
\label{sec:profile1}

The density profile is given by
\eemp
\rho(\VEC{r})
=&\frac{N}{N!\, \mathcal{Z}}\int\dif^{N}\VEC{r}\, \delta\Rpar{\VEC{r}-\VEC{r}_1}\, e^{-\mathcal{H}}
=\frac{N}{N!\, \mathcal{Z}}\Rpar{\frac{R^{2}}{\Xi}}^N\int\dif^{N}y\dif^{N}\theta\, \frac{\delta\Rpar{R\, e^{y/\Xi}-R\, e^{y_1/\Xi}}}{2\pi\, R\, e^{y/\Xi}}\, e^{-\mathcal{H}+\frac{2}{\Xi}\sum_{j}y_j}\\
=&\frac{\Xi}{2\pi\, R^2\, e^{2y/\Xi}}\Rpar{\frac{2\pi\, R^{2}}{\Xi}}^Ne^{-E_0}\frac{N}{N!\, \mathcal{Z}}\int\dif^{N}y\, \delta\Rpar{y-y_1}\, e^{-\mathcal{H}^\prime}\\
=&\frac{\Xi}{2\pi R^2e^{2y/\Xi}}\rho_{y}
\label{eq:rho1}
\ffin
where
\emp
\rho_{y}\defeq N\Mean{\delta\Rpar{y-y_1}}_{\{y_j\}}\quad \text{with}\quad \Mean{\delta\Rpar{y-y_1}}_{\{y_j\}} \defeq \Rpar{\frac{2\pi\, R^{2}}{\Xi}}^Ne^{-E_0}\frac{1}{N!\, \mathcal{Z}}\int\dif^{N}y\, \delta\Rpar{y-y_1}\, e^{-\mathcal{H}^\prime}
\fin
which speaks equally for $\rho$.

The average for $\rho_y$ is easily calculated using the procedure for the partition function considering that in the partitioning of the phase space $\Mean{\delta{(y-y_1)}}$ takes $(N-1)!$ permutations with $y_1$ at a given position in the group of $\{y_j\}$'s. Since each arrangement is obtained through a series of permutations of the {\bf [BO]} ($y_1<y_2<\cdots<y_N$ denoted by {\bf [BO]}) then the average yields that
\emp
N\Mean{\delta\Rpar{y-y_1}}_{\{y_j\}}=\Mean{\delta\Rpar{y-y_1}}_{\{y_j\}}^{T}+\Mean{\delta\Rpar{y-y_2}}_{\{y_j\}}^{T}+\cdots+\Mean{\delta\Rpar{y-y_N}}_{\{y_j\}}^{T},
\label{eq:rho_spanned}
\fin
where the $T$ stands for a truncation of the integration to the subregion of the phase space delimited by the {\bf [BO]} leading to,
\eemp
\Mean{\delta\Rpar{y-y_k}}_{\{y_j\}}^{T}=&\Rpar{\frac{2\pi\, R^2}{\Xi}}^N\frac{e^{-E_0}}{\mathcal{Z}}\int_{\textbf{[BO]}}\dif^{N}y\, \delta\Rpar{y-y_k}e^{-\mathcal{H}^\prime}\\
=&\Rpar{\frac{2\pi\, R^2}{\Xi}}^N\frac{e^{a_N\Xi\Delta-E_0}}{\mathcal{Z}}\int_{\textbf{[BO]}}\dif^{N}y\, \delta\Rpar{y-y_k}e^{-\sum_{j=0}^{N}a_j(y_{j+1}-y_j)}.\\
=&\Rpar{\frac{2\pi\, R^2}{\Xi}}^N\frac{e^{a_N\Xi\Delta-E_0}}{\mathcal{Z}}\underbrace{\int_{\textbf{[BO]}}\dif^{N}y\, \delta\Rpar{y-y_k}\prod_{j=0}^{N}f_j(y_{j+1}-y_j)}_{\text{recalled as }\delta_{k}^{T}}.\\
\ffin

The $N!$ steps in to account for all permutations of any given arrangement of the $\{y_j\}$'s deriving from the $(N-1)!$ intrinsic permutations times $N$ from the average of the density. Let us evaluate $\delta_{k}^{T}$ term using the Laplace transformation. From the steps followed in \cref{eq:finZ2},%
\footnotesize
\eemps
\mathcal{T}_{\Kpar{\delta_{k}^{T}}}^{\Bpar{y}}(s)&=\int_{0}^{\infty}\dif y e^{-s\, y}\Kpar{\int_{0}^{\Xi\Delta}\dif y_N\int_{0}^{y_N}\dif y_{N-1}\dots\int_{0}^{y2}\dif y_1\, \Rpar{\prod_{j=0}^{N}f_j\Rpar{y_{j+1}-y_j}}\delta\Rpar{y-y_k}}\\
&=\int_{0}^{\Xi\Delta}\dif y_N\int_{0}^{y_N}\dif y_{N-1}\dots\int_{0}^{y2}\dif y_1\, \Rpar{\prod_{j=0}^{N}f_j\Rpar{y_{j+1}-y_j}}\Kpar{\prod_{l=0}^{k-1}e^{-s(y_{l+1}-y_l)}}.\\
\ffins
\normalsize
Defining $g_j(x;s)=e^{-s\, x}f_j(x)$,
\footnotesize
\eemps
\mathcal{T}_{\Kpar{\delta_{k}^{T}}}^{\Bpar{y}}(s)
&=\Bpar{f_N\otimes f_{N-1}\otimes\cdots\otimes f_k\otimes g_{k-1}\otimes g_{k-2}\otimes\cdots\otimes g_1\otimes g_0}(\Xi\Delta),\\
\ffins
\normalsize
which again, admits a solution via the Laplace transformation on $\Xi\Delta$; therefore,
\footnotesize
\eemps
\mathcal{T}_{\Kpar{\mathcal{T}_{\Kpar{\delta_{k}^{T}}}^{\Bpar{y}}}}^{\Bpar{\Xi\Delta}}(t)
&=\Rpar{\prod_{j=0}^{k-1}\frac{1}{t+s+a_j}}\Rpar{\prod_{j=k}^{N}\frac{1}{t+a_j}}\\
&=\Rpar{\sum_{j=0}^{k-1}\frac{d_{j,k}(s)}{t+s+a_j}+\sum_{j=k}^{N}\frac{d_{j,k}(s)}{t+a_j}},\\
\ffins
\normalsize
with
\emps
d_{j,k}(s)=\begin{cases}
\Rpar{\prod_{l=0,l\neq j}^{k-1}\frac{1}{-a_j+a_l}}\Rpar{\prod_{l=k}^{N}\frac{1}{-a_j-s+a_l}},&\mbox{if } j < k\\
\Rpar{\prod_{l=0}^{k-1}\frac{1}{-a_j+s+a_l}}\Rpar{\prod_{l=k,l\neq j}^{N}\frac{1}{-a_j+a_l}},&\mbox{if } j \geq k.\\
\end{cases}
\fins
The presentation of the previous constants is not convenient for the final inversion. Expanding the second and first products in, respectively, the lower and upper bounds of $j$ we obtain,
\eemp
d_{j,k}(s)
=&\begin{cases}
C_{0,k-1;j}\Rpar{-\sum_{m=k}^{N}C_{k,N;m}\frac{1}{s+a_j-a_m}},&\mbox{if } j < k\\
C_{k,N;j}\Rpar{\sum_{m=0}^{k-1}C_{0,k-1;m}\frac{1}{s-a_j+a_m}},&\mbox{if } j \geq k,
\end{cases}
\ffin
with the constant $C_{m,n;k}=\prod_{l=m,l\neq k}^{n}\frac{1}{a_l-a_k}$ ($m,n\in\mathcal{N}\suchthat m<n\, \wedge\, k\in[m,n]$), evaluated equally as $C_k$ in \cref{eq:coef_Cj}, given by,
\eemp
C_{m,n;k}=\frac{2(-1)^{k-m}}{(k-m)!\Rpar{n-k}!}\frac{\Rpar{k-\Bpar{\frac{\xi-1}{\Xi}+\frac{1}{2}}}\, \gammafun{m+k-2\Bpar{\frac{\xi-1}{\Xi}+\frac{1}{2}}}}{\gammafun{n+1+k-2\Bpar{\frac{\xi-1}{\Xi}+\frac{1}{2}}}}
\ffin
The first inversion yields,
\footnotesize
\eemps
\mathcal{T}_{\Kpar{\delta_{k}^{T}}}^{\Bpar{y}}(s)=&\sum_{j=0}^{k-1}d_{j,k;s}\, e^{-(s+a_j)\Xi\Delta}+\sum_{j=k}^{N}d_{j,k;s}\, e^{-a_j\Xi\Delta}\\
=&-\sum_{j=0}^{k-1}\sum_{m=k}^{N}C_{0,k-1;j}\, C_{k,N;m}\, \frac{e^{-(s+a_j)\Xi\Delta}}{s+a_j-a_m}+\sum_{j=k}^{N}\sum_{m=0}^{k-1}C_{k,N;j}\, C_{0,k-1;m}\, \frac{e^{-a_j\Xi\Delta}}{s-a_j+a_m},\\
\ffins
\normalsize
which is invertible noticing that the first term is proportional to $\Theta(y-\Xi\Delta)$, the Heaviside step function,\footnote{$$
\Theta(x)=\left\{\begin{array}{cc}
0 & x\leq0\\
1 & x>0
\end{array}\right.
$$
} hence trivial since $y\leq\Xi\Delta$. Finally, the truncated density reads,
\footnotesize
\eemp
\Mean{\delta\Rpar{y-y_k}}_{\{y_j\}}^{T}
=&\frac{\sum_{j=k}^{N}\sum_{l=0}^{k-1}C_{k,N;j}\, C_{0,k-1;l}\, e^{-a_j\Xi\Delta-(a_l-a_j)y}}{\sum_{j=0}^{N}C_{j}\, e^{-a_j\Xi\Delta}}\\
=&\frac{\Kpar{\sum_{j=k}^{N}C_{k,N;j}\, e^{-a_j(\Xi\Delta-y)}}\Kpar{\sum_{j=0}^{k-1}C_{0,k-1;j}\, e^{-a_jy}}}{\sum_{j=0}^{N}C_{j}\, e^{-a_j\Xi\Delta}}\\
\label{eq:exact_rho_T}
\ffin
\normalsize

The previous formulation for the density teaches us many things about the behavior of the density close to $R$ and $D$. First of all, inspecting the above relationship we find that,
\eemp
\Mean{\delta\Rpar{y-y_k}}_{\{y_j\}}^{T}\propto&\frac{\mathcal{Z}[\Xi,\xi,k-1,\Delta-y/\Xi]\, \times\, \mathcal{Z}[\Xi,\xi,N-k,y/\Xi]}{\mathcal{Z}[\Xi,\xi,N,\Delta]},
\label{eq:rho_T_parts}
\ffin
the truncated profile at a given position equates the product of partition functions corresponding to $k-1$ particles before that position and $N-k$ particles beyond the latter. This trait is characteristic of decorrelated fluids as presumed in the mean field regime.

Summarizing, if we define $\widetilde{\rho}(\VEC{r})=2\pi\,
R^2\rho(\VEC{r})/(N\xi)$, the density profile is given by
\begin{equation}
	\widetilde{\rho}(\VEC{r})= \frac{\Xi}{N\xi}\Rpar{\frac{R}{r}}^2\sum_{k = 1}^N\frac{\Kpar{\sum_{j=k}^{N}C_{k,N;j}\, \Rpar{\frac{D}{r}}^{-a_j\Xi}}\Kpar{\sum_{j=0}^{k-1}C_{0,k-1;j}\, \Rpar{\frac{r}{R}}^{-a_j\Xi}}}{\sum_{j=0}^{N}C_{j}\, \Rpar{\frac{D}{R}}^{-a_j\Xi}}.
  \label{eq:rhotilde-exact-Gl2}
\end{equation}
From the previous relationship we can extract the leading behavior of the weakly coupled regime which turns out to be precisely that of mean field; $\rho\propto1-2x/\mu$ with $x$ the perpendicular distance from the disk \juan{and $\mu$ the Gouy length.}

On the other hand, as $\Xi\Delta$ becomes large, the functional form of the profile simplifies close to both boundaries. Knowing that the first term (that which sums from $k$ to $N$) will be non--{\it{zero}}, or, the least, relevant, if $k\leq j^\star$ comparing to the partition function which scales as $e^{-a_{j^\star}\Xi\Delta}$ from \cref{eq:approx_Z_non-deg}. Conversely, for the outer shell, or close to $D$, the situation is that which the second term is non--{\it{zero}} if $k>j^\star$. This provides an interesting parallel rationale to condensation. Those particles which contribute to the profile near the surface, i.e. the condensed population, are $j^\star$. On the other hand, the remaining $N-j^\star$ contribute to the profile at the outer shell. Approximating from \cref{eq:exact_rho_T} and looking at the profile near $R$ we are examining the truncated profiles for the case of $k\leq j^\star$ since all profiles of index beyond $j^\star$ will correspond to evaporated counter-ions. Hence,
\small
\eemp
\Mean{\delta\Rpar{y-y_k}}_{\{y_j\}}^{T}
\simeq&\sum_{j=0}^{k-1}\frac{C_{0,k-1;j}}{C_{0,k-1;j^\star}}\, e^{-(a_j-a_{j^\star})y}\\
&+\Bpar{\sum_{j=0}^{k-1}\frac{C_{0,k-1;j}C_{k,N;j^\star\pm1}}{C_{0,N;j^\star}}\, e^{-(a_j-a_{j^\star\pm1})y}-\frac{C_{j^\star\pm1}}{C_{j^\star}}\sum_{j=0}^{k-1}\frac{C_{0,k-1;j}}{C_{0,k-1;j^\star}}\, e^{-(a_j-a_{j^\star})y}}\, e^{-(a_{j^\star\pm1}-a_{j^\star})\Xi\Delta}\\
&+\mathcal{O}\Rpar{e^{-2\Xi\Delta}}.
\label{eq:approx_rho_T_R}
\ffin
\normalsize
Conversely the contribution to the exterior shell reads for $k>j^\star$,
\small
\eemp
\Mean{\delta\Rpar{y-y_k}}_{\{y_j\}}^{T}
\simeq&\sum_{j=k}^{N}\frac{C_{k,N;j}}{C_{k,N;j^\star}}e^{-(a_j-a_{j^\star})(\Xi\Delta-y)}\\
&+\Bpar{\sum_{j=k}^{N}\frac{C_{0,k-1;j^\star\pm1}C_{k,N;j}}{C_{0,N;j^\star}}e^{-(a_j-a_{j^\star\pm1})(\Xi\Delta-y)}-\frac{C_{j^\star\pm1}}{C_{j^\star}}\sum_{j=k}^{N}\frac{C_{k,N;j}}{C_{k,N;j^\star}}e^{-(a_j-a_{j^\star})(\Xi\Delta-y)}}\, e^{-(a_{j^\star\pm1}-a_{j^\star})\Xi\Delta}\\
&+\mathcal{O}\Rpar{e^{-y-\Xi\Delta}},\\
\label{eq:approx_rho_T_D}
\ffin
\normalsize
sharing the same functional form at both edges. Notice that the density, according to the procedure we have followed, is written as an expansion of powers of the radial distance.

\psfrag{AX}{$r/R$}
\psfrag{BY}{$\widetilde{\rho}$}
\ImT{0.5}{fig6}{profile_method1}{The density profile $\widetilde{\rho}$ near the charged disk for different values of the Manning parameter for $N=10$. The data displays the results for different box sizes. The dashed curves represent the analytic prediction \cref{eq:exact_rho_T} into \cref{eq:rho1}. The values for the Manning parameter are chosen for small $\Xi$ and very close to unity where the theory is no longer valid.}

Now we turn to \cref{fig:profile_method1} where the comparison between the analytic prediction and the simulation results is displayed. From the results, the deviations from the data, as expected, increase with higher coupling (see $\xi=9$ in the plot). Notice that despite the differences the contact density as the profile approaches to $r=R$ matches that reached by the analytic profile.

\subsection{Integrated charge}
\label{sec:charge1}

The integrated charge amounts to the number of condensed counter-ions
from the density profile \cref{eq:exact_rho_T}. The integrated charge
reads in terms of the centrifugal variables $\{y_j=\Xi\log({r}_j/{R})\}$ as follows,

\eemp
Q(r)=&\int_{0}^{y}{\rho_{y}}({y^\prime})\, \dif {y^\prime}
=\sum_{k=1}^{N} Q_k(y)
,
\ffin
that split into contributions of each truncated density reads,
\eemp
Q_k(r)
=&\frac{\sum_{j=k}^{N}\Kpar{C_{k,N;j}\, e^{-a_j\Xi\Delta}\Bpar{\sum_{l=0}^{k-1}\frac{C_{0,k-1;l}}{a_l-a_j}\, \Rpar{1-e^{-(a_l-a_j)y}}}}}{\sum_{j=0}^{N}C_{j}\, e^{-a_j\Xi\Delta}},
\label{eq:Q_k_R}
\ffin
with particular emphasis on a functional form suited for distances close to $R$, or,
\eemp
Q_k(r)
=&1-\frac{\sum_{j=0}^{k-1}\Kpar{e^{-a_j\Xi\Delta}\Bpar{C_j-C_{0,k-1;j}\sum_{l=k}^{N}\frac{e^{-(a_l-a-j)(\Xi\Delta-y)}C_{k,N;l}}{a_l-a_j}}}}{\sum_{j=0}^{N}C_{j}\, e^{-a_j\Xi\Delta}},
\label{eq:Q_k_D}
\ffin
corresponding to that closest to $D$. Notices how the two functional
forms consistently show that when $y\to0$ (close to $R$) $Q_k\to0$ and
$y\to\Xi\Delta$ (close to $D$) $Q_k\to 1$ as expected since the truncated densities corresponds to the contribution of a single particle in the {\bf [BO]}.

In order to understand how many ions condense, let us look at
\cref{eq:Q_k_R} and notice that when $\Delta$ is large compared to
$y$, $Q_k\to 0$ if $k>j^\star$ and so the condensed counter-ions will
equate $j^\star$, an argument consistent with what we have assumed
constructing the density profile for infinite box sizes. With regards
to condensation, it means that the fraction of condensed ions is
$f=\frac{j^\star}{N}$, with special attention to a minor, yet
relevant, detail of the density profile. For simplicity we have
omitted the degeneracy issue here considering that the pressure is a
continuous function. Therefore, the density profiles should not
exhibit any particular behavior at the troublesome values. As a matter
of fact, there is one problem when $a_{j^\star}=a_{j^\star-1}$, as
discussed earlier, where the profile behaves as $\rho\sim1/r^2$ and
the corresponding integrated quantity $Q_{j^\star}(r)\simeq
y/(\Xi\Delta)$. This form tells us that the {\it condensing}
counter-ion is \emph{evenly} distributed in the box (using $y$
coordinates). An important rationale then for condensation is that
this particle is neither condensed nor evaporated and, thus, the
integrated charge will exhibit a constant slope form when this
occurs. This situation does not take place in the three-dimensional
case
\citep{PhysRevLett.95.185703,PhysRevE.73.056105,doi:10.1021/jp311873a}.

With this matter clarified, the fraction of condensed counter-ions corresponds then to,
\emp
\juan{f=\frac{1}{N}\Ceil{\frac{\xi-1}{\Xi}}=\frac{1}{N}\Ceil{\frac{f_MN}{1+\frac{\xi_B}{\xi}}}}.
\fin

\psfrag{AX}{$\log (r/R)$}
\psfrag{BY}{$N-Q(r)$}
\ImT{0.5}{fig7}{charge_method1}{The integrated charge $N-Q({r})$ as a function of the logarithmic distance for $\Delta=10^2$, $\xi_B=0$ and $N=10$ for various $\Xi$. The plots read for the coupling parameter from top to bottom $\Xi=\frac{2}{5},\frac{10}{21},\frac{1}{2},\frac{10}{19},\frac{2}{3},\frac{10}{11},1,\frac{10}{9}$ and $2$.}

The numerical results from Monte Carlo simulations are presented in \cref{fig:charge_method1} highlighting the evaporated counter-ions as indicated by the plateau. We observe that, at the transitions, the integrated charge displays a constant slope form associated to the counter-ion which lies at the borderline of condensation (in \cref{fig:charge_method1}, $\Xi=1/2,\, 1$). Away from the transitions the plateau is flat similar to the three dimensional case.

The onset for condensation in two dimensions coincides with the
well-known result $\xi=1$. This signals a transition between the
regime of full evaporated to partially condensed
counter-ions. However, the details of the behavior beyond this point
are unique to the two dimensional problem. Unlike the situation in
three dimensions, the thermodynamic limit ($N\to\infty$), at a fixed
Manning parameter and no charge in the exterior boundary ($\xi_B=0$),
takes the coupling to zero, which is the mean field regime
investigated thoroughly in previous works. In such a case, $f\to
f_M=1-\frac{1}{\xi}$.

The situation maintaining a fixed, non vanishing, coupling constant
$\Xi$ and $N\to\infty$ will imply an infinite Manning parameter
$\xi\to\infty$, and consequently a strongly bound set of
counter-ions. However, despite the situation, the number of condensed
counter-ions increases with $N$ while the remaining evaporated remains
all the same. This is easy to see from \cref{eq:jstar} where the
number of evaporated ione is, 
\emp
N_{\text{evap}}=\Whole{\frac{1+\xi_B}{\Xi}}.  
\fin

In other words, evaporation, looking at $\xi>1$ is dominated by the coupling parameter. This is the reason for the division of the two dimensional case between $\Xi<1$, $\Xi=1$, and $\Xi>1$; \juan{assuming that $\xi_B=0$ if $\Xi<1$ there is evaporation, $\Xi>1$ full condensation, leaving $\Xi=1$ the critical value for the coupling parameter where there is only one unbound counter-ion}. From the earlier considerations, the problem with $\Xi=1$ will have, exactly, one free ion while the problem for greater couplings is equivalent to full condensation. This ultimate remark helps to find an analytic route towards the profile for large couplings and the contact theorem.

\juan{One question that rises with the previous result is what is the effective charge of the center disk and the ion cloud. For simplicity we assume that $Q_1>0$. The condensed ions will screen the charge of the center disk thus reducing the mean potential interaction between the center region and the ions in the outer region of the disk. In other words, what is the value of $\xi_{\eff}\defeq\xi-N_c\Xi$?
\emp
\xi_{\eff}=\Xi\Whole{\frac{1+\xi_B}{\Xi}}-\xi_B,
\fin
which tells us that $\xi_{\eff}\leq1$ for any coupling and exterior charge $Q_2$; also, if $\Xi\leq1$ then $\xi_{\eff}\geq0$; lastly, if $\xi_B=0$ then $\xi_{\eff}\geq0$. However, if $\Xi>1$ the effective charge could be negative only if $Q_2\neq0$! This striking effect, similar to charge inversion, is only possible at strong couplings.}

\subsection{Contact density}
\label{sec:contact1}

For the contact density, we turn to \cref{eq:exact_rho_T} at $y=0$ and $y=\Delta$. In both cases, we encounter that the truncated densities have vanishing values except for two particular cases. For $y=0$,
\emp
\left.\Mean{\delta\Rpar{y-y_k}}_{\{y_j\}}^{T}\right\vert_{y=0}=\left\{\begin{array}{cc}
\frac{\sum_{j=1}^{N}C_{1,N;j}\, e^{-a_j\Xi\Delta}}{\sum_{j=0}^{N}C_{j}\, e^{-a_j\Xi\Delta}} & k=1\\
0 & k>0
\end{array}\right.,\\
\label{eq:contact_rho_R}
\fin
likewise for $y=\Xi\Delta$,
\emp
\left.\Mean{\delta\Rpar{y-y_k}}_{\{y_j\}}^{T}\right\vert_{y=\Xi\Delta}=\left\{\begin{array}{cc}
0 & k<N\\
\frac{\sum_{j=0}^{N-1}C_{0,N-1;j}\, e^{-a_j\Xi\Delta}}{\sum_{j=0}^{N}C_{j}\, e^{-a_j\Xi\Delta}} & k=N
\end{array}\right.,\\
\label{eq:contact_rho_D}
\fin
because
\emps
\sum_{j=l}^{m}C_{l,m;j}=\sum_{j=l}^{m}\Bpar{\prod_{k=l,k\neq j}^{m}\frac{1}{a_k-a_j}}=\delta_{l,m}
\fins
that tells us which terms contribute to the contact densities. Although, this is not surprising since the arrangement of the {\bf [BO]} intuitively truncates the average to contributions to both contacts coming from the first and last \emph{particles}.

Taking the large box limit, \cref{eq:contact_rho_R,eq:contact_rho_D} yield,
\footnotesize
\emp
\widetilde{\rho}(R)=\frac{1}{N\, \xi}(a_0-a_{j^\star})=\Rpar{f_M-\Bpar{f_M-\frac{j^\star}{N}}}\Rpar{f_M+\Bpar{f_M-\frac{j^\star}{N}}+\frac{1}{N}},
\label{eq:contact_rho_R_DeltaLimit}
\fin
\normalsize
and,
\footnotesize
\eemp
e^{2\Delta}\widetilde{\rho}(D)=\frac{1}{N\, \xi}(a_N-a_{j^\star})=\Rpar{\frac{1}{\xi}-\Bpar{f_M-\frac{j^\star}{N}}-\frac{1}{N}}\Rpar{\frac{1}{\xi}+\Bpar{f_M-\frac{j^\star}{N}}},\\
\label{eq:contact_rho_D_DeltaLimit}
\ffin
\normalsize
recovering mean field's result by taking $N\to\infty$ where
$\widetilde{\rho}(R)=f_{M}^{2}$ and
$e^{2\Delta}\widetilde{\rho}(D)=(1-f_{M})^{2}$. Additionally,
note that the value of the density at contact ($r=R$) is non-zero for $\xi>1$ coinciding with the onset for condensation discussed earlier. As expected when all ions condense, or $j^\star/N\to1$, the density at the exterior shell vanishes.

\psfrag{AX}{$\Xi$}
\psfrag{BY}{$\widetilde{\rho}\vert_{r=R}$}
\psfrag{BY2}{$\xi^2e^{2\Delta}\widetilde{\rho}\vert_{r=D}$}
\IImT{0.38}{fig8a}{contact_smallD_R}{$\rho(R)$}{0.38}{fig8b}{contact_smallD_D}{$\rho(D)$}{contact_smallD}{The density $\widetilde{\rho}$ at contact in $r=R$ as a function of the Coupling parameter $\Xi$ with $\xi_B=0$; here, $\Delta$ varies and $N=10$. The range of $\Xi$ extends below the onset for condensation up to high couplings. The dashed curves represent the exact contact density taken from evaluating the density from \cref{eq:rho1}. We considered the analytic exact result since $\Delta$ is small. The arrows indicate the location of $\Xi=1$, the borderline to strong coupling.}
\psfrag{AX}{$\Xi$}
\psfrag{BY}{$\widetilde{\rho}\vert_{r=R}$}
\psfrag{BY2}{$\xi^2e^{2\Delta}\widetilde{\rho}\vert_{r=D}$}
\IImT{0.38}{fig9a}{contact_smallD_R_log}{$\rho(R)$}{0.38}{fig9b}{contact_smallD_D_log}{$\rho(D)$}{contact_smallD_log}{The density $\widetilde{\rho}$ at contact in $r=R$ as a function of the Coupling parameter $\Xi$ with $\xi_B=0$; here, $\Delta$ varies and $N=10$. The range of $\Xi$ extends below the onset for condensation up to high couplings. The dashed curves represent the exact contact density taken from evaluating the density from \cref{eq:rho1}. We considered the analytic exact result since $\Delta$ is small. The arrows indicate the location of $\Xi=1$, the borderline to strong coupling.}

The values compare very well with the simulation data from
\cref{fig:contact_smallD_R,fig:contact_smallD_D} considering a small
$\Delta$. Notice that the contact densities are continuous functions;
for $r=R$ the function does not present any appreciable or qualitative
changes while its exterior counterpart shows a succession of bumps
coming from the evaporated counter-ions. Seen that increasing $\Delta$
bolsters the transitions we plot the contacts at $\Delta=10^2$ in
\cref{fig:contact_method1_D100}.

\psfrag{AX}{$\Xi$}
\psfrag{BY}{{\color{OliveGreen} $\widetilde{\rho}\vert_{r=R}$}}
\psfrag{BY2}{{\color{red} $\xi^2e^{2\Delta}\widetilde{\rho}\vert_{r=D}$}}
\ImT{0.38}{fig10}{contact_method1_D100}{The density $\widetilde{\rho}$ at contact in $r=R$ (red) and $r=D$ (olive green) as a function of the Manning parameter; here, $\Delta=100$ and $N=10$. The range of $\xi$ extends below the onset for condensation up to high couplings. The dashed curves represent the $\Delta\to\infty$ formulation from \cref{eq:contact_rho_R_DeltaLimit,eq:contact_rho_D_DeltaLimit}. The arrows indicate the location of $\Xi=1$, the borderline to strong coupling.}

The prediction gives a very accurate estimate of the contact densities seen in the figures at all ranges. Observe that the data displayed considers values of the Manning parameter which fall out of the scope of the present course; at $\Xi=1$ ($\Gamma=2$) we have the drastic change of behavior to strong coupling, as indicated in \cref{fig:contact_smallD,fig:contact_smallD_log,fig:contact_method1_D100}.

\subsection{Energy}
\label{sec:energy1}

The energy for the system can be found with the derivative of the partition function from \cref{eq:approx_Z_non-deg} with respect to $\xi$. Then,
\eemps
\widetilde{E}=&-\frac{\xi}{N}\parde{}{\Rpar{\log\mathcal{Z}}}{\xi},\\
\ffins
which simplifies in the \emph{large} $\Delta$ limit to,
\eemp
\widetilde{E}\simeq\xi\Delta\Rpar{\frac{N+1-j^\star}{N}}\Rpar{\frac{N-j^\star}{N}}.
\label{energy1}
\ffin
This equation for the energy has a saw-like shape, as was acknowledged by \citet{PhysRevE.73.056105,PhysRevLett.95.185703} and explored via the aforementioned procedure for $\Xi<1$ by \citet{PhysRevE.73.010501}, due to the transitions at temperatures below the critical temperature ($\xi>1$) when $N/\xi$ becomes a whole number. \Cref{fig:energy_method1} presents the numerical results for two values of $\Delta$ displaying, as anticipated, the effect of the box size to the transitions. Notice that for $\Xi>1$ ($\Gamma>2$) the energy goes below zero due to the interaction among the counter-ions.

\psfrag{AX}{$\xi$}
\psfrag{BY}{$\widetilde{E}/\Delta$}
\ImT{0.38}{fig11}{energy_method1}{The energy as a function of the Manning parameter; here $N=10$ and $\Delta=20,10^2$. The dashed curve corresponds to the $\Delta\to\infty$ prediction from \cref{energy1}.}

Related to the problem of condensation and minimal free energy, the energy shift at any transition is given for the change of energy when an unbound ion is condensed\footnote{The Manning parameter at which the transitions will occur are,
\emps
\xi_j=1+\Xi(j-1),
\fins
for $j\in\{1,2,\dots,N\}$.
}. Then,
\eemp
\mu_{evap}^{(2D)}=&N\Bpar{\xi_{j^\star}\Delta\Rpar{\frac{N+1-j^\star}{N}}\Rpar{\frac{N-j^\star}{N}}-\xi\Delta\Rpar{\frac{N+1-j^\star-1}{N}}\Rpar{\frac{N-j^\star-1}{N}}}\\
=&2\Delta,
\ffin
which, unsurprisingly, coincides with the entropy cost for binding a
free ion, a discussion thoroughly stripped by
\citet{manning:924}.

\section{The $\Xi=1$ case}
\label{sec:method2}

The case when $\Xi=1$, or $\Gamma=2$, represents the borderline before full
condensation and also could be understood as the limiting situation
before the strong coupling regime. This situation admits an exact
formulation that is interesting to explore. To begin with, let us look at the Boltzmann factor of the Hamiltonian \cref{eq:H_reduced_units} as it reads,

\eemp
e^{-\mathcal{H}}=&e^{-E_0}\prod_{j=1}^{N}\left\vert\frac{\VEC{r}_j}{R}\right\vert^{-2\xi}\,
\, \prod_{1\leq j<k\leq
  N}\left\vert\frac{\VEC{r}_j-\VEC{r}_k}{R}\right\vert^{2\Xi}
\,.
\label{eq:Z_gamma2_case}
\ffin

For this Hamiltonian, it is always convenient to refer vectors into a complex number in order to simplify the evaluation of the partition function, and from it the correlation functions. That way, we will define
\eemp
z_j\defeq&\frac{r_j}{R}\,e^{i\theta_j}.
\ffin
The partition function of the system (\cref{eq:Z_gamma2_case}) is rewritten as,
\eemp
\mathcal{Z}=&\frac{R^{2N}e^{-E_0}}{N!}\int\left[\prod_{1\leq j<k\leq N}\left|z_j-z_k\right|^2\right]^{\Xi}\prod_{j=1}^{N}|z_{j}|^{-2\xi+1}\dif{|z_j|}\dif{\theta_j}\\
\ffin

Appealing to the value of the coupling, it is clear from the previous form that $\Xi=1$ simplifies the calculation. Indeed, the procedure to solve the previous integration has been widely used to solve two dimensional systems for $\Gamma=2$ \citep[see][]{PhysRevA.20.2631,PhysRevLett.46.386} and for $\Gamma=2n$ described by \citet{PhysRevE.49.5623} and \citet{tellez1999exact}.

Holding $\Xi=1$ the calculation concerns the Vandermonde determinant in complex variables. Here,
\eemp
\Mat{V}{N\times N}=&\prod_{1\leq j<k\leq N}(z_j-z_k)\\
=&\text{Det}\left[\begin{array}{ccccc}
1 & 1 & 1 & \dots & 1 \\
z_1 & z_2 & z_3 & \dots & z_N\\
z_{1}^{2} & z_{2}^{2} & z_{3}^{2} & \dots & z_{N}^{2} \\
. & . & . & \dots & . \\
. & . & . & \dots & . \\
. & . & . & \dots & . \\
z_{1}^{N-1} & z_{2}^{N-1} & z_{3}^{N-1} & \dots & z_{N}^{N-1} \\
\end{array}\right]\\
=&\sum_{P}\sigma({P})\prod_{j=1}^{N}z_{j}^{P(j)}=\sum_{P}\sigma({P})\prod_{j=1}^{N}|z_{j}|^{P(j)}e^{iP(j)\theta_j},
\ffin
where $P$ and $P^\prime$ are permutations of $\{0,1,2,...,N-1\}$, and $\sigma(P)$ the respective permutation signature ($\sigma(P)=\pm1$). Directly to the partition function,
\eemp
\mathcal{Z}=&\frac{R^{2N}e^{-E_0}}{N!}\sum_{P,P^\prime}\sigma({P})\sigma({P^\prime})\prod_{j=1}^{N}\underbrace{\int_{0}^{2\pi} \dif{\theta_j}e^{i(P(j)-P^\prime(j))\theta_j}}_{{ 2\pi\delta_{P(j),P^\prime(j)}}}\int_{e^{-\Delta}}^{1}|z_{j}|^{P(j)+P^\prime(j)-2(\xi-1)-1}\dif|z_j|,\\
\ffin
which defining $\gamma(\Delta,\xi,j)$
\eemp
\gamma(\Delta,\xi,j)=&2\int_{1}^{e^{\Delta}}t^{2(j-(\xi-1))-1}\dif t\\
=&\left\{\begin{array}{lr}
2\Delta & \text{for }j=\xi-1\\
\frac{1-e^{2(j-(\xi-1))\Delta}}{(\xi-1)-j} & \text{for }j\neq \xi-1\\
\end{array}\right.,\\
\label{gamma}
\ffin
reads,
\eemps
\mathcal{Z}=&(\pi R^{2})^{N}e^{-E_0}\prod_{j=1}^{N}\gamma(\Delta,\xi,j-1)\underbrace{\Bpar{\frac{1}{N!}\sum_{P,P^\prime}\sigma({P})\sigma({P^\prime})\delta_{P,P^\prime}}}_{\color{blue}1}.\\
\ffins

Finally, simplifying yields,
\eemp
\mathcal{Z}=&(\pi R^{2})^{N}e^{-E_0}\prod_{j=1}^{N}\gamma(\Delta,\xi,j-1).
\label{Z2}
\ffin

\subsubsection{Large $\Delta$ limit}
\label{sec:large_delta_gamma}

The large box size is characterized by the effect of $\Delta$ in the solutions. We can observe that to the present case it enters directly in $\gamma$ in such a way that if $\Delta\to\infty$,
\eemp
\gamma(\Delta,\xi,j)\simeq&\left\{\begin{array}{lr}
2\Delta & \text{for }j=\xi-1\\
\frac{1}{(\xi-1)-j} & \text{for }j <  \xi-1\\
\frac{e^{2(j-(\xi-1))\Delta}}{j-(\xi-1)} & \text{for }j > \xi-1\\
\end{array}\right..\\
\label{large_delta_gamma}
\ffin

This teaches us that only those values of $j<\xi-1$ are independent of the box size and consequently are responsible for the condensed counter-ions. \juan{Countrarywise, the remaining entail the behavior of the evaporated population. A priori,} by inspection of \cref{Z2}, the total condensed counter-ions are $\Whole{\xi}-1$; ergo, for the particular $\xi_B=0$ case, neutrality imposes $\xi=N$, which dictates that $N-1$ particles condense and only one particle is unbound, consistent with the results obtained in the previous section for $\Xi<1$.

On the other hand, the evaporated counterpart has two different solutions: one proportional to $\Delta$ and that which grows exponentially with the box size. Related to the problem of condensation, we acknowledged full condensation beyond $\Xi=1$ ($\Gamma=2$) which tells us that this solution stands at the borderline. Then, according to the analysis on the profile the functional form that best suites this situation is that where the density decays like $r^{-2}$.

\subsection{Density profile}
\label{sec:profile2}

The $n$-body distribution functions can be obtained by a procedure
used in random matrix theory~\cite{mehta2004random} for the circular
unitary ensemble. To this end, let us introduce the kernel
\emp
K(z_{j},z_{k})=\sum_{l=1}^{N}\frac{|z_{j} z_{k}|^{l-1-\xi}
e^{i(l-1)(\theta_k-\theta_j)}}{\gamma(\Delta,\xi,l-1)},
\fin
Then, following~\cite{mehta2004random} (Chap. 5 and 10), it follows
that
\begin{equation}
  \rho(\VEC{r})=\frac{1}{\pi R^2} K(z,z)
\end{equation}
and finally,
\emp
\rho(\VEC{r})=\frac{1}{\pi R^2}\Bpar{\frac{R}{r}}^2\sum_{l=1}^{N}\frac{\left(\frac{r}{R}\right)^{2(l-\xi)}}{\gamma(\Delta,\xi,l-1)},
\label{profile2}
\fin
where, for  $\widetilde{\rho}=2\pi\, R^2\rho/(N\xi)$, this yields,
\emp
\widetilde{\rho}(r)=\frac{2}{N\xi}\Bpar{\frac{R}{r}}^2\sum_{l=1}^{N}\frac{\left(\frac{r}{R}\right)^{2(l-\xi)}}{\gamma(\Delta,\xi,l-1)}.
\label{eq:profile2_tilde}
\fin
The profile is presented in \cref{fig:profile_method2} comparing to the results from Monte Carlo simulations with excellent agreement with the above exact result. Notice the profile exhibits a sequence of terms with power shape. This resembles the situation obtained for $\Xi<1$ ($\Gamma<2$) referred in \cref{sec:profile1}.

Particularly for $\xi_B=0$ the Manning parameter equates the number of ions, then,
\eemp
\widetilde{\rho}(r)&=\frac{2}{N^2}\Bpar{\frac{R}{r}}^2\Kpar{\sum_{m=1}^{N-1}\frac{1}{\gamma(\Delta,N,N-m-1)}\left(\frac{R}{r}\right)^{2m}+\frac{1}{N\xi\, \Delta}}\\
&\juan{\,\overset{\Delta\to\infty}{=}\,\frac{2}{N^2}\Bpar{\frac{R}{r}}^2\Kpar{\frac{1}{N\xi\, \Delta}+\left(\frac{R}{r}\right)^{2}\frac{1-N\left(\frac{R}{r}\right)^{2(N-1)}+(N-1)\left(\frac{R}{r}\right)^{2N}}{\Rpar{1-\left(\frac{R}{r}\right)^{2}}^2}}}\\
\label{eq:profile2_tilde_Delta_large}
\ffin

Earlier in the situation for weak couplings for $\Xi=1$ we have $j^\star=N$; then, it determines that $a_m-a_{j^\star}=(N-m)(N-m-1)$ a quantity that is always a whole number and particularly for $m=N-1$ is zero. This tells us that the contribution to $\rho$ of one of the summands comes as $r^{-2}$ (see \cref{eq:exact_rho_T}). Additionally, the density from \cref{eq:approx_rho_T_R} is written in terms of a sum of powers of the ratio between the radial distance and the radius of the cylinder, a result that holds at $\Xi=1$ ($\Gamma=2$). Although the two problems match qualitatively, a quantitative comparison demonstrates the limitations of the procedure followed for weak couplings in \cref{sec:method1}, the angular uncorrelated fluid limit.

\psfrag{AX}{$r/R$}
\psfrag{BY}{$\widetilde{\rho}$}
\ImT{0.38}{fig12}{profile_method2}{The density profile $\widetilde{\rho}$ near the charged disk for different values of the Manning parameter for $N=\xi$; here $\Delta=20$. The dashed curves represent the exact profile from \cref{eq:profile2_tilde}. The axis are in logarithmic to show the trend at short and long distances.}

\subsection{Integrated charge}
\label{sec:charge2}

\psfrag{AX}{$\log (r/R)$}
\psfrag{BY}{$Q(r)/N$}
\ImT{0.38}{fig13}{charge_method2}{The integrated charge $Q(r)/N$ as
  a function of the logarithmic distance for $\Delta=20$ and
  $\xi=3,6,9$. The exact result from integrating the density
  superimposes with the Monte Carlo curve in all cases. Notice the
  linear trend of the integrated charge for $\log(r/R)$ above $3$.
  \label{fig:QG2}
}

As for the charge,\juan{
\eemp
Q({r})=&\frac{N}{\xi}\int_{\widetilde{R}}^{\widetilde{r}}\widetilde{\rho}(\widetilde{r}^\prime)\, \widetilde{r}^\prime\dif\widetilde{r}^\prime=\frac{2}{\xi^2}\sum_{l=1}^{N}\frac{1}{\gamma(\Delta,\xi,l-1)}\int_{\widetilde{R}}^{\widetilde{r}}\left(\frac{\widetilde{r}^\prime}{\widetilde{R}}\right)^{2((l-1)-\xi)}\, \widetilde{r}^\prime\dif\widetilde{r}^\prime\\
=&\sum_{l=1}^{N}\frac{\gamma\Rpar{\log\frac{r}{R},\xi,l-1}}{\gamma(\Delta,\xi,l-1)}
\ffin
}which for neutral systems, or $\xi=N$, gives,
\eemp
Q({r})=&\frac{1}{\Delta}\log\frac{r}{R}+\sum_{l=1}^{N-1}\frac{\gamma\Rpar{\log\frac{r}{R},N,l-1}}{\gamma(\Delta,N,l-1)},\\
\ffin
an anticipated expression for the integrated charge since the last transition, located at $\Gamma=2$ (or $\Xi=1$), tells us that the profile of the condensing counter-ion behaves as $r^{-2}$ which to the above quantity indicates a linear term in a logarithmic scale as evidenced from the first term. The remaining terms can be approximated for large $\Delta$ to,\juan{
\eemp
Q({r})
\simeq&\frac{1}{\Delta}\log\frac{r}{R}+N-2+\frac{1-\left(\frac{R}{r}\right)^{2N}}{1-\left(\frac{R}{r}\right)^{2}},\\
\\
\label{int_charge_2}
\ffin
%
}which agrees with the fact that a cylinder is only able to bind
$(N-1)$ charges when $\Delta\to\infty$. Notice that the form of the
integrated profile for the condensed counter-ions (second plus third
terms in \cref{int_charge_2}) is zero at $r=R$, then quickly converges
to the value $N-1$ which is the number of condensed ions. The
evaporated counterpart (first term in \cref{int_charge_2}), as
mentioned earlier, amounts for a linear (in $\log r$ scale) behavior
of the function, as shown in \cref{fig:charge_method2} comparing to
the numeric results from Monte Carlo.

Continuing with the discussion on the large $\Delta$ limit at the end of \cref{sec:large_delta_gamma}, the decay of the profile tells us that, indeed, the system permits the evaporation of exactly one counter-ion; the profile that describes the evaporated particle behaves as $1/r^2$ (see the $j=N$ term in \cref{profile2}) agreeing with the notion that drove us to the same conclusion and represented by a linear trend in the integrated charge as shown in \cref{fig:charge_method2}.

\subsection{Contact density}
\label{sec:contact2}

The value of the density at contact is evaluated from \cref{eq:profile2_tilde} resulting in
\emp
\widetilde{\rho}(R)=\frac{2}{N\xi}\sum_{l=1}^{N}\frac{1}{\gamma(\Delta,\xi,l-1)}.
\label{contact2_R}
\fin
Taken a large box size, the only relevant contributions come from $l<\xi$, or, strictly speaking $l\leq\Whole{\xi}$ coinciding with the number of condensed counter-ions found in \cref{sec:contact1} at $\Xi=1$; hence,
\eemps
\widetilde{\rho}(R)\simeq&\frac{\Whole{\xi}}{N}\Rpar{\frac{\xi-1}{\xi}+\frac{\xi-\Whole{\xi}}{\xi}}.
\label{approx_contact2_R}
\ffins
which simplifies for $\xi_B=0$ to,
\eemp
\widetilde{\rho}(R)\simeq&\frac{\xi-1}{\xi}.
\label{approx2_contact2}
\ffin
The contact value equates the Manning fraction differing from its mean field counterpart, and that corresponding to strong coupling in the three dimensional case. Additionally, it determines the onset for condensation supported by \cref{approx_contact2_R} for $\xi=1$.

On the other hand, the value for the contact at the outer shell
diverges in the limit of large box sizes when $N>\xi$. The opposite situation, with $N<\xi$ tends to zero because in this case there is no evaporation. Henceforth, for $\xi=N$,
\emp
e^{2\Delta}\widetilde{\rho}(\widetilde{D})\simeq\frac{1}{N\xi\Delta}\simeq\frac{1}{\xi^2\Delta},
\label{approx_contact2_D}
\fin
a result that differs once again from the mean field limiting behavior of $1/\xi^2$.

\section{Intermediate couplings: The case $\Xi$ integer}
\label{sec:interm}

In this section, we consider the case when the coupling $\Xi$ is an
integer. The simplest case is when $\Xi=1$, which has been treated in
the previous section, where explicit analytic expressions for the
partition function and the density of counter-ions can be
obtained. When $\Xi=2, 3, \ldots$, some exact results can also be
obtained, by using an expansion of the powers of the Vandermonde
determinant in symmetric ($\Xi$ even) or antisymmetric ($\Xi$ odd)
polynomials, a technique that has been used in the study of the
two-dimesional one-component plasma~\cite{PhysRevE.49.5623,
  vsamaj1995functional, tellez1999exact, vsamaj2004two,
  tellez2012expanded} and the fractionary quantum Hall
effect~\cite{doi:10.1142/S0217751X94001734,doi:10.1142/S0217979293003838,
  0305-4470-27-12-026, PhysRevLett.103.206801}.

\subsection{Partition function}

Suppose $\Xi$ is even. To compute the partition function
\begin{equation}
  {\cal Z}=\frac{e^{-E_0}}{N!}\int \prod_{k=1}^N{d^2\r_k}
  \prod_{j=1}^{N} |z_j|^{-2\xi}
  \prod_{1\leq k < j \leq N} |z_j-z_k|^{2\Xi}
\end{equation}
it is useful to expand the power of the Vandermonde determinant
\begin{equation}
  \label{eq:vandermonde-exp}
  \prod_{1\leq k < j \leq N} (z_j-z_k)^{\Xi}=\sum_{\mu\leq\kappa^{(N)}}
  c_{\mu}^{(N)}(\Xi)\,
  m_{\mu} (z_1,\ldots, z_N)
\end{equation}
in monomial symmetric functions
\begin{equation}
  m_{\mu} (z_1,\ldots, z_N)=\frac{1}{\prod_i m_i!}\sum_{\sigma\in S_N}
  z_{\sigma(1)}^{\mu_1}\cdots z_{\sigma(N)}^{\mu_N}
\end{equation}
corresponding to a partition
$\mu=(\mu_1,\ldots,\mu_N)$ of $|\mu|=\sum_{k=1}^{N} \mu_k=\Xi N (N-1)/2$ such that
\begin{equation}
  \label{eq:part-cond}
  0\leq\mu_N\leq\cdots\leq \mu_1\leq (N-1)\Xi
\end{equation}
where $S_N$ is the permutation group of $N$ elements. The partition can
also be represented by the ocupation numbers $m_i$, that is the
frequency of the integer $i$ in the
partition $\mu$. As remarked in~\cite{PhysRevLett.103.206801}, the
expansion~(\ref{eq:vandermonde-exp}) only involves partitions $\mu$
that are dominated~\cite{macdonald1998symmetric} by the root partition defined as
$\kappa^{(N)}=((N-1)\Xi,\ldots, 2\Xi,\Xi,0)$. The coefficients
$c_\mu^{(N)}(\Xi)$ of the expansion satisfy some recurrence
relations~\cite{macdonald1998symmetric,vsamaj2004two,PhysRevLett.103.206801,tellez2012expanded} which can be used to compute them
numerically.

Using the orthogonality relation $\int_{[0,2\pi]^N} 
m_{\mu} (z_1,\ldots, z_N) m_{\mu'} (z_1,\ldots, z_N) \prod_{k=1}^N
d\theta_k =0$ if $\mu\neq\mu'$, one obtains
\begin{equation}
  {\cal Z}=A(\Xi,\xi,N,\Delta) Z(\Xi,\xi,N,\Delta) 
\end{equation}
where
\begin{equation}
  \label{eq:Zmu}
  Z(\Xi,\xi,N,\Delta)=\sum_{\mu\leq \kappa^{(N)}}
  \frac{c_{\mu}^{(N)}(\Xi)^2}{\prod_i m_i!} 
  \prod_{k=1}^{N} \gamma(\Delta,\xi,\mu_k)
\end{equation}
with 
\begin{eqnarray}
  A(\Xi,\xi,N,\Delta) &=& \left(\frac{R}{L}\right)^{-\Xi N}
  e^{-\beta Q^2_2\Delta/2} 
  (\pi R^2)^N\nonumber\\
  &=& \left(\frac{R}{L}\right)^{-\Xi N}
  e^{-\Xi(N-\frac{\xi}{\Xi})^2\Delta} 
  (\pi R^2)^N
\end{eqnarray}
and the function $\gamma$ is defined in (\ref{gamma})
\begin{equation}
  \label{eq:gamma}
  \gamma(\Delta,\xi,\mu_k)=
  \begin{cases}
  \frac{1-e^{2(\mu_k+1-\xi)\Delta}}{\xi-\mu_k-1} &  \text{if\ }\mu_k\neq
  \xi-1\\
  2\Delta &  \text{if\ }\mu_k=
  \xi-1\\
  \end{cases}
\end{equation}
We recall that $\Delta=\log(D/R)$. In the case where $\Xi$ is odd, the
power of the Vandermonde determinant should be expanded in monomial
antisymmetric functions, but the final result~(\ref{eq:Zmu}) still
holds. Notice that the simplest case $\Xi=1$ is included in this
general formalism. In that case there is only one partition, the root
partition $\kappa^{(N)}=(N-1,N-2,\ldots,1,0)$ with coefficient
$c_{\kappa^{(N)}}^{(N)}(1)=1$, and~(\ref{eq:Zmu}) reduces to (\ref{Z2}).

\subsection{Density profile}

With the same expansion of the power of Vandemonde determinant, one
can also compute the density profile~\cite{tellez1999exact}
\begin{equation}
  \label{eq:rho}
  \rho(r)=\frac{1}{\pi R^2}\frac{1}{Z(\Xi,\xi,N,\Delta)}
  \sum_{\mu\leq \kappa^{(N)}} \frac{c_{\mu}^{(N)}(\Xi)^2}{\prod_i m_i!} 
  \prod_{k=1}^{N} \gamma(\Delta,\xi,\mu_k)
  \sum_{\ell=1}^{N} \frac{(r/R)^{2(\mu_{\ell}-\xi)}}{\gamma(\Delta,\xi,\mu_{\ell})}
\end{equation}
and the integrated charge is
\begin{equation}
  \label{eq:lambda}
  Q(r)=\frac{1}{Z(\Xi,\xi,N,\Delta)}
  \sum_{\mu\leq \kappa^{(N)}} \frac{c_{\mu}^{(N)}(\Xi)^2}{\prod_i m_i!} 
  \prod_{k=1}^{N} \gamma(\Delta,\xi,\mu_k)
  \sum_{\ell=1}^{N}
  \frac{\gamma(\log(r/R),\xi,\mu_{\ell})}{\gamma(\Delta,\xi,\mu_{\ell})}
  \,,
\end{equation}
which is normalized such that $Q(D)=N$.
By regrouping terms with the same dependence on $r$, the previous
expressions can be rewriten as
\begin{equation}
  \label{eq:rho-poly}
  \rho(r)=(r/R)^{-2\xi}\sum_{n=0}^{(N-1)\Xi} \rho_n (r/R)^{2 n}
\end{equation}
and 
\begin{equation}
  \label{eq:Q}
  Q(r)=\sum_{n=0}^{(N-1)\Xi}
  Z_n\,\frac{\gamma(\log(r/R),\xi,n)}{\gamma(\Delta,\xi,n)}
\end{equation}
where the coefficients $\rho_n$ are given in terms of 
\begin{equation}
  \label{eq:Zn}
  Z_n=\frac{1}{Z(\Xi,\xi,N,\Delta)}
  \sum_{\mu\leq\kappa^{(N)}\atop\text{with\ }n\in\mu}
  \frac{c_{\mu}^{(N)}(\Xi)^2}{\prod_i m_i!}  \prod_{k=1}^{N}
  \gamma(\Delta,\xi,\mu_k)
\end{equation}
as
\begin{equation}
  \rho_n=\frac{Z_n}{\pi R^2 \gamma(\Delta,\xi,n)}
  \,.
\end{equation}
In~(\ref{eq:Zn}), the numerator has the same form as the partition
function~(\ref{eq:Zmu}), except that the sum is restricted to partitions
$\mu$ which include the number $n$ in them, whereas in the partition
function~(\ref{eq:Zmu}) the sum run over all partitions.

From~(\ref{eq:rho-poly}) one can derive an interesting relation
between the density at the contact of the inner disk and the outer
disk. Indeed, note that
\begin{equation}
  \rho(R)-(D/R)^2\rho(D)=\sum_{n=0}^{(N-1)\Xi} \rho_n
  \left(1-\left(\frac{D}{R}\right)^{2(n+1-\xi)}\right)
\end{equation}
but $\rho_n=Z_n (\xi-n-1)/[\pi R^2(1-(D/R)^{2(n+1-\xi)})]$ provided that
$\xi\neq n+1$. Then
\begin{eqnarray}
  \rho(R)-(D/R)^2\rho(D)&=&(\pi R^2)^{-1} \sum_{n=0}^{(N-1)\Xi} Z_n (\xi
  -n -1)
  \nonumber\\
  &=&
  \frac{1}{\pi R^2 Z(\Xi,\xi,N,\Delta)}
  \sum_{\mu\leq\kappa^{(N)}}
  \frac{c_{\mu}^{(N)}(\Xi)^2}{\prod_i m_i!}  \prod_{k=1}^{N}
  \gamma(\Delta,\xi,\mu_k) \sum_{\ell=1}^{N} (\xi-\mu_{\ell}-1)
  \,.
  \nonumber\\
\end{eqnarray}
Using the fact that $\sum_{\ell=1}^N \mu_{\ell} = N(N-1)\Xi/2$, this
simplifies to
\begin{equation}
\pi R^2\rho(R)-\pi D^2\rho(D)= N \left[\xi - \frac{(N-1)\Xi}{2} -1\right]
\,.
\end{equation}
This relationship has been derived on more general grounds and its
consequences explored in~\cite{MTT14}.

\psfrag{AX}{$r/R$}
\psfrag{BY}{$\widetilde{\rho}(\widetilde{r})$}
\ImT{0.38}{fig14}{rho2D_SC_G4_N13_D100}{The density
  $\widetilde{\rho}$ as a function of the radial distance $r/R$ for
  $\Xi=2$, $\Delta=100$, $N=13$, and different $\xi$ as indicated. The
  dashed curves represent the exact result from \cref{eq:rho} as
  compared to numerical results from Monte Carlo simulations (symbols)}


\subsection{Counter-ion condensation}

\subsubsection{Case $Q_2=0$} 

In this section we wish to study the behavior of the density profile
and the integrated charge when $\Delta=\log(D/R)\to\infty$.  Let us
consider first the case where the outer shell in not charged $Q_2=0$,
then the electroneutrality condition imposes $\xi=N\Xi$. In the
expansions of the density and the integrated charge in terms of
partitions, each particion should satisfy~(\ref{eq:part-cond}), that
is $\mu_k\leq (N-1)\Xi = \xi -\Xi < \xi$, then $\mu_k+1-\xi\leq
\Xi-1$. If $\Xi\geq 2$, then the function $\gamma(\Delta,\xi,\mu_k)$
defined in~(\ref{eq:gamma}) has a finite limit when $\Delta\to \infty$
\begin{equation}
  \lim_{\Delta\to\infty} \gamma(\Delta,\xi,\mu_k) =
  \frac{1}{\xi-\mu_k-1}
  \,.
\end{equation}
The partition function also has a finite limit
\begin{equation}
  \label{eq:Zinfty}
  Z(\Xi,\xi,N,\infty)=\sum_{\mu\leq \kappa^{(N)}}
  \frac{c_{\mu}^{(N)}(\Xi)^2}{\prod_i m_i!} 
  \prod_{k=1}^{N} (\xi-\mu_k-1)^{-1}
  \,.
\end{equation}
We can notice that the density in~(\ref{eq:rho-poly}) is a sum of
terms of the form $r^{-2(\xi-n)}$ with $0\leq n\leq (N-1)\Xi$. Since
$\xi=N\Xi$, this means that $-\xi\leq -(\xi-n) \leq -\Xi$.  Therefore,
the density decays as
\begin{equation}
\rho(r) \mathop\sim_{r\to\infty} \rho_{(N-1)\Xi}\ r^{-2\Xi}
\,.
\end{equation}
The prefactor is 
\begin{equation}
  \label{eq:pref-rho-prelim}
  \rho_{(N-1)\Xi}  =  \frac{1}{\pi R^2 Z(\Xi,\xi,N,\infty)}
  \sum_{\mu\leq\kappa^{(N)}\atop\text{with\ }\mu_1=(N-1)\Xi}
  \frac{c_{\mu}^{(N)}(\Xi)^2}{\prod_i m_i!}  \prod_{k=2}^{N}
  (\xi-\mu_k-1)^{-1}
  \,.
\end{equation}
The above sum involves only partitions
$\mu=(\mu_1,\mu_2,\ldots,\mu_N)$ such that $\mu_1=(N-1)\Xi$. Consider
the partition $\tilde{\mu}=(\mu_2,\ldots,\mu_N)$, it is a partition
with $N-1$ elements with $|\tilde{\mu}|=(N-1)(N-2)\Xi/2$. One can
think of $\mu$ as being composed by two partitions, one with one
element and the other one with $N-1$ elements:
$\mu=((N-1)\Xi,\tilde{\mu})$. Using a factorization property of the
coefficients of the partitions shown in~\cite{PhysRevLett.103.206801} one has
$c_{\mu}^{(N)}(\Xi)= c_{\tilde\mu}^{(N-1)}(\Xi)$. Therefore, the
numerator of~(\ref{eq:pref-rho-prelim}) is the partition function of a
system with $N-1$ particles. Then,
\begin{equation}
\rho(r) \mathop\sim_{r\to\infty} 
\frac{1}{\pi R^2} \frac{Z(\Xi,\xi,N-1,\infty)}{Z(\Xi,\xi,N,\infty)}
\ r^{-2\Xi}
\,.
\label{eq:rho-asympt-r-infty}
\end{equation}

Since we are considering only the case $\Xi>1$, this means that the
density decays faster that $r^{-2}$ when $r\to\infty$. This behavior
is different from the prediction of the Poisson--Boltzmann equation:
the mean field regime does not applies even at large distances from
the inner disk. This can be contrasted to the situation for a three
dimensional system of a charged cylinder, where at large distances the
density profile behaves as the mean field prediction~\cite{doi:10.1021/jp311873a}
\juan{a fact that can be understood as a sign that the coupling is independent
of the density in two dimensions}.

The decay of the density faster than $r^{-2}$ is also an indication
that all counter-ions condense into the inner charge disk. Indeed, the
integrated charge~(\ref{eq:Q}) has a limit for $\Delta\to\infty$
\begin{equation}
  \label{eq:lambda_infty}
  Q_{c}^{(N)}(r)=\sum_{n=0}^{(N-1)\Xi} Z_n^{\infty}\  (1-(r/R)^{-2(\xi-n-1)})
\end{equation}
with
\begin{equation}
  Z_n^{\infty}=\frac{1}{Z(\Xi,\xi,N,\infty)} 
  \sum_{\mu\leq\kappa^{(N)}\atop\text{with\ }n\in\mu}
  \frac{c_{\mu}^{(N)}(\Xi)^2}{\prod_i m_i!}  \prod_{k=1}^{N}
  (\xi-\mu_k-1)^{-1}
\end{equation}
which has the limit $Q_{c}^{(N)}(r)\to N$, when $r\to\infty$:
all counter-ion are condensed. The notation $Q_{c}^{(N)}$ with
the subscript $c$ (condensed) and the superscript $(N)$ has been
choosen to recall that this is the integrated charge of a system with $N$
condensed ions.

Another indication of the complete condensation of counter-ions when
$\Xi>1$ and $Q_2=0$ can be noticed in the absence of an inflexion
point in the curve of $Q(r)$ as a function of $\log r$. Indeed,
returning to the case $D<\infty$, we have
\begin{equation}
  \frac{\partial^2 Q(r)}{(\partial \log(r/R))^2}
  =
  \frac{1}{Z(\Xi,\xi,N,\Delta)} 
  \sum_{\mu\leq \kappa^{(N)}} \frac{c_{\mu}^{(N)}(\Xi)^2}{\prod_i m_i!} 
  \prod_{k=1}^{N} \gamma(\Delta,\xi,\mu_k)
  \sum_{\ell=1}^{N}
  \frac{4(\xi-\mu_{\ell}-1)^2 (r/R)^{2(\mu_{\ell+1-\xi})}
  }{1-e^{-2(\xi-\mu_{\ell}+1)\Delta}}
  \,,
  \label{eq:inflex}
\end{equation}
Since $\xi>\mu_{\ell}+1$ when $\Xi>1$, then
$\gamma(\Delta,\xi,\mu_k)>0$ and we notice that each term in the
sum~(\ref{eq:inflex}) is positive. Therefore $\frac{\partial^2
  Q(r)}{(\partial \log(r/R))^2}>0$ and never vanishes: the curve
$Q(r)$ {\em vs.\/} $\log r$ does not have an inflexion point.

\subsubsection{Case $Q_2>0$: unbinding of one counter-ion} 

If the outer shell is charged $Q_2>0$, then one can choose
independently the charge of the inner disk $\Xi$ and the number of
counter-ions $N$, these are no longer restricted by the relation
$\xi=N\Xi$. The global electroneutrality now reads $N \Xi -\xi= \Xi
Q_2/q$. Starting from the situation of the previous section $\xi=\Xi
N$, we will see that if $\xi$ decreases keeping fixed $N$, some
counter-ions will progresively start to unbind from the inner disk.

The existence of unbound ions can be traced back to a divergence
in the partition function when $D\to\infty$. In the
expression~(\ref{eq:Zmu}) for the partition function we notice that $Z$
will diverge if at least one of the functions $\gamma(\Delta, \xi,
\mu_k)$ diverges when $\Delta\to\infty$. This occurs if it exists at
least one $\mu_k$ such that $\mu_k + 1 \geq \xi$. By construction,
the largest possible value for $\mu_k$ is $(N-1) \Xi$. Therefore
unbinding of ions will occur as soon as $(N-1) \Xi + 1 \geq \xi$,
ie.~$N\geq \frac{\xi-1}{\Xi}+1$. 

We can use this argument as a basis for the definition of the number
of condensed ions $N_c$. Let us define $N_c$ as the number of
ions such that the partition function of $N_c$ ions converges,
$\lim_{\Delta\to\infty} Z(\Xi,\xi, N_c,\Delta)<\infty$, but the
partition function with $N_c+1$ ions diverges, $\lim_{\Delta\to\infty}
Z(\Xi,\xi, N_c+1,\Delta)=\infty$. Then from the previous analysis,
$N_c$ is the integer such that 
\begin{equation}
  \frac{\xi-1}{\Xi} \leq N_c <   \frac{\xi-1}{\Xi} + 1
  \quad\text{that\ is\ }
    N_c = \ceil{\frac{\xi-1}{\Xi}}
\end{equation}
where $\ceil{x}$ is the ceiling function. Notice that, in general, the
number of condensed ions $N_c$ and unbound ones $N-N_c$ are
different from the number of charges at the inner and outer disk,
\begin{eqnarray}
    N_c = \ceil{\frac{\xi-1}{\Xi}} &\text{\ and\ }
    & \frac{Q_1}{q}=\frac{\xi}{\Xi} \,,\\
    N-N_c = N-\ceil{\frac{\xi-1}{\Xi}} &
    \text{\ and\ }& \frac{Q_2}{q}=N-\frac{\xi}{\Xi}
    \,.
\end{eqnarray}

For a fixed number of counter-ions $N$, as $\xi$ is decreased from the
value $N\Xi$ (situation where $Q_2=0$), the number of condensed
counter-ions will decrease, in a piecewise fashion due to the ceiling
function. 

While $N\Xi +1-\Xi<\xi\leq N\Xi$ all counter-ions remain condensed
$N_c=N$, the partition function of the system has the same form as in
the previous section given by Eq.~(\ref{eq:Zinfty}), in the limit
$\Delta\to\infty$. The first evaporation of one ion will occur when
$\xi=N\Xi +1 -\Xi$, then $N_c=N-1$. Let us consider this case in some
detail. If $\xi=(N-1)\Xi +1$, then for some partitions such that its
largest member is $\mu_1=(N-1)\Xi$, we have $\mu_1=\xi-1$. Then in the
partition function~(\ref{eq:Zmu}), the corresponding function
$\gamma(\Delta,\xi,\mu_1)= 2\Delta$ diverges when
$\Delta\to\infty$. Thus the partition function will diverge when
$\Delta\to\infty$. Its leading order is
\begin{equation}
  Z(\Xi,\xi, N,\Delta)\mathop\sim_{\Delta\to\infty}
  Z_{\infty}(\Xi,\xi, N)=2\Delta
  \sum_{\mu\leq \kappa^{(N)} \text{with}\atop\mu_1=(N-1)\Xi}
  \frac{c_{\mu}^{(N)}(\Xi)^2}{\prod_i m_i!} 
  \prod_{k=2}^{N} ( \xi-\mu_k -1)^{-1} 
  \,.
\end{equation}
Using the same argument that lead to~(\ref{eq:rho-asympt-r-infty}), we
recognize in the above expression the partition function of a system
with $N-1$ particles,
\begin{equation}
  \label{eq:ZinftyDelta}
   Z_{\infty}(\Xi,\xi, N)=2\Delta Z(\Xi,\xi,N-1,\infty)
   \,.
\end{equation}
Then, 
\begin{equation}
  -\log(Z_{\infty}(\Xi,\xi,N))=-\log(2\Delta)-\log(Z(\Xi,\xi,N-1,\infty))
\,.
\end{equation}
From this, we see that the free energy of the system has a dominant
contribution $-k_B T\log(2\Delta)$ due to the free energy from the
unbound ion, plus a subdominant contribution
($-\log(Z(\Xi,\xi,N-1,\infty))$, finite as $\Delta\to\infty$) from the
$N-1$ condensed ions. More precisely, adding the contributions from
the charged boundaries, the total free energy (in units of $k_BT$, $F=-\log
{\cal Z}$) of the system is
\begin{equation}
  F(\Xi,\xi,N,\Delta) = -\log(2\Delta) 
  +\left(\Xi-2\right)\Delta
  + F(\Xi,\xi,N-1,\infty)
  +\Xi\log\frac{R}{L}-\log(\pi R^2)
  +O(\Delta^{-1})
\end{equation}
where $F(\Xi,\xi,N-1,\infty)$ is the free energy of a system with
$N-1=N_c$ condensed particles when $\Delta\to\infty$. 

A similar analysis can be done for the density profile by separating
the contributions of partitions with $\mu_1=(N-1)\Xi$ which are
dominant. We have $\rho(r)\displaystyle\mathop\sim_{\Delta\to\infty}
\rho_{(N)}^{\infty}(r)$ with
\begin{equation}
  \rho_{(N)}^{\infty}(r)=\frac{1}{\pi R^2}\frac{2\Delta}{Z_{\infty}(\Xi,\xi,N)}
  \sum_{\mu\leq \kappa^{(N)} \text{with}\atop\mu_1=(N-1)\Xi}
  \frac{c_{\mu}^{(N)}(\Xi)^2}{\prod_i m_i!} 
  \prod_{k=2}^{N} (\xi-\mu_k-1)^{-1}
  \left[
  \sum_{\ell=2}^{N} \frac{(r/R)^{2(\mu_{\ell}-\xi)}
  }{(\xi-\mu_{\ell}-1)^{-1}}+\frac{(r/R)^{-2}}{2\Delta}
  \right]
\,.
\end{equation}
Introducing the density $\rho_{c}^{(N-1)}(r)$ for a system with $N-1$
particles (all condensed), but with the same values of $\Xi$ and $\xi$
and $\Delta\to\infty$, we notice that
\begin{equation}
  \rho_{(N)}^{\infty}(r)=\rho_{c}^{(N-1)}(r)+\frac{1}{2\pi r^2\Delta}
  \,.
\end{equation}
If $r\ll D$, the last term vanishes as $\Delta\to\infty$, and the
density of the system with $N$ is the same as the one with $N-1$
particles. This is an explicit indication that one counter-ion has
evaporated. The contribution to the density from this evaporated ion
is the term $1/(2\pi r^2 \Delta)$ which vanishes in the limit
$\Delta\to\infty$. 

From the above expression we deduce that the integrated charge has
also one contribution from the $N-1$ condensed ions, plus an
additional contribution from the evaporated ion
\begin{equation}
  Q(r)\mathop{=}_{\Delta\to\infty}
  Q_{c}^{(N-1)}(r)+ \frac{\log(r/R)}{\Delta}
\end{equation}
where $Q_{c}^{(N-1)}(r)$ is the integrated charge of a system
with $N-1$ ions which are all condensed (with $\Delta=\infty$), and it
is given by an expression similar to~(\ref{eq:lambda_infty}) but with
the replacement of $N$ by $N-1$ particles. In particular, we notice
from~(\ref{eq:lambda_infty}) that if $r\gg R$,
$Q_{c}^{(N-1)}(r)= N-1+ O((r/R)^{-2\Xi})$. Therefore, when
$r\gg R$,
\begin{equation}
  \label{eq:lambda-sim}
  Q(r)= N-1 + \frac{\log(r/R)}{\log(D/R)}
  +O\left(e^{-2\Xi\log(r/R)}\right)
\end{equation}
where we recalled the fact that
$\Delta=\log(D/R)$. From~(\ref{eq:lambda-sim}), we see that when
$Q(r)$ is plotted as a function of $\log(r/R)$ it should be a
function which varies fast (in the log scale) from 0 to $N-1$, then
linearly up to the value $N$. This behavior can be observed in \cref{fig:QG2}
for the case $\Xi=1$. In this limiting situation when $(\xi-1)/\Xi$ is
an integer (equal to $N-1$), the unbound ion is in fact ``floating''
between the inner and outer disk. The curve $Q(r)$ {\em vs.\/}
$\log(r)$ does not exhibit yet an inflexion point, but rather a linear
tendency.

Suppose now that we decrease $\xi$ below the previous value
($(N-1)\Xi+1$) when one ion has unbound from the inner disk,
$(N-2)\Xi+1<\xi <(N-1)\Xi+1$. We can repeat the previous analysis,
separating the partitions for which $\mu_{1}=(N-1)\Xi$ from the
rest. For those partitions, we have $\gamma(\Delta,\xi,\mu_1)\sim
\frac{e^{2((N-1)\Xi+1-\xi)\Delta}}{(N-1)\Xi+1-\xi}\to\infty$ when
$\Delta\to\infty$, while for any other element of the partition
$\mu_k$ (with $k\neq 1$), $\gamma(\Delta,\xi,\mu_k)\to
(\xi-\mu_k-1)^{-1}<\infty$. Using this and following similar steps to
the one that lead to~(\ref{eq:ZinftyDelta}) we obtain
\begin{eqnarray}
   -\log(Z(\Xi,\xi, N,\Delta))&\displaystyle
   \mathop{\sim}_{\Delta\to\infty}&
   -\log(Z(\Xi,\xi,N-1,\infty))
   -2((N-1)\Xi+1-\xi)\Delta
   \nonumber\\
   &&
   +\log\left[(N-1)\Xi+1-\xi\right]
\end{eqnarray}
The free energy is 
\begin{equation}
  F_{\infty}(\Xi,\xi,N) = (\Xi-2)\Delta
  +F(\Xi,\xi,N-1,\infty)
  +\Xi\log\frac{R}{L}-\log(\pi R^2)
  +\log\left[(N-1)\Xi+1-\xi\right]
\end{equation}

From this expression we see once again that the leading contribution
to the free energy is given by the unbound ion, which here
contributes with a term $(\Xi-2)\Delta$, plus some
subleading terms from the $N-1$ condensed ions.

Similarly as before, the density profile appears as a sum of a
contribution from the $N-1$ condensed ions and the unbound ion
\begin{equation}
  \rho(r)=\rho_{c}^{(N-1)}(r)
  + \frac{(N-1)\Xi+1-\xi}{\pi D^2} 
  \left(\frac{r}{D}\right)^{2((N-1)\Xi-\xi)}
  \,,
\end{equation}
when $\Delta\to\infty$.
Close to the inner disk, $r\ll D$, the second term is negligible and
$\rho(r)\simeq \rho_{c}^{(N-1)}(r)$. The second
term becomes important only close to the outer disk when $r\to D$.

In the limit $\Delta\to\infty$, the integrated charge is
\begin{equation}
  Q(r)=
  Q_{c}^{(N-1)}(r)+
  e^{-2((N-1)\Xi+1-\xi)(\Delta-\log(r/R))}
  +O(e^{-2((N-1)\Xi+1-\xi)\Delta})
\end{equation}
When $R\ll r\ll D$, we have
\begin{equation}
Q(r)=
N-1 + O(e^{-2((N-2)\Xi+1-\xi)\log\frac{r}{R}}) +
O(e^{-2((N-1)\Xi+1-\xi)(\Delta-\log\frac{r}{R})})
\,,
\end{equation}
and as $r\to D$ (ie. $\log(r/R)\to \Delta$), the integrated charges
approaches the value $N$, exponentially fast in $\log$ scale, ie. as
$e^{-2((N-1)\Xi+1-\xi)(\Delta-\log(r/R))}$.

\subsubsection{Case $Q_2>0$: unbinding of many counter-ions} 

In this subsection we study the general case where many counter-ions
unbind from the inner disk. Suppose that the charge of the inner disk
$\xi$ is such that $(N-N_u-1)\Xi+1 < \xi < (N-N_u)\Xi+1$, with $N_u$
an integer, $N_u\geq 2$, which is the number of unbound
counter-ions. Indeed, if we recall that the number of condensed ions
is $N_c=\ceil{\frac{\xi-1}{\Xi}} < N-1$, and we have $N_u=N-N_c$.

To put in evidence the counter-ion condensation from the analytical
expressions for the partition function, the density and integrated
charge profiles, it is convenient to recall some properties of the
coeficients $c_{\mu}^{(N)}$ of the
expansion~(\ref{eq:vandermonde-exp}) of the power of the Vandermonde
determinant. The partitions $\mu$ present in the expansion are
dominated by the root partition $\kappa^{(N)}$, that is $\mu$ can be
obtained from $\kappa^{(N)}$ by ``squeezing'' operations: $\mu_j
\mapsto \mu_j + n$ and $\mu_k\mapsto\mu_k -n$ with $j<k$ and
$n>0$. Consider that we divide the root partition into two parts
$\kappa^{(N)}=(\tilde{\kappa}^{(N_u)},\kappa^{(N_c)})$, where
$\kappa^{(N_c)}=((N_c-1)\Xi,(N_c-2)\Xi, \ldots, 0)$ and
$\tilde{\kappa}^{(N_u)}=((N-1)\Xi, (N-2)\Xi, \ldots, (N-N_u)\Xi)$. The
latter can be though as a partition of $N_u$ elements
$\kappa^{(N_u)}=((N_u-1)\Xi,(N_u-2)\Xi, \ldots, 0)$, with all its
parts shifted by $\Xi N_c$:
$\tilde{\kappa}^{(N_u)}_k=N_c\Xi+{\kappa}^{(N_u)}_k$. Suppose that
squeezing operations are performed on $\kappa^{(N_c)}$ to obtain a
partition $\mu^{(N_c)}$ and, separately, squeezings operations are
performed on $\kappa^{(N_u)}$ to obtain $\mu^{(N_u)}$. Define the
shifted partition $\tilde{\mu}^{(N_u)}_{k}=\Xi N_c +\mu^{(N_u)}_{k}$ and
consider the composite partition $\mu=(\tilde{\mu}^{(N_u)},
\mu^{(N_c)})$ of $N=N_u+N_c$ particles. Then it is shown
in~\cite{PhysRevLett.103.206801,MT2015} that the corresponding coefficients of the expansion of
the power $\Xi$ of the Vandermonde determinant of these partitions
satisfy the factorization relation
\begin{equation}
  \label{eq:factorization}
  c_{(\tilde{\mu}^{(N_u)}, \mu^{(N_c)})}^{(N)}(\Xi) =
  c_{\mu^{(N_u)}}^{(N_u)}(\Xi)
  \,
  c_{\mu^{(N_c)}}^{(N_c)}(\Xi)
  \,.
\end{equation}
In the analysis of the previous sections we used a special case of
this factorization property where $N_u=1$:
$c_{((N-1)\Xi,\mu^{(N-1)})}^{(N)}=c_{(0)}^{(1)}c_{\mu^{(N-1)}}^{(N-1)}$ 
with $c_{(0)}^{(1)}=1$.

With the aid of the factorization property~(\ref{eq:factorization}) we
will be able to factorize the leading order, when $\Delta\to\infty$,
of the partition function of $N$ particles into partitions functions
of $N_u$ and $N_c$ particles. Indeed, consider the contribution to the
partition function~(\ref{eq:Zmu}) from partitions
$\mu=(\tilde{\mu}^{(N_u)}, \mu^{(N_c)})$ constructed as explained
earlier. For the parts $\mu_k$ of $\mu$ with $k=1,\ldots,N_u$ that
belong to $\tilde{\mu}^{(N_u)}$, we have
\begin{equation}
  \gamma(\Delta,\xi,\mu_k)\sim \frac{e^{2(\mu_k+1-\xi)\Delta}}{\mu_k+1-\xi}
  \to\infty\quad\text{when\ }\Delta\to\infty
\end{equation}
because $\mu_k+1\geq (N-N_u)\Xi +1 = N_c\Xi+1 > \xi$. Furthermore, by
the nature of the squeezing operations one has that the sum
\begin{equation}
\sum_{k=1}^{N_u}
\mu_k = \sum_{\ell=1}^{N_u} (N-\ell)\Xi = 
\frac{(N-1+N_c)N_u\Xi}{2}
\end{equation}
is fixed and equal to the the same sum for the root partition. Then 
\begin{equation}
  \prod_{k=1}^{N_u} \gamma(\Delta,\xi,\mu_k)
  \mathop\sim_{\Delta\to\infty}
  e^{N_u\left[(N-1)\Xi+N_c\Xi+2(1-\xi)\right]\Delta}
  \prod_{k=1}^{N_u}\frac{1}{\mu_k+1-\xi}
  \,.
\end{equation}
Notice that the coefficient of the exponential is independent of the
partition considered. On the other hand the contribution from the
other parts of the partition are finite when $\Delta\to\infty$
\begin{equation}
  \label{eq:leading-gamma}
  \prod_{k=N_u+1}^{N} \gamma(\Delta,\xi,\mu_k)
  \mathop\to_{\Delta\to\infty}
  \prod_{k=N_u+1}^{N}
  \frac{1}{\xi-\mu_k-1}
  \,.
 \end{equation}

Consider the contribution from a partition $\nu$ that is not
constructed from the mecanism below. This means that at some point a
squeezing operation was performed with a part from the first $N_u$
terms and another part from the remaining $N_c$ terms. For example,
consider the following partition $\nu$ constructed from the following
squeezing of the root partition $\kappa^{(N)}$
\begin{equation}
  \kappa_j^{(N)}\mapsto \kappa_j^{(N)} - n = \nu_j 
  \text{\ and\ }
  \kappa_{\ell}^{(N)}\mapsto \kappa_{\ell}^{(N)} + n = \nu_{\ell}
  \text{\,,\ with\ } 
  j\leq N_u \text{\ and\ } \ell \geq N_u+1
  \,.
\end{equation}
Then, if $\nu_{\ell}< (N-N_u)\Xi$, the leading order contribution of
this partition, when $\Delta\to\infty$, is
\begin{equation}
  \prod_{k=1}^{N_u}\gamma(\Delta,\xi,\nu_k)
  \mathop\sim_{\Delta\to\infty}
  e^{N_u\left[(N-1)\Xi+N_c\Xi+2(1-\xi-n)\right]\Delta}
  \prod_{k=1}^{N_u}\frac{1}{\nu_k+1-\xi}
  \,,
\end{equation}
which is a subdominant contribution compared
to~(\ref{eq:leading-gamma}). In the case where $\nu_{\ell}\geq
(N-N_u)\Xi$ the contribution is
\begin{multline}
  \gamma(\Delta,\xi,\nu_{\ell})\prod_{k=1}^{N_u}\gamma(\Delta,\xi,\nu_k)
  \mathop\sim_{\Delta\to\infty}
  \frac{1}{\nu_{\ell}+1-\xi}\prod_{k=1}^{N_u}\frac{1}{\nu_k+1-\xi}
  \\ \times
  \exp\left[N_u\left[(N-1)\Xi+N_c\Xi+2(1-\xi-(\xi-\kappa^{(N)}_{\ell}+1))\right]
    \Delta\right]
  \,,
\end{multline}
but since $\xi-\kappa^{(N)}_{\ell}+1>0$, this contribution is again subdominant
compared to~(\ref{eq:leading-gamma}). Thus, to leading order in
$\Delta$ only partitions of the form $\mu=(\tilde{\mu}^{(N_u)},
\mu^{(N_c)})$ contribute to the partition function. Then, using the
factorization property of the coefficients~(\ref{eq:factorization}),
the partition function can also be factorized
\begin{equation}
  \label{eq:Zfactor}
  Z(\Xi,\xi,N,\Delta)
  \mathop\sim_{\Delta\to\infty}
  Z_{u}(\Xi,\xi,N_u,\Delta)
  Z(\Xi,\xi,N_c,\infty)
\end{equation}
with
\begin{equation}
  \label{eq:Zu-prelim}
  Z_{u}(\Xi,\xi,N,\Delta)=
  e^{N_u\left[(N-1)\Xi+N_c\Xi+2(1-\xi)\right]\Delta}
  \sum_{\mu^{(N_u)}\leq \kappa^{(N_u)}}
  \frac{c_{\mu^{(N_u)}}^{(N_u)}(\Xi)^2}{\prod_{i\in\mu^{(N_u)}} m_i!} 
  \prod_{k=1}^{N_u} (\tilde{\mu}^{(N_u)}_k+1-\xi)^{-1}
\end{equation}
and $Z(\Xi,\xi,N_c,\infty)$ is given by (\ref{eq:Zinfty}) but for a
system of $N_c$ (condensed) particles. The partition
function~(\ref{eq:Zu-prelim}) gives the contribution from the unbound
ions. Recalling that $\tilde{\mu}_k^{(N_u)}=\Xi N_c+\mu_k^{(N_u)}$, we
notice that this contribution~(\ref{eq:Zu-prelim}) can be rewritten as
\begin{equation}
  \label{eq:Zu}
    Z_{u}(\Xi,\xi,N,\Delta)=
    e^{N_u\left[(N-1)\Xi+N_c\Xi+2(1-\xi)\right]\Delta}
    Z^{*}_u(\Xi,\xi-N_c\Xi,N_u)
    \,.
\end{equation}
with
\begin{equation}
  \label{eq:Zstar}
  Z^{*}_{u}(\Xi,z,n)=\sum_{\mu\leq \kappa^{(n)}}
  \frac{c_{\mu}^{(n)}(\Xi)^2}{\prod_i m_i!} 
  \prod_{k=1}^{n} (\mu_k+1-z)^{-1}
  \,.
\end{equation}
Notice that, formally, $Z^{*}_{u}(\Xi,z,n)=(-1)^n Z(\Xi,z,n,\infty)$
if $Z(\Xi,z,n,\infty)$ is analytically continued using the left hand
side of~(\ref{eq:Zinfty}) when $z<(n-1)\Xi+1$. Thus, in a loose sense,
$Z^{*}_u(\Xi,\xi-N_c\Xi,N_u)$ is the partition function of a system of
$N_u$ particles and the inner disk with its charge reduced by the
$N_c$ charges of the condensed ions. The free energy is
\begin{equation} 
  F_{\infty}(\Xi,\xi,N)=F(\Xi,\xi,N_c,\infty)
  -\log\left[Z_u^*(\Xi,\xi-N_c\Xi,N_u)\right]
  +N_u\left[\Xi\log\frac{R}{L}+(\Xi-2)\Delta
  -\log(\pi R^2)\right]
\end{equation}

Using the factorization of the partition function, one can find that
at leading order in $\Delta$, the density profile is the sum of two
contributions, 
\begin{equation} 
  \rho(r)=\rho_c^{(N_c)}(r)+\rho_u^{(N_u)}(r)
\end{equation}
with the contribution from the condensed ions
\begin{equation}  
  \rho_c^{(N_c)}(r)= \frac{1}{\pi R^2}
  \frac{1}{Z(\Xi,\xi,N_c,\infty)}
  \sum_{\mu^{(N_c)}\leq \kappa^{(N_c)}} 
  \frac{c_{\mu^{(N_c)}}^{(N_c)}(\Xi)^2}{\prod_i m_i^{(N_c)}!} 
  \prod_{k=1}^{N_c} (\xi-\mu_k^{(N_c)}-1)^{-1}
  \sum_{\ell=1}^{N_c} 
  \frac{(r/R)^{2(\mu_{\ell}^{(N_c)}-\xi)}}{(\xi-\mu_{\ell}^{(N_c)}-1)^{-1}}
\end{equation}  
 and the contribution from the unbound the unbound ions
\begin{equation} 
  \rho_u^{(N_u)}(r)= \frac{1}{\pi D^2} \frac{1}{Z_{u}^{*}(\Xi,\xi-N_c\Xi,N_u)}
  \sum_{\mu^{(N_u)}\leq \kappa^{(N_u)}} \frac{c_{\mu^{(N_u)}}^{(N_u)}(\Xi)^2}{\prod_i
    m_i!}  \prod_{k=1}^{N_u} (\mu_k^{(N_u)}+1+N_c\Xi-\xi)^{-1}
  \sum_{\ell=1}^{N_u}
  \frac{(r/D)^{2(\mu_{\ell}^{(N_u)}+N_c\Xi-\xi)}}{(\mu_{\ell}^{(N_u)}+1+N_c\Xi-\xi)^{-1}}
  \,.
\end{equation}  
From this, it follows that the integrated charge is
\begin{equation} 
  Q(r)=Q_c^{(N_c)}(r)+Q_u^{(N_u)}(r)
  \label{eq:lambda-nofloat}
\end{equation}
with
\begin{equation}
  \label{eq:lambdac}
  Q_c^{(N_c)}(r)=N_c-
  \frac{1}{Z(\Xi,\xi,N_c,\infty)}
  \sum_{\mu^{(N_c)}\leq \kappa^{(N_c)}} 
  \frac{c_{\mu^{(N_c)}}^{(N_c)}(\Xi)^2}{\prod_i m_i^{(N_c)}!} 
  \prod_{k=1}^{N_c} \frac{1}{\xi-\mu_k^{(N_c)}-1}
  \sum_{\ell=1}^{N_c} 
  e^{-2(\xi-\mu_{\ell}^{(N_c)}-1)\log(r/R)}
\end{equation}
and
\begin{multline}
  \label{eq:lambdau}
  Q_u^{(N_u)}(r) =
  \frac{1}{Z_{u}^{*}(\Xi,\xi-N_c\Xi,N_u)}
  \sum_{\mu^{(N_u)}\leq \kappa^{(N_u)}} 
  \frac{c_{\mu^{(N_u)}}^{(N_u)}(\Xi)^2}{\prod_i m_i^{(N_u)}!} 
  \prod_{k=1}^{N_u} \frac{1}{\mu_k^{(N_u)}+1+\Xi N_c-\xi}
  \\ \times
  \sum_{\ell=1}^{N_u} 
  \left(
  e^{-2(\mu_{\ell}^{(N_u)}+1+\Xi N_c - \xi)(\Delta-\log\frac{r}{R})}
  - e^{-2(\mu_{\ell}^{(N_u)}+1+\Xi N_c - \xi)\Delta}
  \right)
  \,.
\end{multline}
We notice that if $R\ll r \ll D$, 
\begin{equation}
  \label{eq:lambdac-sim}
  Q_c^{(N_c)}(r)=N_c+O(e^{-2(\xi-(N_c-1)\Xi -1)\log\frac{r}{R}})
\end{equation}
and
\begin{equation}
  \label{eq:lambdau-sim}
  Q_u^{(N_u)}(r)=O(e^{-2((N-1)\Xi+1-\xi)(\Delta-\log\frac{r}{R})})
\,.
\end{equation}
Thus the integrated charge density increases from 0 when $r=R$ to
$Q(r)\simeq N_c$ in that intermediate region $R\ll r \ll
D$. Then, when $r$ is close to $D$,
\begin{equation}
  Q_u^{(N_u)}(r)\mathop\to_{r\to D} N_u
\end{equation}
exponentially fast in log scale, to finally recover the total number
of particles 
\begin{equation}
  Q(r)\mathop\to_{r\to D} N_c+N_u=N
  \,.
\end{equation}

The case when $(\xi-1)/\Xi$ is an integer, is a special limiting
case. The number of condensed counter-ions is $N_c=(\xi-1)/\Xi$. The
dominant terms in the partition function, when $\Delta\to\infty$, are
due to partitions of the form $\mu=(\tilde{\mu}^{(N_u-1)}, N_c \Xi,
\mu^{(N_c)})$ where $\tilde{\mu}^{(N_u-1)}$ is a partition of $N_u-1$
parts squeezed from the corresponding root partition
$\tilde{\kappa}^{(N_u-1)}$. The $N_u$-th part of the partition is
fixed $\mu_{N_u}=N_c \Xi$, and $\mu^{(N_c)}$ is, as before, a
partition of $N_c$ parts squeezed from the root partition
$\kappa^{(N_c)}$. The results~(\ref{eq:Zfactor}) and~(\ref{eq:Zu}) for
the partition function become
\begin{equation}
  \label{eq:Zfactor-log}
  Z(\Xi,\xi,N,\Delta)
  \mathop\sim_{\Delta\to\infty}
  2\Delta e^{N_u(N_u-1)\Xi \Delta}
  Z_{u}^{*}(\Xi,1,N_u-1)
  Z(\Xi,\xi,N_c,\infty)
  \,.
\end{equation}
Thus, the free energy acquires an additional $\log \Delta$ correction
\begin{equation}
  F_{\infty}(\Xi,\xi,N)=F(\Xi,\xi,N_c,\infty)
  -\log\left[Z_u^*(\Xi,1,N_u-1)\right]
  +N_u\left[\Xi\log\frac{R}{L}+(\Xi-2)\Delta
    -\log(\pi R^2)\right]
  -\log(2\Delta)
  \,.
\end{equation}
This $\log\Delta$ correction, which is a contribution coming from the
$N_u$-th part of each partition ($\mu_{N_u}=N_c \Xi$), is the
fingerprint of the existence of a ``floating'' counter-ion. Indeed,
the density profile is now the sum of three contributions, one from
$N_c$ condensed ions $\rho_c^{(N_c)}$, one from $N_u-1$ unbind
counter-ions $\rho_u^{(N_u-1)}$, and an additional contribution from
one floating ion, proportional to $r^{-2}$,
\begin{equation}
  \rho(r)
  \mathop{=}_{\Delta\to\infty}
  \rho_c^{(N_c)}(r)
  +\rho_u^{(N_u-1)}(r)
  +\frac{1}{2\pi r^2 \Delta}
  \,.
\end{equation}
Similarly, the integrated charge can be cast as
\begin{equation}
  Q(r)=Q_c^{(N_c)}(r)+Q_u^{(N_u-1)}(r)+
  \frac{\log(r/R)}{\Delta}
  \label{eq:lambda-float}
\end{equation}
with the integrated charge corresponding to the condensed ions
$Q_c^{(N_c)}(r)$ given by~(\ref{eq:lambdac}) and the one
correspoding to $N_u-1$ unbind ions $Q_u^{(N_u-1)}(r)$ given
by~(\ref{eq:lambdau}) with the replacement $N_u\mapsto N_u-1$. The
charge of the condensed ions converges exponentially fast (in log
scale) to $N_c$ when $r\gg R$ (Eq.~(\ref{eq:lambdac-sim})). The term
$\log(r/R)/\Delta$, in log scale, varies linearly from 0 to 1 when $r$
varies from $R$ to $D$, thus linearly increasing the total integrated
charge from $N_c$ (close to $r=R$) to $N_c+1$ ($r=D$). Close $r=D$,
the charge corresponding to the $N_u-1$ unbind ions varies
exponentially fast (in log scale) from $0$ to $N_u-1$, thus completing
at $r=D$ the total charge $Q(D)=N$. This is illustrated in
\cref{fig:lambdar}, for the case $\Xi=2$. When $\xi=22.8$, the number
of condensed ions is $N_c=11$, and for $\xi=23.2$, $N_c=12$. The
results from \cref{fig:lambdar} for these two cases are compatible
with the approximation \cref{eq:lambda-nofloat}. In the case where
$\xi=23$, there is a ``floating'' ion because $(\xi-1)/\Xi=11$ is an
integer. In that case the integrated charge shown in \cref{fig:lambdar}
follows \cref{eq:lambda-float}.

\psfrag{AX}{$\log(r/R)$}
\psfrag{BY}{$Q(r)$}
\ImT{0.38}{fig15}{lambda2D_SC_G4_N13_D100}{The
  integrated charge $Q(r)$ as a function of the radial distance
  $r/R$ for $\Xi=2$, $\Delta=100$, $N=13$, and different $\xi$ as
  indicated. The solid curves represent the exact result from
  \cref{eq:lambda} as compared to numerical results from Monte Carlo
  simulations (symbols).
  \label{fig:lambdar}}

\section{Strong couplings: The case $\Xi\gg1$}
\label{sec:SC}

\ImTc{0.4}{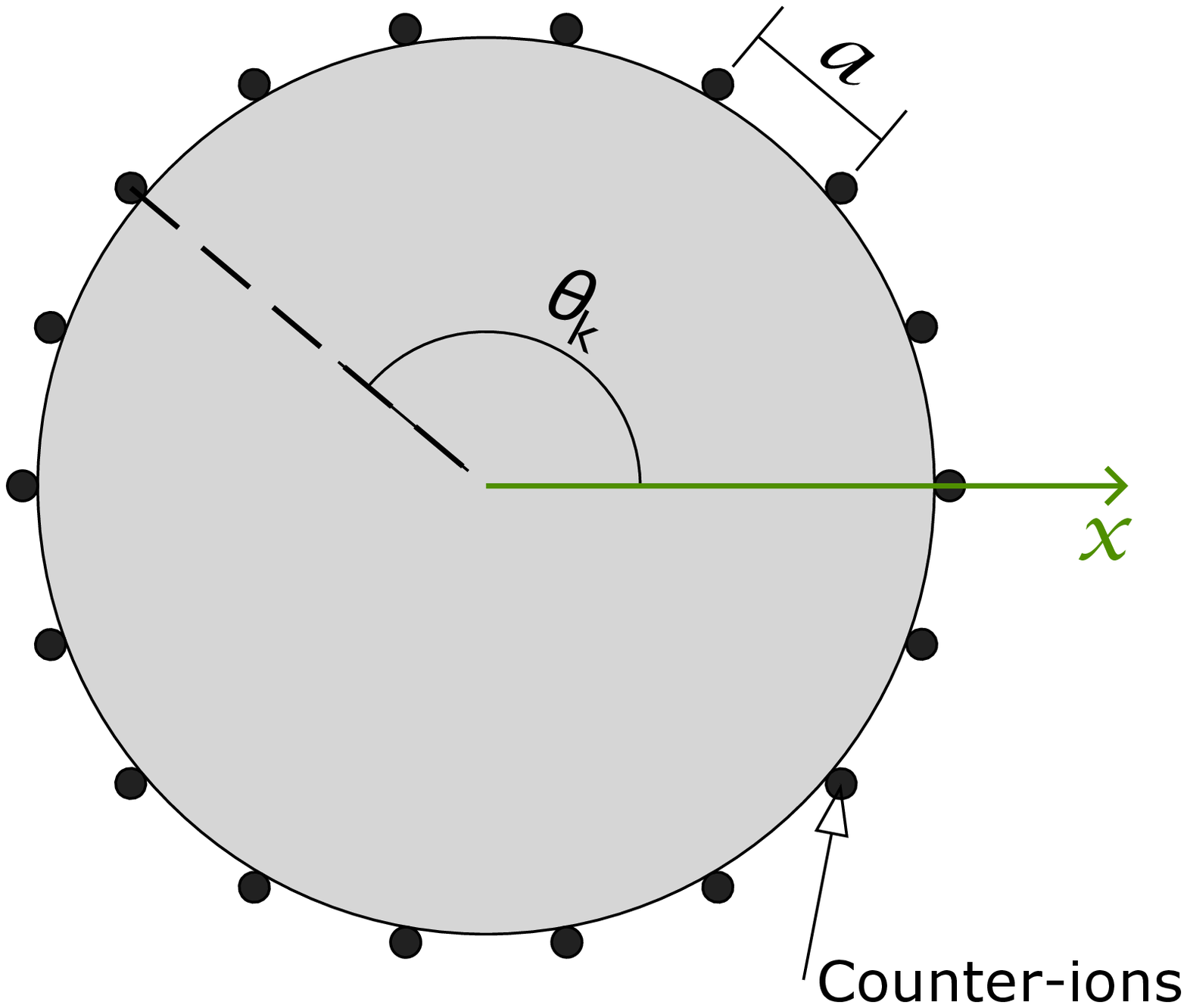}{2D_system_SC}{The ground state for the cylindrical cell model in two dimensions. The counter-ions sit on a position with $\theta_k=2\pi(k-1)/N$.}{The ground state for the two dimensional system.}

The strong coupling regime, indicated for $\Xi\gg1$ ($\Gamma\gg2$) is characterized by full condensation for $\xi_B=0$ as we have seen throughout the previous sections. We have insisted that the behavior at strong couplings is attributed from the small fluctuations of the counter-ions around the ground state \citep{PhysRevLett.106.078301}, a perspective of the minimal energy configuration at zero temperature (as shown in \cref{fig:2D_system_SC}) matches a Wigner crystal like a pebbled necklace with a minimum inter-particle distance $a$; notice that $a=R\sqrt{2-2\cos{\theta_{2}}}$. This has been evidenced in the investigations on strong coupling from \citet{vsamaj2011counter} on the two dimensional case with a charged plate. Then, the ground state positions are
\emp
\VEC{r}_{j}^{(0)}=R\Bpar{\cos\theta_j\ii+\sin\theta_j\jj},
\fin
with,
\emp
\theta_{j}=\frac{2\pi(j-1)}{N}.
\fin

In the $T\to0$ image, the profile is dominated by the leading order, or the contribution to the energy due to the charged disk alone. Corrections to the leading order arise from the interaction between charges which are attributed to their arrangement at the crystalline positions. In this sense, looking at the ground state in \cref{fig:2D_system_SC}, we can write the shift of energy when an ion moves to a new position from its ground state as follows; from \cref{eq:H_reduced_units}, using \cref{eq:log_potential},
\emp
\beta\delta H=2\xi\sum_{j=1}^{N}\log\frac{\VVpar{\VEC{r}_j}}{R}-2\Xi \sum_{1\leq j<k\leq N}\Rpar{\log\frac{\VVpar{\VEC{r}_j-\VEC{r}_k}}{R}-\frac{1}{2}\log\Bpar{2-2\cos\theta_{jk}}},
\fin
with $\VEC{r}_j$ the new position of the $j$\TH\, particle and
$\theta_{jk}\defeq\theta_j-\theta_k$. In the previous equation $\delta
H=H-H_{GS}$ where $H_{GS}$ is the energy of the
ground-state \footnote{Using, from \citet{Gradshteyn2007},  
\emps
\prod_{j=1}^{N-1}\Rpar{2-2\cos\frac{2\pi}{N}j}=N^2
\fins
}
\eemp
\beta H_{GS}=&-\Xi \sum_{1\leq j<k\leq N}\log\Bpar{2-2\cos\theta_{jk}}+E_0\\
=&-\Xi N\log N+E_0.
\label{eq:H-GS}
\ffin
However, the effect of the remaining condensed counter-ions must not be disregarded. In fact, the Hamiltonian admits another factorization that permits to see this. One which takes into consideration further interactions that stem from the ground state due to curvature. Unlike the plate, investigated by \citet{vsamaj2011counter}, the contributions here come inevitably from the disk and the other condensed counter-ions to leading order as any displacement from the ground state will have a strong influence from the remaining counter-ions as well.

In order to see this, let us take \cref{eq:log_potential} that
reformulates the shift of energy as,
\begin{equation}
  \beta \delta H= H_{\text{SC-0}} + H_{\text{SC-1}}
  \,,
  \label{eq:H_SC}
\end{equation}
with
\begin{equation}
  H_{\text{SC-0}}=
  \Rpar{2\xi-\Xi (N-1)}\sum_{j=1}^{N}\log\frac{\VVpar{\VEC{r}_j}}{R}\\
  \,,
  \label{eq:H_SC-0}
\end{equation}
and
\eemp
H_{\text{SC-1}}=&-\Xi \sum_{j<k}\Rpar{\log\Bpar{2\cosh\Rpar{\log\frac{\VVpar{\VEC{r}_j}}{R}-\log\frac{\VVpar{\VEC{r}_k}}{R}}-2\cos\theta_{jk}}-\log\Bpar{2-2\cos\theta_{jk}}}.\\
\label{eq:H_SC-1}
\ffin
Here we can distinguish that the second term is small compared to the
first: $H_{\text{SC-1}}\ll H_{\text{SC-0}}$ . In fact, near
the ground state $\cosh(\log r_j-\log r_k)\approx 1$. At strong couplings ($\xi\to\infty$), the extension of the double layer goes to zero, enforcing the condition mentioned before. Nevertheless, the interaction of an ion with the disk and its neighbors is redefined by an effective one as can be observed in the first term of the energy.

This statement provides the starting point for the analysis of the strong coupling in two dimensions that is unique. Furthermore, it redefines the characteristic length of the diffusive layer, which corresponded to the Gouy-Chapman length like $R/\xi$. The redefined scale, according to the potential energy, is,
\emp
\mu_{\eff}=\frac{R}{\xi-\frac{\Xi }{2}(N-1)},
\label{eq:new_Gouy}
\fin
larger than its predecessor for the equivalent problem. Then, strong couplings at two dimensions reads for $\mu_{\eff}\ll a$ as
\eemp
\Rpar{\xi-\frac{\Xi }{2}(N-1)}\sqrt{2-2\cos\theta_2}\gg1,\\
\ffin
which is ultimately a condition imposed on the number of counter-ions coming from neutrality. Seen as $a<2\pi R/N$, the aforementioned conditions is equivalent to,
\eemps
\frac{2\xi-\Xi (N-1)}{N}\pi\gg1,
\ffins
which is ultimately,
\eemp
\Xi \frac{N-1}{N}\pi\gg1\\
\label{eq:SC_condition}
\ffin
a similar statement was proposed for the strong coupling by \citet{PhysRevLett.95.185703,PhysRevE.73.056105} for the same problem; they considered that the strong coupling regime is reached for small number of counter-ions given that at a constant Manning parameter the coupling is largest at $N=1$. Through this formulation, $N=1$ is not the proper limit for two dimensions since the absence of counter-ions disregards couplings. In fact, one should reconsider $N>1$ in which case \cref{eq:SC_condition} defines adequately the strong coupling regime. This teaches us that the strong coupling limit is achieved, for instance, at \emph{large} couplings alone. Then, deviations from the leading order are expected to be stronger with a small $\Xi$. At fixed $\xi$ it means that $N\to\infty$ suggests small couplings, a similar conclusion held by \citet{PhysRevE.73.010501}, \citet{PhysRevLett.95.185703,PhysRevE.73.056105} and \citet{PhysRevE.85.011119}.

\subsection{Density profile}
\label{sec:profile3}
\subsubsection{Leading order}

The shift of energy to leading order ($H_{\text{SC-0}}=\sum_{j=1}^N
h_{\text{SC-0}}(r_j)$) consists of one-particle decoupled
contributions $h_{\text{SC-0}}(r_j)=(2\xi-(N-1)\Xi)\log (r_j/R)$. Then
the density profile at leading order is given by 
\emp
\widetilde{\rho}_{\text{SC-0}}(\widetilde{r})=
\widetilde{\rho}_0\,e^{-h_{\text{SC-0}}(r)}
=\widetilde{\rho}_0\Rpar{\frac{\widetilde{R}}{\widetilde{r}}}^{2\xi-\Xi(N-1)}
\,.
\label{eq:rho0_SC}
\fin 
We recall that $\widetilde{\rho}(\widetilde{r})=2\pi\,
R^2\rho(r)/(N\xi)$. The proportionality constant $\widetilde{\rho}_0$
is equal to the contact density
($\widetilde{\rho}_0=\widetilde{\rho}(\widetilde{R})$) and it can be determined through
neutrality, since the counter-ions are fully condensed. Thus the
density should satisfy (taking $\Delta\to\infty$)
\begin{equation}
  \int_{\widetilde{R}}^{\infty} \widetilde{\rho}(\widetilde{r})
  \,\widetilde{r}\,\dif\widetilde{r} = \xi
  \,.
  \label{eq:neutrality-tilde}
\end{equation}
Therefore
\emp
\widetilde{\rho}_0=\frac{2(\xi-1)-\Xi(N-1)}{\xi}.
\label{eq:rho_SC_contact_R}
\fin
We will recall this result as the {\bf 2D--SC-0} as a reminder that it is the strong coupling to leading order.

The result (\ref{eq:rho_SC_contact_R}) for the contact density turns out to be  is identical to \cref{eq:contact_rho_R_DeltaLimit} for full condensation obtained earlier for $\Xi<1$ in \cref{sec:contact1} and compared very well with the simulation results shown earlier in \cref{fig:contact_smallD,fig:contact_smallD_log,fig:contact_method1_D100}.

\subsubsection{Corrections to leading order}
\label{sec:order_parameter3}

The corrections to leading order come from the neglected term
$H_{\text{SC-1}}$ in \cref{eq:H_SC}. Notice that the correction can be
considered as an expansion of small differences of logarithmic radial
distances. Since this value is small, conditioned by
\cref{eq:SC_condition}, we can expand the interacting term in
\cref{eq:H_SC} to second order as performed for the three dimensional
case in previous works by
\citet{PhysRevLett.106.078301,doi:10.1021/jp311873a} and in two
dimensions by \citet{vsamaj2011counter}; we will recall this form as
{\bf 2D--SC-1} which corresponds to first correction to leading
order. Expanding, \eemp
\delta\beta H\approx&\Rpar{2\xi-\Xi (N-1)}\sum_{j=1}^{N}\log\frac{\VVpar{\VEC{r}_j}}{R}\\
&-\Xi \sum_{j<k}\Rpar{\frac{\Bpar{\Rpar{\frac{\VVpar{\VEC{r}_j}}{R}-1}-\Rpar{\frac{\VVpar{\VEC{r}_k}}{R}-1}}^2}{2-2\cos\theta_{jk}}}.\\
\label{eq:H_SC_expanded}
\ffin

Evaluating the profile as done before,
\eemp
\widetilde{\rho}_{\text{SC-1}}(\widetilde{r})=&C_0\int\prod_{j=1}^{N}\dif^2\widetilde{r}_j\, \prod_{j=1}^{N}\Rpar{\frac{\widetilde{R}}{\widetilde{r}_j}}^{2\xi-\Xi (N-1)}\delta(\Ttilde{r}-\Ttilde{r}_1)\\
&\times\Kpar{1+\Xi \sum_{j\neq1}\Rpar{\frac{\Bpar{\frac{\VVpar{\Ttilde{r}_1}}{\widetilde{R}}-1}^2-2\Bpar{\frac{\VVpar{\Ttilde{r}_1}}{\widetilde{R}}-1}\Bpar{\frac{\VVpar{\Ttilde{r}_j}}{\widetilde{R}}-1}}{2-2\cos\theta_{j1}}}},
\ffin
with $C_0$ a constant adjusted for proper normalization and we have kept intentionally only the terms which correspond to $\VEC{r}_1$ and the remaining will be integrated and, therefore, accumulated into the constant. Simplifying,
\eemps
\widetilde{\rho}_{\text{SC-1}}(\widetilde{r})=&\widetilde{\rho}_0\Rpar{\frac{\widetilde{R}}{\widetilde{r}}}^{2\xi-\Xi (N-1)}\int\prod_{j=2}^{N}\dif^2\widetilde{r}_j\, \prod_{j=2}^{N}\Bpar{C_j\Rpar{\frac{\widetilde{R}}{\widetilde{r}_j}}^{2\xi-\Xi (N-1)}}\\
&\times\Kpar{1+\Xi \sum_{j\neq1}\Rpar{\frac{\Bpar{\frac{\widetilde{r}}{\widetilde{R}}-1}^2-2\Bpar{\frac{\widetilde{r}}{\widetilde{R}}-1}\Bpar{\frac{\widetilde{r}_j}{\widetilde{R}}-1}}{2-2\cos\theta_{j1}}}},
\ffins
with,
\emp
C_j=\frac{2(\xi-1)-\Xi }{2\pi \widetilde{R}^2},
\fin
chosen such that,
\emp
2\pi C_j\int_{\widetilde{R}}^{\infty}\Rpar{\frac{\widetilde{R}}{\widetilde{r}}}^{2\xi-\Xi (N-1)}\widetilde{r}\, \dif \widetilde{r}=1.
\fin
Therefore, the density reads,\footnote{Using, from \citet{Gradshteyn2007},
\emps
\sum_{j=1}^{N-1}\frac{1}{2-2\cos\frac{2\pi}{N}j}=\frac{1}{12}\Bpar{N^2-1}.
\fins
}
\eemp
\widetilde{\rho}_{\text{SC-1}}(\widetilde{r})=&\widetilde{\rho}_0\Rpar{\frac{\widetilde{R}}{\widetilde{r}}}^{2\xi-\Xi (N-1)}\times\Kpar{1+\frac{1}{N^2\Xi }\Rpar{\Bpar{\widetilde{r}-\Mean{\widetilde{r}_j}_j}^2-\Bpar{\Mean{\widetilde{r}_j}_j-\widetilde{R}}^2}\sum_{j\neq1}\frac{1}{2-2\cos\theta_{j1}}}\\
=&\widetilde{\rho}_0\Rpar{\frac{\widetilde{R}}{\widetilde{r}}}^{2\xi-\Xi (N-1)}
\times\Kpar{1+\frac{1}{12\Xi}\Bpar{1-\frac{1}{N^2}}\Rpar{\Bpar{\widetilde{r}-\Mean{\widetilde{r}_j}_j}^2-\Bpar{\Mean{\widetilde{r}_j}_j-\widetilde{R}}^2}},
\label{eq:rho1_SC}
\ffin
with
\eemp
\Mean{\widetilde{r}_j}_j=&\frac{2(\xi-1)-\Xi (N-1)}{\widetilde{R}}\int_{\widetilde{R}}^{\infty}\Bpar{\frac{\widetilde{r}_j}{\widetilde{R}}}\Rpar{\frac{\widetilde{R}}{\widetilde{r}_j}}^{2\xi-\Xi (N-1)}\widetilde{r}_j\dif \widetilde{r}_j\\
=&\widetilde{R}\Rpar{1+\frac{1}{2(\xi-1)-(N-1)-\Xi-1}}.
\label{eq:SC_mean_r}
\ffin
The average restricts $\Xi$ in such a way that the denominator of \cref{eq:SC_mean_r} is positive or, equivalently,
\emp
\Xi >\frac{3}{N+1}.
\fin
In order to evaluate the constant $\widetilde{\rho}_0$, one should use
the normalization condition \cref{eq:neutrality-tilde} imposed by the
electroneutrality. Alternatively, one could use the contact theorem
derived for the two dimensional case \citep{MTT14}. Through it, the
contact density $\widetilde{\rho}_0$ gives the same value as in the 
leading order \cref{eq:rho_SC_contact_R}.


\psfrag{AX}{\fontsize{7}{0}\selectfont $r/R$}
\psfrag{BY}{\fontsize{7}{0}\selectfont $\widetilde{\rho}_{\text{SC-1}}-\widetilde{\rho}_{\text{SC-0}}$}
\begin{figure}[htp!]
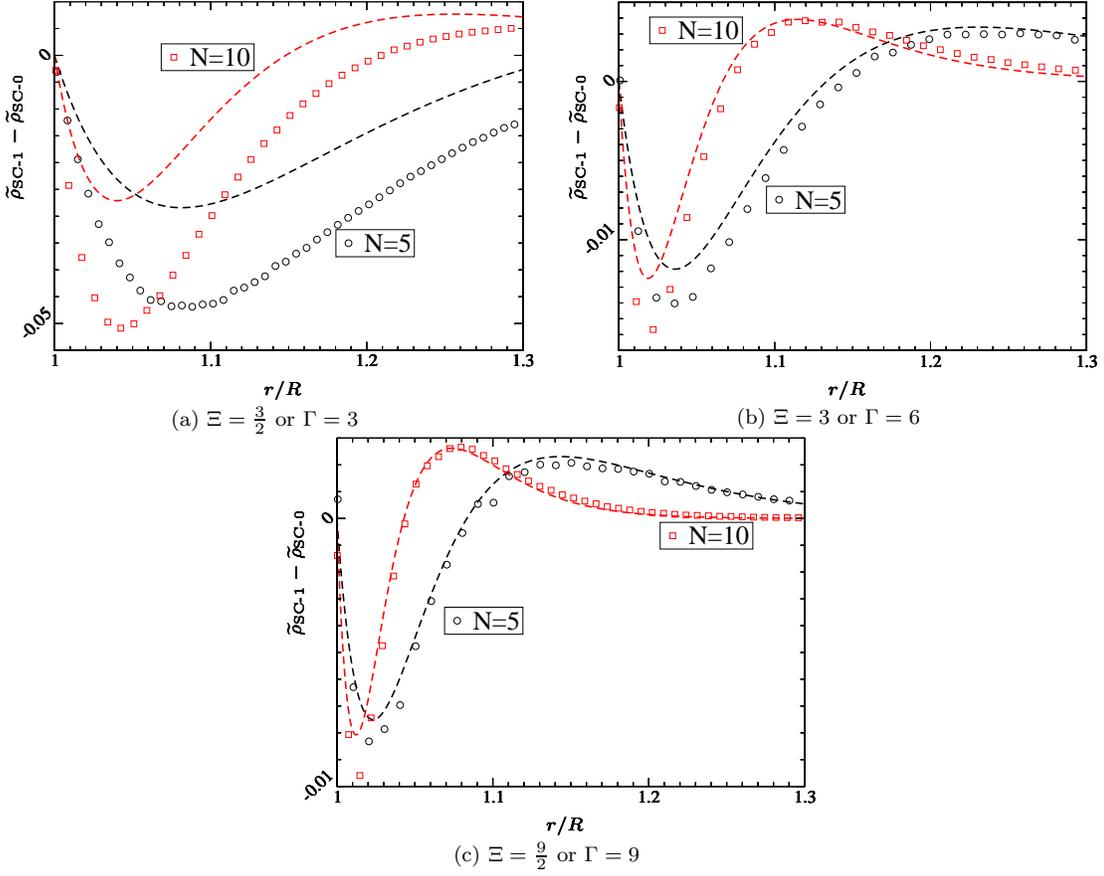

\centering
\subfloat[][$\Xi =\frac{3}{2}$ or $\Gamma=3$]{\includegraphics[width=.4\linewidth]{fig17a.eps}}%
\quad%
\subfloat[][$\Xi =3$ or $\Gamma=6$]{\includegraphics[width=.4\linewidth]{fig17b.eps}}%
\quad
\subfloat[][$\Xi =\frac{9}{2}$ or $\Gamma=9$]{\includegraphics[width=.4\linewidth]{fig17c.eps}}%
\caption[Density profile for $\Xi>1$.]{The density profile
  $\widetilde{\rho}_{\text{SC-1}}$ (\cref{eq:rho1_SC}) as it differs
  with the leading order {\bf 2D--SC-0}
  $\widetilde{\rho}_{\text{SC-0}}$ (\cref{eq:rho0_SC}) as a function
  of the distance near the charged disk for different values of the
  coupling $\Xi$ for $N=5,10$; here $\Delta=20$. The  dashed represent
  the corrected profile {\bf 2D--SC-1} from \cref{eq:rho1_SC}. The
  symbols are numerical results from Monte Carlo simulations.}
\label{fig:profile_method3}
\end{figure}

Our theoretical predictions for the density profile are compared to
Monte Carlo simulations results in \cref{fig:profile_method3} where we
have chosen to plot the difference between {\bf 2D--SC-1}
(\cref{eq:rho1_SC}) and {\bf 2D--SC-0} (\cref{eq:rho0_SC}). We observe
that the prediction {\bf 2D--SC-1} (\cref{eq:rho1_SC}) is increasingly
accurate with higher coupling. We also notice the value of the contact
density is identical in both analytic and numerical profiles as
anticipated from the contact theorem and corroborated from the
extracted contact density value shown in
\cref{fig:contact_smallD,fig:contact_smallD_log,fig:contact_method1_D100}
for the region beyond $\Xi=1$.

\subsubsection{Alternative approach}
\label{sec:alt_rho_SC}

An alternative approach to evaluate the profile, similar to the single
particle variant \citep{PhysRevE.84.041401,doi:10.1021/jp311873a},
stems from a quasi-effective potential assuming that the interaction
between the 1\ST\ particle and the j\TH\ particle can be spanned as the interaction of the former and a particle sitting at $\Mean{\log(r_j/R)}_j$. This statement is compatible with the minimum observed in \cref{fig:profile_method3} that tells that the counter-ions will sit preferentially near it. In other words, the shift of energy due to the j\TH\, charge gives,
\eemp
\delta\beta H_j&=\Xi \log\Bpar{2\cosh\Rpar{\log\frac{\VVpar{\VEC{r}_j}}{R}-\log\frac{\VVpar{\VEC{r}_1}}{R}}-2\cos\theta_{j1}}\\
&\approx\Xi \log\Bpar{2\cosh\Rpar{\Mean{\log\frac{\VVpar{\VEC{r}_j}}{R}}_j-\log\frac{\VVpar{\VEC{r}_1}}{R}}-2\cos\theta_{j1}},\\
\ffin
with the average given by
\eemp
\Mean{\log\frac{\widetilde{r}_j}{\widetilde{R}}}_j=&\frac{2(\xi-1)-\Xi (N-1)}{\widetilde{R}^2}\int_{\widetilde{R}}^{\infty}\log\frac{\widetilde{r}_j}{\widetilde{R}}\Rpar{\frac{\widetilde{R}}{\widetilde{r}_j}}^{2\xi-\Xi (N-1)}\widetilde{r}_j\dif \widetilde{r}_j\\
=&\frac{1}{2(\xi-1)-\Xi(N-1)}.
\label{eq:SC_mean_log}
\ffin

\begin{figure}[htp!]
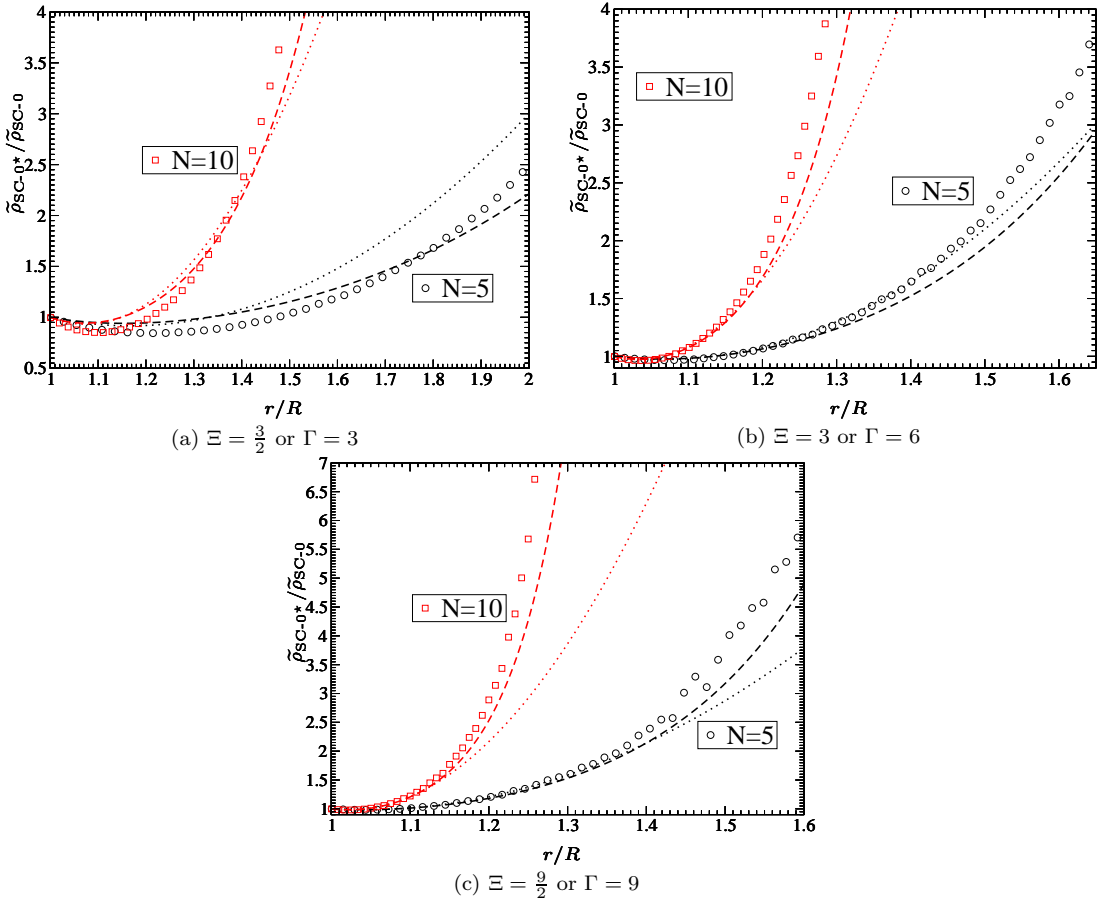

\centering
\subfloat[][$\Xi =\frac{3}{2}$ or $\Gamma=3$]{\includegraphics[width=.4\linewidth]{fig18a.eps}}%
\quad%
\subfloat[][$\Xi =3$ or $\Gamma=6$]{\includegraphics[width=.4\linewidth]{fig18b.eps}}%
\quad
\subfloat[][$\Xi =\frac{9}{2}$ or $\Gamma=9$]{\includegraphics[width=.4\linewidth]{fig18c.eps}}%
\caption[Density profile for $\Gamma>2$ or $\Xi>1$.]{The ratio of density profile $\widetilde{\rho}_{\text{SC-0}^\star}$ (\cref{eq:rho1star_SC}) and the leading order {\bf 2D--SC-0} as a function of the distance near the charged disk for different values of the coupling $\Xi$ for $N=5,10$; here $\Delta=20$. The dotted curves represent the corrected analytic term {\bf 2D--SC-1} from \cref{eq:rho1_SC} while the dashed represent the heuristic alternative {\bf 2D--SC-$0^\star$} from \cref{eq:rho1star_SC}.}
\label{fig:profile_method3_ratio}
\end{figure}

Then, the density gives,
\eemp
\widetilde{\rho}_{\text{SC-0}^\star}(\widetilde{r})=&\widetilde{\rho}_0\Rpar{\frac{\widetilde{R}}{\widetilde{r}}}^{2\xi-\Xi(N-1)}
e^{\Xi \sum_{j\neq1}\Kpar{\log\Bpar{2\cosh\Rpar{\log\frac{\widetilde{r}}{\widetilde{R}}-\Mean{\log\frac{\widetilde{r}}{\widetilde{R}}}}-2\cos\theta_{j1}}-\log\Bpar{2\cosh\Rpar{\Mean{\log\frac{\widetilde{r}}{\widetilde{R}}}}-2\cos\theta_{j1}}}},
\label{eq:rho1star_SC}
\ffin
with the contact density $\widetilde{\rho}_0$ at $r=R$ given by \cref{eq:rho_SC_contact_R}. The model compares quite good to the Monte Carlo data shown in \cref{fig:profile_method3_ratio} intentionally drawn as the ratio of the {\bf 2D--SC-$0^\star$} (\cref{eq:rho1star_SC}) and {\bf 2D--SC-0} (\cref{eq:rho0_SC}) for different couplings to bolster the deviations at long ranges. We observe from the figure that the profiles depart from the leading order and tend to {\bf 2D--SC-$0^\star$} as expected since the practical alternate approach is a construct proposed to match better the behavior for large distances. This approach has been used in previous occasions in the works by \citet{0295-5075-98-3-36004,0295-5075-100-5-56005,PhysRevB.85.205131,doi:10.1021/jp311873a} successfully.

\psfrag{AX}{\fontsize{7}{0}\selectfont $r/R$}
\psfrag{BY}{\fontsize{7}{0}\selectfont $\widetilde{\rho}_{\text{SC-0}^\star}/\widetilde{\rho}_{\text{SC-0}}$}

\subsection{Crystallization and energy}
\label{sec:crystal_energy_SC}

\psfrag{AX}{$\theta$}
\psfrag{BY}{$g_{\theta}$}
\psfrag{BY2}{$g_{\theta}$}
\ImTc{0.38}{fig19}{profile_method3_theta}{The correlation
  function $g_{\theta}$ as a function of the angle $\theta$; here, the
  number of particles for the upper plot is $N=5$ and $N=10$ for the
  other. 
  The data was obtained by means of Monte Carlo simulations.
The normalization is enforced with $g_\theta\to1$ for weakly correlated as in the small coupling cases.}{Angular correlation function for $\Gamma>2$ or $\Xi>1$ in log scale.}

The success of the hypothesis we constructed to derive the profile was build upon the existence of the ground state, which can be further viewed through the angular correlation function from \cref{fig:profile_method3_theta}. We can appreciate that counter-ions sit at equally spaced angular distances which is more pronounced for larger couplings as was anticipated from the constraint (\cref{eq:SC_condition}).

A guarantee to further stress the conditions comes from the energy. Since the system tends to a crystalline state, we are able to determine the energy of the system enforcing the ground state plus a contribution from the disk. In other words,
\eemp
\beta H\approx&\Rpar{2\xi-\Xi (N-1)}\sum_{j=1}^{N}\log\frac{\VVpar{\VEC{r}_j}}{R}-\Xi N\log N+E_0\\
\ffin
coming from \cref{eq:H-GS,eq:H_SC}, which gives for the partition function

\emp
\mathcal{Z}_N\approx\Bpar{\frac{e^{\Xi\log N-E_0/N}}{2(\xi-1)-\Xi (N-1)}}^N,
\fin
that tells us that the energy per particle behaves as,
\emp
\widetilde{E}
=
\frac{\langle \cal H \rangle}{N}
=\frac{2\xi-\Xi (N-1)}{2(\xi-1)-\Xi (N-1)}-\Xi \log \frac{N L}{R}.
\label{eq:energy3}
\fin

In order to compare we turn to \cref{fig:energy_method3} where the agreement is quite good despite all the simplifications. The fact that the energy is well described validates the original hypothesis of the ground state which helped us factorize the Hamiltonian.

\psfrag{AX}{$\xi$}
\psfrag{BY}{$\widetilde{E}/\Delta$}
\ImTc{0.38}{fig20}{energy_method3}{The energy as a function
  of the Manning parameter; here $N=5,10$ and $\Delta=20,10^2$. The
  dashed curve corresponds to the prediction from \cref{eq:energy3}
  and the symbols where obtained from Monte Carlo simulations.}{The energy as a function of $\xi$ for $N=5,10$ and $\Delta=20,10^2$ at strong couplings.}

\section{Conclusion}

Condensation in two dimensions differs greatly from its three
dimensional counterpart. In three dimensions the fraction of condensed
ions can be larger than the ideal fraction $f_M=1-1/\xi$ at strong
couplings for small box sizes ($10\leq \Delta \leq 100$)
\citep{doi:10.1021/jp311873a}. On the contrary, in two dimensions the
fraction of condensed ions is not affected by $\Delta$. Here, the phenomenon relies on the ability to bind a number of ions such that the effective charge of the ion cloud and the disk is greater than unity (in the same dimensionless scale as $\xi$). This means that if a candidate to condense reduces the effective charge to unity or less, the cloud is unable to bind it. Naturally this shows that in two dimensions counter-ions condense at specific temperatures. The number of unbound ions can be determined precisely as $\Floor{(1+\xi_B)/\Xi}$. This result, surprisingly, is valid for arbitrary couplings even though we concluded such effect starting from the weakly coupled case, or $\Xi<1$.

Besides condensation, there are other interesting features of these systems such as the ion density profile. It has been known for a long time that the general problem cannot be solved analytically for all couplings; however,  following Burak and Orland's \citep{PhysRevE.73.010501}, we were able to estimate the behavior pertaining the weakly coupled regime. For instance,  the leading behavior is given by that of mean field at short distances, \emph{i.e.} $\rho\sim 1-2x/\mu$ with $x$ the perpendicular distance from the disk, different from that of strong couplings where the behavior is mediated by an effective Gouy length $\mu_{\eff}=(\xi+\Xi)/2$ as a result of the screening of the neighbouring ions. Even so, we recovered the mean field infinite dilution profile proceeding from the approximations.

It was interesting to see the non-mean field-like behavior at large distances unlike in three dimensions \citep{doi:10.1021/jp311873a}. A trait that follows from the logarithmic potential in two dimensions. At the onset for condensation an ion is bound creating a shift of energy that equates to $2\Delta$; similar to the shift of entropy for confining a particle. We found that this ion at the critical point is neither bound nor free. The profile that best describes the behavior of such ion is that of an ion interacting with a disk of unit charge (or $\xi = 1$) decaying in a powerlaw-shape as $1/r^2$; hence, to the integrated charge profile appears as a line with slope $1/\Delta$ (see \cref{fig:charge_method1,fig:charge_method2,fig:lambda2D_SC_G4_N13_D100}).

%
In addition, we verified the results for the contact theorem
\citep{MTT14} for the corresponding two dimensional system. Presuming
of the validity of the approximation in the weakly coupled case, it
was possible to anticipate the behavior of the value of the density at
contact for small box sizes as well as recovered the expressions that
correspond to $\Delta\to\infty$.

For intermediate and particular values of the coupling ($\Xi$
integer), we where able to obtain exact analytic results for the
partition function and density profile of the system. These results
support some predictions observed in the low coupling regime, and
provide a bridge between the low and strong coupling regimes. The
analytic structure of the density profile shows some interesting
features in which the separation between condensed and unbind ions
could be clearly observed.

Ultimately, we addressed the strong coupling regime. Unlike in the line case \citep{vsamaj2011counter} where the leading behavior came from the interaction between the wall and the ion, for disks or curved surfaces, as discussed in \citep{MTT14}, the structure of the double layer is the crucial. Such arrangement, that in the disk resembles a pebbled necklace, contributes in such a way that it modifies the scale at which the profile decays by a factor of 2.

Appropriately,  the strong coupling regime here reads as $\Xi\gg1$ with $N>1$. As should be expected,  this structure effect reproduces appropriately the planar limit and agrees with the contact theorem. Furthermore, the corrections steming from fluctuations of the ground-state,  the milestone for the strong coupling Wigner approach,  both show very good agreement with the numerical results and known results for the line \citep{vsamaj2011counter}.

The authors would like to thank L. \v{S}amaj and E. Trizac for their
support and comments preceding this work. We acknowledge partial
financial support from ECOS-Nord/COLCIENCIAS-MEN-ICETEX and from Fondo
de Investigaciones, Facultad de Ciencias, Universidad de los Andes.

\appendix
\section{Partition function in the degenerate cases}
\label{sec:PF_deg}

The form of the $a_j$ suggests that we can study the problem approaching to a degenerate scenario. In simplified form $a_j=(j-x)^2$ with $x$ a variable containing the parameters of the system ($\xi$, $\Xi$, $\xi_B$ and $N$). If at a value $x$ the set displays degeneracy then let us look at $x^\prime=x+\delta x$. Hence,
\emps
a_{k}^{\prime}=(k-x^\prime)^2=(k-x)^2-2\delta x(k-x)+{\delta x}^2 \overset{\delta x\to0}{=} a_k-2\delta x(k-x)+\mathcal{O}({\delta x}^2),
\fins
where $\{a_{j}^{\prime}\}$ is non-degenerate but $\{a_{j}\}$ is; the prime notation will proceed throughout referring to quantities in the non-degenerate case. Let $\{j^\dag\}$ represent the set of degenerate indeces of $\{a_{j}\}$. \Cref{eq:laplace_Z_non-deg} reads,
\eemp
\mathcal{T}_{\Kpar{f_N\otimes\cdots\otimes f_0}}^{\Bpar{\Xi\Delta}}(s)&=\sum_{k=0}^{N}C_{k}^{\prime}\frac{1}{s+a_{k}^{\prime}}=\sum_{k\notin\{j^\dag\}}C_{k}^{\prime}\frac{1}{s+a_{k}^{\prime}}+{\sum_{k\in\{j^\dag\}}}^\star \Kpar{C_{k}^{\prime}\frac{1}{s+a_{k}^{\prime}}+C_{k^\dag}^{\prime}\frac{1}{s+a_{k^\dag}^{\prime}}},\\
\label{eq:laplace_Z_non-deg-demo1}
\ffin
where,
\eemp
\frac{1}{s+a_{k}^{\prime}}\overset{\delta x\to0}{=}\frac{1}{s+a_k}\Bpar{1+\frac{2\delta x(k-x)}{s+a_k}+\mathcal({\delta x}^2)},
\ffin
and if $k\in\{j^\dag\}$
\eemp
C_{k}^{\prime}&=\prod_{l=0,l\neq k}^{N}\frac{1}{a_{l}^{\prime}-a_{k}^{\prime}}
=\frac{1}{2\delta x(k-k^\dag)}\Kpar{\prod_{l=0,l\neq k,k^\dag}^{N}\frac{1}{a_{l}-a_{k}+2\delta x(k-l)}}\\
&\overset{\delta x\to0}{=}\frac{1}{2\delta x(k-k^\dag)}\Kpar{\prod_{l=0,l\neq k,k^\dag}^{N}\frac{1}{a_{l}-a_{k}}}\Bpar{1-2\delta x\sum_{l=0,l\neq k,k^\dag}^{N}\frac{k-l}{a_l-a_k}+\mathcal{O}({\delta x}^2)}\\
&\overset{\delta x\to0}{=}C_{k,k^\dag}\Bpar{\frac{1}{2\delta x(k-k^\dag)}-\frac{1}{k-k^\dag}\sum_{l=0,l\neq k,k^\dag}^{N}\frac{k-l}{a_l-a_k}+\mathcal{O}({\delta x})}.
\ffin

Substituting in \cref{eq:laplace_Z_non-deg-demo1},
\eemps
\mathcal{T}_{\Kpar{f_N\otimes\cdots\otimes f_0}}^{\Bpar{\Xi\Delta}}(s)
\overset{\delta x\to0}{=}&\sum_{k\notin\{j^\dag\}}C_{k}\Rpar{\frac{1}{s+a_{k}}+\mathcal{O}(\delta x)}\\
&+{\sum_{k\in\{j^\dag\}}}^\star\left\{\frac{C_{k,k^\dag}}{s+a_k}\Bpar{\frac{1}{2\delta x(k-k^\dag)}-\frac{1}{k-k^\dag}\sum_{l=0,l\neq k,k^\dag}^{N}\frac{k-l}{a_l-a_k}+\frac{k-x}{(k-k^\dag)(s+a_k)}+\mathcal{O}({\delta x})}+\underset{k\leftrightarrow k^\dag}{\text{i.d.}}\right\},\\
\ffins
simplifies to the following, knowing that $a_k=a_{k^\dag}$,
\eemp
\mathcal{T}_{\Kpar{f_N\otimes\cdots\otimes f_0}}^{\Bpar{\Xi\Delta}}(s)
\overset{\delta x\to0}{=}&\sum_{k\notin\{j^\dag\}}C_{k}\Rpar{\frac{1}{s+a_{k}}+\mathcal{O}(\delta x)}+{\sum_{k\in\{j^\dag\}}}^\star\frac{C_{k,k^\dag}}{s+a_k}\Bpar{\frac{1}{s+a_k}-S_{k,k^\dag}+\mathcal{O}({\delta x})},
\label{eq:laplace_Z_non-deg-demo2}
\ffin
proving that the Laplace transform function in the degenerate cases are a limit of the non-degenerate case. An identical procedure can be followed for the partition function yielding that,
\emp
\mathcal{Z}\overset{\delta x\to0}{\propto}\sum_{k\notin\{j^\dag\}}C_ke^{-a_k\Xi\Delta}\Rpar{1+\mathcal{O}(\delta x)}+{\sum_{k\in\{j^\dag\}}}^\star C_{k,k^\dag}\, \Bpar{\Xi\Delta-S_{k,k^\dag}+\mathcal{O}(\delta x)}\, e^{-a_k\Xi\Delta},
\fin
which applies to both degenerate cases.

\section{Mean field limit}
\label{sec:MF1}

In the analysis of \cref{sec:method1}, when $\Xi<1$, an important result that
should be recovered is that of mean field; met only when $\Xi\to0$
that for constant $\xi$ and $\xi_B$ equates to $N\to\infty$. Our
solution to the profiles predicted for weak couplings must be
consistent with that of mean field, presented for infinite dilution by
\citet{PhysRevE.73.056105,doi:10.1021/jp311873a},
i.e. $\Delta\to\infty$, as follows, \emp
\widetilde{\rho}_{MF}(\widetilde{r})=f_{M}^{2}\Rpar{\frac{\widetilde{R}}{\widetilde{r}}}^2\frac{1}{\Rpar{1+(\xi-1)\log\frac{\widetilde{r}}{\widetilde{R}}}^2}.
\label{eq:MF_DeltaInf}
\fin

From \cref{eq:rhotilde-exact-Gl2,eq:approx_rho_T_R}, with $y^\prime=\log(r/R)$ (remember that $y=\Xi\log(r/R)$),
\eemp
\widetilde{\rho}(y^\prime)=&\frac{\Xi\, e^{-2y^\prime}}{N\xi}\sum_{k=1}^{j^\star}\Kpar{\sum_{j=0}^{k-1}\frac{C_{0,k-1;j}}{C_{0,k-1;j^\star}}}e^{-(a_j-a_{j^\star})\Xi y^\prime}\\
=&\frac{e^{-2y^\prime}}{\xi}\frac{\Xi}{N}\sum_{j=0}^{j^\star-1}{\Kpar{\sum_{k=j+1}^{j^\star}\frac{C_{0,k-1;j}}{C_{0,k-1;j^\star}}}e^{-(a_j-a_{j^\star})\Xi y^\prime}}.\\
\label{eq:meanfield-rho-start}
\ffin

The thermodynamic limit, as mentioned earlier, corresponds to $N\to\infty$, which by transitive definition is equally inherited by $j^\star$. However the way we approach this limit should avoid any divergences coming from degeneracies or in other words we should avoid the case of $a_{j^\star}=a_{j^\star-1}$, corresponding to $2(1+\xi_B)/\Xi$ odd (\cref{subsec:degeneracy_case2}). For that matter the simplest of all approaches consists of the case where $a_{j^\star}=0$ or $\Xi=(\xi-1)/(j^\star-1/2)$. Through this approach, $a_j=(j-j^\star)^2$, allowing us to simplify the previous equation for the density to,
\eemp
\widetilde{\rho}(y^\prime)
=&\frac{e^{-2y^\prime}}{\xi}\frac{\Xi}{N}\sum_{j=0}^{j^\star-1}{\underbrace{\Kpar{\sum_{k=j}^{j^\star-1}\Rpar{\prod_{l=0,l\neq j}^{k}\frac{a_l-a_{j^\star}}{a_l-a_j}}}}_{\sum_{k=j}^{j^\star-1}\Rpar{\prod_{l=0,l\neq j}^{k}\frac{(l-j^\star)^2}{(l-j)(l+j-2j^\star)}}}(a_j-a_{j^\star})\, e^{-(j-{j^\star})^2\Xi y^\prime}},\\
\ffin
that changing $j\to m $ such that $j=j^\star-m$,
\eemp
\widetilde{\rho}(y^\prime)
=&\frac{e^{-2y^\prime}}{\xi}\frac{\Xi}{N}\sum_{m=1}^{j^\star}{\Kpar{\sum_{k=j^\star-m}^{j^\star-1}\Rpar{\prod_{l=0,l\neq j^\star-m}^{k}\frac{(l-j^\star)^2}{(l+m-j^\star)(l-m-j^\star)}}}m^2\, e^{-m^2\Xi y^\prime}},\\
\ffin
$k\to n$ such that $k=j^\star-n$,
\eemp
\widetilde{\rho}(y^\prime)
=&\frac{e^{-2y^\prime}}{\xi}\frac{\Xi}{N}\sum_{m=1}^{j^\star}{\Kpar{\sum_{n=1}^{m}\Rpar{\prod_{l=0,l\neq j^\star-m}^{j^\star-n}\frac{1}{1-\frac{m^2}{(l-j^\star)^2}}}}m^2\, e^{-m^2\Xi y^\prime}},\\
\ffin
and $l\to p$ such that $l=j^\star-p$,
\eemp
\widetilde{\rho}(y^\prime)
=&\frac{e^{-2y^\prime}}{\xi}\frac{\Xi}{N}\sum_{m=1}^{j^\star}{\Kpar{\sum_{n=1}^{m}\Rpar{\prod_{p=n,p\neq m}^{j^\star}\frac{1}{1-\frac{m^2}{p^2}}}}m^2\, e^{-m^2\Xi y^\prime}}.
\ffin

Analyzing separately each term we learn that in the thermodynamic limit the sum approaches to an infinite series whose behavior can be determined asymptotically. For instance,
\emp
\sum_{j=1}^{m}\Rpar{\prod_{k=j,k\neq m}^{n}\frac{1}{1-\frac{m^2}{k^2}}}\overset{n\to\infty}{=}\, 2m\, e^{-m^2/n}.
\fin
Substituting the coupling in this limit -- i.e. $\Xi\simeq(\xi-1)/j^\star$ -- yields,
\eemp
\widetilde{\rho}(y^\prime)
\overset{j^\star\to\infty}{=}&\, f_M\frac{j^\star}{N}\, e^{-2y^\prime}\Bpar{\frac{2}{(j^\star)^2}\sum_{m=1}^{j^\star}{m^3\, e^{-m^2/j^\star(1+(\xi-1)y^\prime)}}},
\ffin
which has, once more, a careful convergence given by,
\eemp
\lim_{n\to\infty}\frac{2}{n^2}\sum_{j=1}^{n}{j^3\, e^{-(j^2/n)\, x}}=\frac{1}{x^2},
\ffin
into the final form for the density as,
\eemp
\widetilde{\rho}(\widetilde{r})=&\frac{f_{M}^2}{1+\frac{\xi_B}{\xi}}\Rpar{\frac{\widetilde{R}}{r}}^2\frac{1}{\Rpar{1+(\xi-1)\log\frac{\widetilde{r}}{\widetilde{R}}}^2},
\ffin
a result identical to the predicted, but general to arbitrary charge at the exterior boundary.

\bibliographystyle{plainnat}
\bibliography{biblio_art}

\end{document}